\documentclass[preprint,noshowpacs,noshowkeys,floatfix,aps,pra]{revtex4-2}
\usepackage{graphicx}
\usepackage[dvipsnames]{xcolor}
\usepackage{amssymb}
\usepackage{amsmath}
\usepackage{stmaryrd}
\usepackage{tikz}
\usetikzlibrary{shapes.geometric,shapes.misc,positioning,calc}
\definecolor{Blue}  {RGB}{  0, 100, 200}
\definecolor{Green} {RGB}{150, 200,   0}
\definecolor{Orange}{RGB}{250, 150,   0}
\usepackage{caption}
\usepackage{subcaption}

\newtheorem{definition}{Definition}

\usepackage{mathtools}
\DeclarePairedDelimiter\ceil{\lceil}{\rceil}
\DeclarePairedDelimiter\floor{\lfloor}{\rfloor}

\newcommand{\rank}[1]{\text{rank}(#1)}

\begin{document}

\title{Quantum dynamics of coupled excitons and phonons in chain-like systems: tensor train approaches and higher-order propagators}

\author{Patrick Gel\ss{}}
\affiliation{
Institut f\"{u}r Mathematik, Freie Universit\"{a}t Berlin \\ Arnimallee 3--9, D-14195 Berlin, Germany}
\affiliation{Zuse-Institut Berlin, Takustraße 7, D-14195 Berlin, Germany}

\author{Sebastian Matera}
\affiliation{
Institut f\"{u}r Mathematik, Freie Universit\"{a}t Berlin \\ Arnimallee 3--9, D-14195 Berlin, Germany}
\affiliation{Fritz-Haber-Institut der Max-Planck-Gesellschaft, Faradayweg 4--6, D-14195 Berlin, Germany}

\author{Rupert Klein}
\affiliation{
Institut f\"{u}r Mathematik, Freie Universit\"{a}t Berlin \\ Arnimallee 3--9, D-14195 Berlin, Germany}

\author{Burkhard Schmidt}
\email{burkhard.schmidt@fu-berlin.de}
\affiliation{
Institut f\"{u}r Mathematik, Freie Universit\"{a}t Berlin \\ Arnimallee 3--9, D-14195 Berlin, Germany}
\affiliation{Weierstra\ss -Institut f\"{u}r Angewandte Analysis und Stochastik, Mohrenstra\ss e 39, D-10117 Berlin, Germany}

\date{\today}

\begin{abstract}
We investigate tensor-train approaches to the solution of the time-dependent Schr\"{o}dinger equation for chain-like quantum systems with on-site and nearest-neighbor interactions only.
Using efficient low-rank tensor train representations, we aim at reducing memory consumption and computational costs.
As an example, coupled excitons and phonons modeled in terms of Fr\"{o}hlich-Holstein type Hamiltonians are studied here.
By comparing our tensor-train based results with semi-analytical results, we demonstrate the key role of the ranks of the quantum state vectors.
Typically, an excellent quality of the solutions is found only when the maximum number of ranks exceeds a certain value. 
One class of propagation schemes builds on splitting the Hamiltonian into two groups of interleaved nearest-neighbor interactions commutating within each of the groups.
In particular, the 4-th order Yoshida-Neri and the 8-th order Kahan-Li symplectic composition yield results close to machine precision.
Similar results are found for 4-th and 8-th order global Krylov scheme.
However, the computational effort currently restricts the use of these four propagators to rather short chains which also applies to propagators based on the time-dependent variational principle, typically used for matrix product states.
Yet another class of propagators involves explicit, time-symmetrized Euler integrators.
Especially the 4-th order variant is recommended for quantum simulations of longer chains, even though the high precision of the splitting schemes cannot be reached.
Moreover, the scaling of the computational effort with the dimensions of the local Hilbert spaces is much more favorable for the differencing than for  splitting or variational schemes.

\end{abstract}

\maketitle

\section{Introduction}
\label{sec:introduction}
Excitonic energy transport (EET), i.e., the transport of electronic energy without net charge is of paramount importance in many biological and technological systems.
In particular, the efficiency of the conversion of light into chemical or electric energy crucially depends on it~\cite{Mikhnenko2015,Kranz2016}.
For example, in photovoltaic devices the transport of energy from the region of primary excitation to the charge separating interfaces has to be fast enough to compete with radiative and non-radiative decay channels, most prominently the recombination of electrons and holes.
The efficiency of EET is strongly affected by the coupling of electronic excitations to (intramolecular and/or intermolecular) vibrational degrees of freedom~\cite{Schroter2015,Zhugayevych2015}.
The simplest theoretical approaches to this exciton-phonon coupling (EPC) are based on the assumption of chain-like, quasi-1D arrangements.
Examples relevant for semiconducting materials are lattice vibrations in 1D molecular crystal models or various stretching, bending, and/or torsional motions in conjugated polymer chains.
The "dressing" of excitons by vibrations in these classes of systems is often described in terms of Holstein or Fr\"{o}hlich type Hamiltonians.
Originally developed for polarons~\cite{Devreese2009}, these model Hamiltonians, or extensions thereof, are routinely used in a variety of applications, ranging from vibrational energy transport in helical proteins~\cite{Davydov1985,Scott1992,Georgiev2019} up to electronic energy transport in conjugated polymer chains~\cite{Binder2018,DiMaiolo2020}.
Despite of the apparent simplicity of the Fr\"{o}hlich-Holstein type systems, their quantum dynamics still poses many open questions, as noted in the abstract of Ref.~\cite{Devreese2009}: 
"It is remarkable how the Fr\"{o}hlich polaron ... has resisted full analytical or numerical solution ... since $\sim$1950, when its Hamiltonian was first written.
The field has been a testing ground for analytical, semi-analytical and numerical techniques, such as path integrals, strong-coupling perturbation expansion, advanced variational, exact diagonalization (ED) and quantum Monte Carlo (QMC) techniques."
Hence, the study of Fr\" {o}hlich-Holstein type systems as models for exciton dynamics in the presence of EPC remains an intriguing and timely research field.

Essentially, the problems of theoretical studies of EPC are due to the quantum dynamical treatment required for the above-mentioned models, i.e., solutions of time-dependent Schr\"{o}dinger equations (TDSE) in many dimensions.
Quantum dynamical simulations to solve the TDSE with conventional methods eventually become unfeasible when increasing the number of dimensions~\cite {Kosloff1988}. 
One way to circumvent this so-called \textit{curse of dimensionality} is to resort to mixed (or hybrid) quantum-classical methods.
There, only the electronic degrees of freedom are treated quantum-mechanically, while the nuclear (ionic) degrees of freedom are treated classically, typically justified by the small electron to ion mass ratio~\cite{Lenz1951}.
The most basic of these hybrid methods is the quantum-classical Ehrenfest approach which plays an important role, e.g., in the Davydov soliton theory for EET in helical proteins~\cite{Davydov1985,Scott1992,Georgiev2019}.
Apart from the classical approximation, the main limitation of Ehrenfest dynamics is that it builds on a mean-field coupling of the two sub-systems~\cite{Bornemann1996,Nettesheim1996,Choi2021}.
This bottleneck is overcome in surface hopping trajectory approaches, where non-adiabatic transitions are approximately treated by stochastic hopping of trajectories between different (diabatic or adiabatic) energy states~\cite{Tully1990}.
Even though surface hopping trajectory techniques have lately been used a lot in studying the photo-induced physics and chemistry of EET~\cite{Xie2020,Nelson2020,Freixas2021}, the validity of the underlying approximations of quantum-classical dynamics is not always clear.
In particular, for EET in organic semiconductors, there is sometimes no clear distinction between slow and fast degrees of freedom~\cite{Zhugayevych2015}. 

The focus of the present work is on methods for fully quantum-mechanical approaches to treat coupled excitons and phonons.
In the literature, reports of such simulations are scarce.
One example are quantum dynamical investigations of the related problem of  formation of polarons in crystals where variational techniques for representing quantum state vectors were used~\cite{Trugman2004,Ku2007}.
In more recent work, structure and dynamics of excitons in $\pi$-conjugated polymers were investigated, with emphasis on coupling to the torsional dynamics of the chain~\cite{Binder2018,DiMaiolo2020}.
In those studies, the multi-configurational time-dependent Hartree (MCTDH) method and its multi-layer (ML) extension were used, which are known as highly efficient algorithms for quantum dynamics in high dimensions~\cite{Beck2000,Meyer2009}.

In recent years, the interest in low-rank tensor decompositions has been growing steadily. 
The potential of different tensor formats like the canonical format~\cite{Hitchcock1927}, the Tucker format~\cite{Tucker1964}, and particularly the tensor-train (TT, aka matrix product states, MPS) format~\cite{Oseledets2011} for mitigating the curse of dimensionality has been shown in various application areas.
So far, tensor decomposition techniques have been used in the fields of, e.g., quantum mechanics~\cite{Orus2014, Borrelli2016, Bose2021, Gelss2022b}, neuro-imaging~\cite{Karahan2015, Chatzichristos2019, Erol2022}, system identification~\cite{Klus2018, Gelss2019, Kargas2020a}, and supervised learning~\cite{Stoudenmire2016, Klus2019, Kargas2020b}, just to name a few.
In previous work, we proposed TT-based techniques for numerically solving the time-independent Schr\"{o}dinger equations (TISE) for coupled excitons and phonons~\cite{Gelss2022a}. 
There, it has been demonstrated that high-dimensional quantum systems comprising only NN interactions may be simulated and analyzed without an exponential scaling of the memory consumption and/or computational effort by using so-called SLIM decompositions~\cite{Gelss2017}.
Note that the underlying TT representations are a special case of the hierarchical Tucker format~\cite{Hackbusch2012}, which is equivalent to the wave function representation in the multi-layer (ML) variant of MCTDH. 
In recent comparisons of the efficiency of both methods for calculating ground states of chains of molecular rotors, it has been found, that  TT/MPS techniques are less time- and memory-consuming than the state-of-the-art implementation of ML-MCTDH~\cite{Mainali2021,Serwatka2022}.

In the present work, we want to exploit low-rank TT decompositions of quantum state vectors and Hamiltonian matrices for the case of the time-dependent Schr\"{o}dinger equation (TDSE).
In addition to previously published time integrators for TTs using projectors onto the tangent space of the tensor manifold and/or building on the time-dependent  variational  principle~\cite{Lubich2014,Lubich2015,Haegeman2016}, also symmetric Euler differencing schemes~\citep{Choi2019} and split operator techniques~\citep{Feit1982,Roulet2019} have been suggested recently for use with TT representations~\citep{Orus2014,Volokitin2019,Paeckel2019}.
Being explicit and easy to implement, these two integrators have been in use since the early days of numerical quantum dynamics~\citep{Leforestier1991}.
Our goal is to investigate them along with their higher-order variants, see Sec.~\ref{sec:numerics}.
There, the various time integrators are subjected to rigorous tests for realistic models of excitons, phonons, and exciton-phonon-coupling.
To that end, we are not only considering their ability to conserve norm and energy but also how accurately they can reproduce semi-analytic propagations which are available for special cases, see Sec.~\ref{sec:results}.
On the one hand, our focus will be on the scaling of the numerical effort with the chain length and the Hilbert space dimensions, thus assessing the propagators for use in simulations of large systems. 
On the other hand, we will elucidate on the use of higher-order methods in comparison to lower-order ones  in the context of low-rank TT representations of high-dimensional wave functions. 
In particular, we will address the question whether the increased costs for each propagation step will be compensated by the possibility to use larger step sizes, and, hence, a smaller number of steps, while obtaining solutions with the same accuracy.
Note that implicit schemes such as the trapezoidal rule or the midpoint rule are not considered here. 
Even though they offer the advantage of conserving energy, their computational effort is unfavorable because they involve solutions of large-scale linear systems of equations.
While this is in principle possible for TTs by virtue of the alternating linear scheme~\cite{Holtz2012}, which is also used in our previous TISE study~\citep{Gelss2022a}, it is computationally too expensive to do this in every time step.

\section{Model Hamiltonians}
\label{sec:model}
In the present work, we use Fr\"{o}hlich-Holstein type Hamiltonians to describe the dynamics of excitons and phonons.
With the coupling of the two subsystems assumed to be linear, these Hamiltonians can be considered as the simplest models of EPC in organic semiconductors~\cite{Schroter2015,Zhugayevych2015}.
Originally developed for electrons in a polarizable lattice \cite{Devreese2009},
these models, partly with minor modifications, are also known as Huang-Rhys, or Peierls models.
There is also a close analogy to Davydov models describing the interaction of the amide I vibrations with hydrogen bond stretching in $\alpha$-helices of proteins~\cite{Scott1992,Georgiev2019}. 

Here, we restrict ourselves to the consideration of effectively one-dimensional, linear chains comprising $N$ exciton-supporting sites assuming NN interactions only, see also our previous work~\cite{Gelss2022a}.
Such models are suitable, e.~g., for the description of two classes of systems~\cite{Mikhnenko2015,Hedley2016} in the context of photovoltaics applications.
First, in conjugated polymer chains without major kinks or turns, the excitonic sites interact with each other mainly via a Dexter mechanism ("through bond")~\citep{Dexter1953}.
Second, in crystals of polycyclic aromatic molecules, the coupling of excitons occurs mainly in the direction perpendicular to the planes of the stacked molecules, governed by the F{\"o}rster mechanism ("through space")~\citep{Dexter1969}.

The total Hamiltonians for the coupled excitons and phonons can be written as
\begin{equation}
	H = H^{\mathrm{(ex)}} \otimes \mathbb{I}^{\mathrm{(ph)}} 
	+ \mathbb{I}^{\mathrm{(ex)}} \otimes H^{\mathrm{(ph)}} 
	+ H^{\mathrm{(epc)}} 
	\label{eq:H_total}
\end{equation}
with the superscripts (ex), (ph), and (epc) standing for excitons, phonons and their coupling, respectively, while the $\mathbb{I}$ are the respective identity operators.
Note that atomic units with $\hbar=1$ are used throughout this work.

Restricting ourselves to a chain of two-state-systems, the excitonic Hamiltonian for a (linear or cyclic) system of $N$ (identical or not) sites is given in terms of (bosonic) exciton raising, $b_i^\dagger$, and lowering, $b_i$, operators
\begin{equation}
	H^{\mathrm{(ex)}} = \sum_{i=1}^N \alpha_i
	                    b_i^\dagger b_i
	                  + \sum_{i=1}^{N} \beta_i
										  \left(
											b_i^\dagger b_{i+1} + 
											b_i b_{i+1}^\dagger
											\right)
	\label{eq:H_ex}
\end{equation}
with local ("on site") excitation energy $\alpha_i$ for site $i$.
The nearest-neighbor (NN) coupling energy $\beta_i$ between site $i$ and $i+1$, also known as transfer or hopping integral, determines the exciton delocalization and mobility.
Here and throughout the following, the last summand ($i=N$) of the NN coupling term (with indices $i+1$ replaced by 1) is intended for systems with periodic boundary conditions only and is deleted otherwise.

The vibrational (phononic) Hamiltonian for a one-dimensional lattice can be written in terms of masses $m_i$, displacement coordinates $R_i$, and conjugated momenta $P_i$ in harmonic approximation
\begin{equation}
	H^{\mathrm{(ph)}} = \frac{1}{2}            \sum_{i=1}^N \frac{P_i^2}{m_i}
										+ \frac{1}{2} \sum_{i=1}^N m_i \nu_i^2        R_i^2
	                  + \frac{1}{2} \sum_{i=1}^N \mu_i \omega_i^2 \left( R_i-R_{i+1} \right)^2
	\label{eq:H_ph0}
\end{equation}
where each site $i$ is restrained to oscillate around its equilibrium position by a harmonic potential with frequency $\nu_i$ and where the NN interactions between sites $i$ and $i+1$ are modeled by harmonic oscillators with frequency $\omega_i$ and corresponding reduced masses $\mu_i=m_im_{i+1}/(m_i+m_{i+1})$.
Note that the frequencies $\nu_i$ and $\omega_i$ are assumed to be independent of the excitonic state of the system.

In parallel with the treatment of the excitons, we introduce second quantization also for the phononic Hamiltonian of Eq.~(\ref{eq:H_ph0}) resulting in
\begin{equation}
	H^{\mathrm{(ph)}} = \sum_{i=1}^N \tilde{\nu}_i
	                    \left(c_i^\dagger c_i + \frac{1}{2}\right)
	                  - \sum_{i=1}^N \tilde{\omega}_i
										\left( c_i^\dagger+c_i \right)
										\left( c_{i+1}^\dagger+c_{i+1} \right)
	\label{eq:H_ph2}
\end{equation}
where raising and lowering operators of (local) vibrations of site $i$ are indicated by $c_i^\dagger$ and $c_i$, respectively.
For the effective frequencies of single site and NN pair vibrations the following expressions
\begin{eqnarray}
\tilde{\nu}_i&=&\sqrt{\nu_i^2+\frac{m_{i-1}}{m_i+m_{i-1}}\omega_{i-1}^2+\frac{m_{i+1}}{m_i+m_{i+1}}\omega_{i}^2} 
\label{eq:nu_E}\\
\tilde{\omega}_i&=&\frac{\mu_i\omega_{i}^2}{2\sqrt{m_i \tilde{\nu}_i m_{i+1} \tilde{\nu}_{i+1}}}
\label{eq:omg_E}
\end{eqnarray}
have been found for cyclic chains.
For linear systems without periodic boundary conditions, the second or third term under the square root of Eq.~\eqref{eq:nu_E} has to be omitted for the first ($i=1$) or last ($i=N$) site, respectively.

Linear models for the coupling of excitons and phonons can be formulated in terms of a Fr\"{o}hlich-Holstein type  Hamiltonian \cite{Devreese2009}
\begin{equation}
	H^{\mathrm{(epc)}} = 
	    \sum_{i=1}^N \sigma_i
	    b_i^\dagger b_i \otimes 
			\left[ R_{i+1} - R_{i-1} \right] = 
			\sum_{i=1}^N 
	    b_i^\dagger b_i \otimes \left[
		\bar{     \sigma}_i	\left( c_{i+1}^\dagger+c_{i+1} \right) -
		\bar{\bar{\sigma}}_i	\left( c_{i-1}^\dagger+c_{i-1} \right)
	\right]
	\label{eq:H_couple}
\end{equation}
where the coupling constants $\sigma_i=\mathrm{d}\alpha_i / \mathrm{d}(R_{i+1}-R_{i-1})$ give the linearized dependence of the exciton energies $\alpha_i$ on the sum of the NN distances to the left and to the right of each site.
The bar notation is used here to convert the EPC constants to second quantization, $\bar{\sigma}_i=\sigma_i/\sqrt{2m_{i+1}\tilde{\nu}_{i+1}}$ and $\bar{\bar{\sigma}}_i=\sigma_i/\sqrt{2m_{i-1}\tilde{\nu}_{i-1}}$.
Note that in our previous work~\cite{Gelss2022a}, the distinction between EPC constants with bars and double bars was missing, which, however, is not required for the homogeneous systems investigated there.

The symmetric formulation \eqref{eq:H_couple} of the EPC is suitable in case of a mirror symmetry of the individual sites frequently occurring in semiconductor materials whereas one-sided expressions are typically used for excitons in protein helices where an amide I vibration couples much more strongly to the directly adjacent H-bond \cite{Scott1991,Scott1992,Georgiev2019}.
For other model assumptions, including Peierls type models based on an $R$--dependence of the NN coupling parameters $\beta_i$, see our previous work~\cite{Gelss2022a}.

\section{Tensor train decompositions}
\label{sec:numerics}

\subsection{Statement of the problem}
\label{sec:TDSE}

Quantum dynamics of non-relativistic, closed systems (i.e., without dissipation or dephasing) is given in terms of the time-dependent Schr{\"o}dinger equation (TDSE)
\begin{equation}
 \mathrm{i}\frac{\mathrm{d}}{\mathrm{d}t} \Psi(t)=H\Psi(t),
\label{eq:TDSE}
\end{equation}
where $t$ is the time and where we restrict ourselves to time-independent Hamiltonians $H$, in our case the Hamiltonian given in Eq.~(\ref{eq:H_total}). 
Note that atomic units with $\hbar=1$ are used throughout this work.
The initial quantum state is given by $\Psi(t=0)=\Psi_0$, which in typical photophysical or photochemical simulation scenarios describes a state prepared by the interaction with light. 
Let us assume that a coupled excitonic-phononic state vector for each of the constituent sites will be given by a vector in the product Hilbert space
\begin{equation}
\psi_i \in \mathcal{H}_i := \mathcal{H}_i^{\mathrm{(ex)}} \otimes \mathcal{H}_i^{\mathrm{(ph)}} \cong \mathbb{C}^{d_i}
\label{eq:psi_coup}
\end{equation}
with dimension $d_i:=d_i^\mathrm{(ex)}d_i^\mathrm{(ph)}$, where $d_i^\mathrm{(ex)}$ and $d_i^\mathrm{(ph)}$ are the dimensions of the Hilbert spaces of excitonic and vibrational state vectors, respectively. 
Then, time-dependent quantum state vectors $\Psi(t)$ of a complete chain (or ring) comprising $N$ sites can be understood as tensors of order $N$, 
\begin{equation}
\Psi  \in 
\mathcal{H} := \mathcal{H}_1 \otimes \mathcal{H}_2 \otimes \cdots \otimes \mathcal{H}_N \cong 
\mathbb{C}^{D}
\label{eq:psi_total}
\end{equation}
with a total dimension of $D:=\prod_{i=1}^N d_i$, i.e., the numerical discretization of $\Psi$ is a tensor in $\mathbb{C}^{d_1 \times \dots \times d_N}$. 
In analogy, the tensorized representation $H$ of the (real-valued, symmetric) Hamiltonian given in Eq.~\eqref{eq:H_total} is a tensor in $\mathbb{R}^{(d_1 \times d_1) \times \dots \times (d_N \times d_N)}$. 

Because the formal time evolution of quantum states solving the TSDE \eqref{eq:TDSE}
\begin{equation}
\Psi(t)=\exp(-\mathrm{i}tH)\Psi_0
\label{eq:Evolution}
\end{equation}
is not computationally available for large $N$, our aim is to numerically integrate the TDSE~\eqref{eq:TDSE} in terms of tensor representations by employing time integrators. 
In general, however, the storage consumption of (non-sparse) tensors of the form \eqref{eq:H_total} or \eqref{eq:psi_total} grows exponentially with the order $N$. 
Due to this so-called \textit{curse of dimensionality}, storing the considered tensors in full format may be not feasible for large $N$. 
Therefore, special tensor representations are required.
As in~\cite{Gelss2022a}, we will focus on the \textit{tensor train format} (TT format)~\cite{Oseledets2009a, Oseledets2009b}. This type of tensor decomposition is not only tailor-made for chain-like topologies such as our Hamiltonians with NN interactions only, but it is also a preferred choice for representing high-dimensional tensors in terms of storage consumption and computational robustness.
We will also introduce properly designed numerical methods based on tensor trains which we will use to solve evolution equations such as the TDSE~\eqref{eq:TDSE} in high-dimensional tensor spaces which otherwise would be extremely expensive or even impossible with today's computational resources.

\subsection{Tensor trains and SLIM decomposition}

\begin{figure}[htbp]
\centering
\begin{subfigure}[b]{0.49\textwidth}
\centering
\begin{tikzpicture}
\draw[black] (0,0) -- node [label={[shift={(0,-0.15)}]$r_1$}] {} ++ (1,0) ;
\draw[black] (1,0) -- node [label={[shift={(0,-0.15)}]$r_2$}] {} ++ (1,0) ;
\draw[black] (2,0) -- node [label={[shift={(0,-0.15)}]$r_3$}] {} ++ (1,0) ;
\draw[black] (3,0) -- node [label={[shift={(0,-0.15)}]$r_4$}] {} ++ (1,0) ;
\draw[black] (0,0) -- node [label={[shift={(0,-1.1)}]$d_1$}] {} ++ (0,-0.7) ;
\draw[black] (1,0) -- node [label={[shift={(0,-1.1)}]$d_2$}] {} ++ (0,-0.7) ;
\draw[black] (2,0) -- node [label={[shift={(0,-1.1)}]$d_3$}] {} ++ (0,-0.7) ;
\draw[black] (3,0) -- node [label={[shift={(0,-1.1)}]$d_4$}] {} ++ (0,-0.7) ;
\draw[black] (4,0) -- node [label={[shift={(0,-1.1)}]$d_5$}] {} ++ (0,-0.7) ;
\node[draw,shape=circle,fill=Gray, scale=0.65] at (0,0){};
\node[draw,shape=circle,fill=Gray, scale=0.65] at (1,0){};
\node[draw,shape=circle,fill=Gray, scale=0.65] at (2,0){};
\node[draw,shape=circle,fill=Gray, scale=0.65] at (3,0){};
\node[draw,shape=circle,fill=Gray, scale=0.65] at (4,0){};
\end{tikzpicture}
\caption{}
\end{subfigure}
\hfill
\begin{subfigure}[b]{0.49\textwidth}
\centering
\begin{tikzpicture}
\draw[black] (0,0) -- node [label={[shift={(0,-0.15)}]$r_1$}] {} ++ (1,0) ;
\draw[black] (1,0) -- node [label={[shift={(0,-0.15)}]$r_2$}] {} ++ (1,0) ;
\draw[black] (2,0) -- node [label={[shift={(0,-0.15)}]$r_3$}] {} ++ (1,0) ;
\draw[black] (3,0) -- node [label={[shift={(0,-0.15)}]$r_4$}] {} ++ (1,0) ;
\draw[black] (0,0) -- node [label={[shift={(0,-1.1)}]$d_1$}] {} ++ (0,-0.7) ;
\draw[black] (1,0) -- node [label={[shift={(0,-1.1)}]$d_2$}] {} ++ (0,-0.7) ;
\draw[black] (2,0) -- node [label={[shift={(0,-1.1)}]$d_3$}] {} ++ (0,-0.7) ;
\draw[black] (3,0) -- node [label={[shift={(0,-1.1)}]$d_4$}] {} ++ (0,-0.7) ;
\draw[black] (4,0) -- node [label={[shift={(0,-1.1)}]$d_5$}] {} ++ (0,-0.7) ;
\draw[black] (0,0) -- node [label={[shift={(0,0.2)}]$d_1$}] {} ++ (0,0.7) ;
\draw[black] (1,0) -- node [label={[shift={(0,0.2)}]$d_2$}] {} ++ (0,0.7) ;
\draw[black] (2,0) -- node [label={[shift={(0,0.2)}]$d_3$}] {} ++ (0,0.7) ;
\draw[black] (3,0) -- node [label={[shift={(0,0.2)}]$d_4$}] {} ++ (0,0.7) ;
\draw[black] (4,0) -- node [label={[shift={(0,0.2)}]$d_5$}] {} ++ (0,0.7) ;
\node[draw,shape=circle,fill=Gray, scale=0.65] at (0,0){};
\node[draw,shape=circle,fill=Gray, scale=0.65] at (1,0){};
\node[draw,shape=circle,fill=Gray, scale=0.65] at (2,0){};
\node[draw,shape=circle,fill=Gray, scale=0.65] at (3,0){};
\node[draw,shape=circle,fill=Gray, scale=0.65] at (4,0){};
\end{tikzpicture}
\caption{}
\end{subfigure}
\caption{Graphical representation of tensor trains: 
Tensors are depicted as circles with different arms indicating the set of free indices. 
(a) Tensor of order 5 in TT format, the first and the last core are matrices, the other cores are tensors of order 3. 
(b) Linear operator in TT format, the first and the last core are tensors of order 3, the other cores are tensors of order 4. 
In both cases, arms corresponding to first and last TT ranks are omitted since $r_0=r_5=1$.}
\label{fig: tensor trains}
\end{figure}
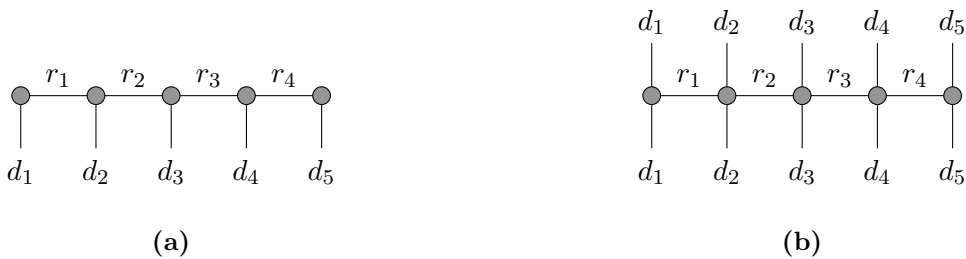

In this work, the TT format is used to represent a quantum state $\Psi$. 
That is, we decompose a tensor in $\mathbb{C}^{d_1 \times \dots \times d_N}$ into $N$  component tensors which are coupled in a chain, see Fig.~\ref{fig: tensor trains} and Def.~\ref{def: TT}.

\begin{definition}\label{def: TT}
 A tensor $\Psi \in \mathbb{C}^{d_1 \times d_2 \times \dots \times d_N}$ is said to be in the TT format if 
 \begin{equation}
  \Psi = \sum_{k_0=1}^{r_0} \cdots \sum_{k_N=1}^{r_N} \Psi^{(1)}_{k_0, :, k_1} \otimes \cdots \otimes \Psi^{(N)}_{k_{N-1}, :, k_N}.
  \label{eq:TT_def}
 \end{equation}
 The tensors $\Psi^{(i)} \in \mathbb{C}^{r_{i-1} \times d_i \times r_i}$ of order $3$ are called TT cores and the numbers $r_i$ are called TT ranks. It holds that $r_0=r_N = 1$ and $r_i \geq 1$ for $i = 1, \dots, N-1$.
\end{definition}

To reduce the memory consumption as well as the computational costs, we are especially interested in low-rank TT decompositions. 
In Ref.~\cite{Gelss2022a}, we used the TT format to represent stationary quantum states, i.e., solutions of the TISE.
Here, we are interested in the numerical solution of Eq.~\eqref{eq:TDSE} in form of a time series $(\Psi(t_0), \Psi(t-1), \dots, \Psi(T))$ of TT representations.
That is, if the quantum states at the considered time points can be approximated accurately enough by tensor trains with ranks of manageable size, we are able to mitigate the curse of dimensionality. 
For more details about the TT format, we refer to Refs.~\cite{Oseledets2009a, Oseledets2009b, Oseledets2011}.
 
Since the efficiency of the tensor-based integrators described in the following sections also depends on the ranks of the operator $H \in \mathbb{R}^{(d_1 \times d_1) \times \dots \times (d_N \times d_N)}$, we want to construct low-rank TT representations of the Hamiltonian tensor as well. 
All terms of the Hamiltonians introduced in Sec.~\ref{sec:model} either act locally on a single site or on two neighboring sites. 
Thus, as we explained in Ref.~\cite{Gelss2022a}, $H$ can be expressed in the TT format by using a so-called SLIM decomposition~\cite{Gelss2017}:
\begin{equation}
\begin{split} H=
    \left\llbracket\begin{matrix}
        S_1 & L_1 & I_1 & M_1 
    \end{matrix}\right\rrbracket
    \otimes 
    \left\llbracket\begin{matrix}
        I_2 & 0            & 0            & 0            \\
        M_2 & 0            & 0            & 0            \\       
        S_2 & L_2 & I_2 & 0            \\
        0            & 0            & 0            & J_2
    \end{matrix}\right\rrbracket
    \otimes \dots \otimes
    \left\llbracket\begin{matrix}
        I_{N-1} & 0                & 0                & 0                \\
        M_{N-1} & 0                & 0                & 0                \\       
        S_{N-1} & L_{N-1} & I_{N-1} & 0                \\
        0                & 0                & 0                & J_{N-1}
    \end{matrix}\right\rrbracket
    \otimes
    \left\llbracket\begin{matrix}
        I_{N} \\
        M_{N} \\       
        S_{N} \\
        L_{N}
    \end{matrix}\right\rrbracket.
    \end{split}
		\label{eq:SLIM_3}
\end{equation}
Above, we use the core notation of the TT format, cf.~Refs.~\cite{Kazeev2013, Gelss2017, Gelss2022a}. 
The single-site components $S_i$ for $i=1, \dots, N$, see Eqs.~(\ref{eq:H_ex},\ref{eq:H_ph2}), are given by
\begin{equation}
	S_i = \alpha_i b_i^\dagger b_i 
	+ \tilde{\nu}_i \left(c_i^\dagger c_i + \frac{1}{2}\right)
\label{eq:SLIM_S}
\end{equation}
The NN interactions are represented by the components $L_i$ and $M_{i+1}$ defined as
\begin{equation*}
    \begin{alignedat}{2}
        \llbracket L_i \rrbracket 
            & =
            \left\llbracket \begin{matrix} L_{i,1} &  \dots & L_{i,\xi_i} \end{matrix}\right\rrbracket
            && \in \mathbb{R}^{1 \times d_i \times d_i \times \xi_i }, \\
        \llbracket M_{i+1} \rrbracket 
            & =
            \left\llbracket\begin{matrix} M_{i+1,1}  & \dots & M_{i+1,\xi_i}  \end{matrix}\right\rrbracket^\mathbb{T}
            && \in \mathbb{R}^{\xi_i \times d_{i+1} \times d_{i+1} \times 1},
    \end{alignedat}
\end{equation*}
for $ i = 1, \dots, N-1 $, and
\begin{equation}
    \begin{alignedat}{2}
        \llbracket L_N \rrbracket 
            & =
            \left\llbracket\begin{matrix} L_{N,1}  & \dots & L_{N,\xi_N}  \end{matrix}\right\rrbracket^\mathbb{T}
            && \in \mathbb{R}^{ \xi_N \times d_N \times d_N \times 1}, \\
        \llbracket M_1 \rrbracket 
            & =
            \left\llbracket\begin{matrix} M_{1,1}  & \dots & M_{1,\xi_d}  \end{matrix}\right\rrbracket
            && \in \mathbb{R}^{1 \times d_1 \times d_1 \times \xi_N },
    \end{alignedat}
\end{equation}
for the case of cyclic systems only.

The two-site contributions to the total Hamiltonian, see Eqs.~(\ref{eq:H_ex},\ref{eq:H_ph2},\ref{eq:H_couple}), are expressible in terms of five operators $L_{i,\lambda}$ acting on site $i$ and five corresponding operators $M_{i+1, \lambda}$ acting on the neighboring site $i+1$:
\begin{equation}
\begin{tabular}{lcl}
 $L_{i,1} = \beta_i b_i^\dagger$, & \hspace{3cm} & $M_{i+1,1} = b_{i+1}$, \\
 $L_{i,2} = \beta_i b_i$, &  & $M_{i+1,2} = b_{i+1}^\dagger$,\\
 $L_{i,3} = - \tilde{\omega}_i\left( c_i^\dagger+c_i \right)$, & & $M_{i+1,3} = c_{i+1}^\dagger+c_{i+1} $, \\
 $L_{i,4} = \bar{\sigma}_i b_i^\dagger b_i$, & & $M_{i+1,4} = c_{i+1}^\dagger+c_{i+1} $, \\
 $L_{i,5} = - \left( c_i^\dagger+c_i \right)$, & & $M_{i+1,5} = \bar{\bar{\sigma}}_{i+1} b_{i+1}^\dagger b_{i+1}$.
\end{tabular}
\label{eq:SLIM_LM}
\end{equation}
If certain coefficients vanish, e.g., if we consider purely excitonic or phononic Hamiltonians, the numbers $\xi_i$ of two-site interactions may even be smaller, leading to reduced TT ranks of the Hamiltonian. 
A detailed derivation of the above components was given in Ref.~\cite{Gelss2022a}. 
Besides the fact that the storage consumption scales only linearly with the number of sites if we consider systems with a fixed/bounded number of NN contributions, the repeating pattern of the SLIM decomposition allows us to easily increase or decrease the number of sites. 
We only have to insert or remove certain TT cores, respectively.

\subsection{Higher-order differencing schemes}
\label{sec:differencing}

The simplest way to integrate the TDSE \eqref{eq:TDSE} forward by one time step $\Delta t$ is to employ an explicit Euler method 
\begin{equation}
\Psi(t+\Delta t) = \Psi(t) - \mathrm{i} \Delta t \mathbf{H} \Psi(t) + \mathcal{O}(\Delta t^2)
\label{eq:expl_Euler}
\end{equation}
which is known to be unstable because neither the time reversal symmetry nor the symplectic structure of the TDSE is conserved.
However, when combining Euler steps forward and backward in time, one arrives at the second order \textit{symmetric Euler} (S2) method
\begin{equation}
\Psi(t+\Delta t) = \Psi(t-\Delta t) - 2 \mathrm{i} \Delta t \mathbf{H} \Psi(t) + \mathcal{O}(\Delta t^3)
\label{eq:symm_Euler}
\end{equation}
which has been in use for more than 40 years in numerical quantum dynamics where it is known as \textit{second order differencing} (SOD) scheme~\cite{Askar1978,Leforestier1991}.
Not requiring a special structure of the Hamiltonian (such as the separability, see Sec.~\ref{sec:splitting}), the S2 scheme is universal and very simple to implement.
In passing, we note the analogy with the Verlet integrator often used for  numerical integration of Newton's classical equation of motion~\cite{Verlet1967}. 
Even though this explicit propagation scheme has the advantages of being time-reversible and conditionally stable, it is neither symplectic nor strictly unitary nor energy conserving~\citep{Choi2019}.
In practice, however, norm and energy conservation are approximately realized  for small enough time steps $\Delta t$.
Since the above method is a two step method, it requires two initial conditions: (i) $\Psi(t)$, which is given, and (ii) $\Psi(t-\Delta t)$, which we obtain by propagating backwards in time.
A common way to achieve this is to start by a first-order scheme \eqref{eq:expl_Euler} for half a time step, and then use the S2 scheme \eqref{eq:symm_Euler} to propagate another half step~\cite{Leforestier1991}.

Higher order schemes are based on Euler methods of (odd) order $L$ forward ($t+\Delta t$) and backward ($t-\Delta t$) in time 
\begin{equation}
\Psi(t\pm\Delta t) = \sum_{\ell=0}^{L} 
\frac{1}{\ell!}
\left( \mp\mathrm{i}\Delta t \mathbf{H} \right)^\ell \Psi(t) + \mathcal{O}(\Delta t^{L+1}) 
\label{eq:euler}
\end{equation}
Combining the forward and backward equation and setting $K = (L+1)/2$ then yields time reversible schemes
\begin{equation}
\Psi(t + \Delta t) = \Psi(t - \Delta t) + \underbrace{\sum_{k=1}^{K} 
\frac{2}{(2k-1)!}
\left( -\mathrm{i} \Delta t \mathbf{H} \right)^{2k-1}}_{=:\mathbf{A}}\Psi(t) + \mathcal{O}(\Delta t^{2 K + 1})
\label{eq:euler:symm}
\end{equation}
which offer the advantage of eliminating all even powers of $\Delta t$, in particular the term with $\mathcal{O}(\Delta t)^{2K}$. 
Hence, the resulting schemes are of order $2K$, and they will be referred to as S2, S4, S6, S8 for $K=1, 2, 3, 4$ throughout Sec.~\ref{sec:results}.

In our practical work using TT representations of state vectors and (Hamiltonian) operators, solving Eq.~\eqref{eq:symm_Euler} or Eq.~\eqref{eq:euler:symm} is straightforward: given $\Psi(t)$, $\Psi(t-\Delta t)$ and $\mathbf{H}$ in the TT format, we can obtain $\Psi(t+\Delta t)$ by simple tensor additions and multiplications. 
However, since TT ranks add under addition and multiply under multiplication, we have 
\begin{equation}\label{eq: rank of S2}
	\rank{\Psi(t+\Delta t)}=\rank{\Psi(t-\Delta t)} + \rank{\mathbf{H}} \cdot \rank{\Psi(t)}
\end{equation}
already for the second order scheme of Eq.~\eqref{eq:symm_Euler}. 
For the higher-order schemes of \eqref{eq:euler:symm}, the rank of $\mathbf{H}$ in Eq.~\eqref{eq: rank of S2} has to be replaced by 
\begin{equation}\label{eq: rank estimation}
	\rank{\mathbf{A}} \leq \sum_{k=1}^{K} \rank{\mathbf{H}}^{2k-1} = \sum_{k=0}^{K-1} \rank{\mathbf{H}}^{2k+1} = \rank{\mathbf{H}} \frac{\rank{\mathbf{H}}^{2 K} -1}{\rank{\mathbf{H}}^2  -1},
\end{equation}
where $\mathbf{A}$ is defined in the underbrace of Eq.~\eqref{eq:euler:symm}.
Hence, the rank of our approximate $\Psi(t)$ would grow exponentially with the number of time steps and would become exceedingly large already after very few time steps. 
Therefore, one needs to perform a rank reduction after every time step, i.e., we approximate the high-rank $\Psi(t+\Delta t)$ resulting from the propagation scheme with a low-rank TT.
That is, given a core of $\Psi(t+\Delta t)$, we reshape it into a matrix and then apply a truncated singular value decomposition (SVD). Depending on the direction of the orthonormalization, the left- or right-singular vectors then define the updated cores and the non-orthonormal part is contracted with the next core, see~\cite{Oseledets2011, Gelss2022a} for details.
Applying this procedure successively to each core from one end to the other, we obtain a low-rank approximation which is then used for the next time step.
The rank reduction requires $N-1$ singular value decompositions on matrices of size $s \times d \cdot s^\prime$ where $d$ is the local Hilbert space dimension and $s$ and $s^\prime \leq s$ are the maximum TT ranks of $\Psi(t + \Delta t)$ before and after the orthonormalization, respectively.
The total costs for each time step can be estimated by $\mathcal{O}(N R^2 r^2 d^2  + N R^3 r^3 d )$, where $r$ and $R$ are the maximum TT ranks of $\Psi(t)$ and $\mathbf{A}$, respectively, cf.~Eq.~\eqref{eq: rank estimation}.
The first term results from estimating the contraction of $\mathbf{A}$, see Eq.~\eqref{eq:euler:symm}, with $\Psi(t)$, whereas the second term corresponds to the truncation/orthonormalization of the tensor cores of $\Psi(t-\Delta t) + \mathbf{A} \Psi(t)$.

\subsection{Higher order splitting schemes}
\label{sec:splitting}

Since the early days of numerical quantum dynamics, split operator (SPO) approaches have been widely used for the matrix exponential~\eqref{eq:Evolution} governing the formal solution of the TDSE \cite{Fleck1976,Feit1982,Leforestier1991}. 
These methods are suitable wherever the quantum-mechanical Hamiltonian can be decomposed into (two or more) exactly (or, at least, easily) solvable parts, e.g., the kinetic and potential energy operator.
Because these parts of the Hamiltonian typically don't commute, deriving approximations to the Baker--Campbell--Hausdorff formula has lead to a hierarchy of exponential splitting methods for use as integrators in solving the time discretized TDSE which are explicit and easy to implement.
Based on the classical first order Lie-Trotter (LT) and second order Strang-Marchuk (SM) schemes, higher-order methods can be obtained as compositions of the basic methods~\cite{Yoshida1990,Suzuki1990,Bandrauk1992,Blanes2006,Blanes2015}.
For a systematic overview of such composition methods, the reader is referred to section V.3 of Ref.~\cite{Hairer2006}.
Common advantages of the exponential splitting methods are that they typically are unitary, symplectic and time-reversible ~\cite{Lubich2008,Roulet2019}.

In what follows, we construct various splitting schemes for Hamiltonians restricted to NN interactions in quasi one-dimensional chains.
Moreover, we assume that all state vectors and Hamiltonian operators are given in TT format as SLIM decompositions. 
Splitting methods have been considered for operators in TT format, see, e.g.,~\citep{Orus2014}, where the case of Lie-Trotter splitting is explained in detail. 
In the closely related context of \textit{density matrix renormalization group} (DMRG), TT/MPS decompositions are used to construct \textit{time-evolving block decimation} (TEBD) schemes for propagations of quantum state vectors~\citep{Volokitin2019, Paeckel2019}, which resemble the techniques described below.

\begin{figure}[htbp]
\centering
\begin{tikzpicture}

\draw[draw=Gray, dashed] (-3,0.5) --++ (12.5,0);
\draw[draw=Gray, dashed] (-3,1.5) --++ (12.5,0);
\draw[draw=Gray, dashed] (-3,2.5) --++ (12.5,0);
\node[anchor=west] at (-3,1) {\textcolor{Gray}{$\exp\left(H_\text{odd} \frac{\Delta t}{2}\right)$}};
\node[anchor=west] at (-3,2) {\textcolor{Gray}{$\exp\left(H_\text{even} \Delta t\right)$}};
\node[anchor=west] at (-3,3) {\textcolor{Gray}{$\exp\left(H_\text{odd} \frac{\Delta t}{2}\right)$}};
\node[anchor=west] at (-3,0) {\textcolor{Gray}{$\Psi(t)$}};

\draw[] (0,-0.15) --++ (4,0);
\draw[dotted] (4,-0.15) --++ (1,0);
\draw[] (5,-0.15) --++ (4,0);

\draw[dotted] (3.25,1) --++ (2.5,0);
\draw[dotted] (4.25,2) --++ (0.5,0);
\draw[dotted] (3.25,3) --++ (2.5,0);

\draw[] (0,-0.15) --++ (0,0.6);
\draw[] (1,-0.15) --++ (0,0.6);
\draw[] (2,-0.15) --++ (0,0.6);
\draw[] (3,-0.15) --++ (0,0.6);
\draw[] (6,-0.15) --++ (0,0.6);
\draw[] (7,-0.15) --++ (0,0.6);
\draw[] (8,-0.15) --++ (0,0.6);
\draw[] (9,-0.15) --++ (0,0.6);

\node[draw,shape=circle,fill=Gray, scale=0.65] at (0,-0.15){};
\node[draw,shape=circle,fill=Gray, scale=0.65] at (1,-0.15){};
\node[draw,shape=circle,fill=Gray, scale=0.65] at (2,-0.15){};
\node[draw,shape=circle,fill=Gray, scale=0.65] at (3,-0.15){};
\node[draw,shape=circle,fill=Gray, scale=0.65] at (6,-0.15){};
\node[draw,shape=circle,fill=Gray, scale=0.65] at (7,-0.15){};
\node[draw,shape=circle,fill=Gray, scale=0.65] at (8,-0.15){};
\node[draw,shape=circle,fill=Gray, scale=0.65] at (9,-0.15){};

\def\x{0.5}
\def\y{1}
\draw[] (\x-0.45,\y-0.45) --++ (0.9,0.9);
\draw[] (\x+0.45,\y-0.45) --++ (-0.9,0.9);
\node[draw,shape=circle,fill=Blue, scale=0.65] at (\x,\y){};
\def\x{2.5}
\def\y{1}
\draw[] (\x-0.45,\y-0.45) --++ (0.9,0.9);
\draw[] (\x+0.45,\y-0.45) --++ (-0.9,0.9);
\node[draw,shape=circle,fill=Blue, scale=0.65] at (\x,\y){};
\def\x{6.5}
\def\y{1}
\draw[] (\x-0.45,\y-0.45) --++ (0.9,0.9);
\draw[] (\x+0.45,\y-0.45) --++ (-0.9,0.9);
\node[draw,shape=circle,fill=Blue, scale=0.65] at (\x,\y){};
\def\x{8.5}
\def\y{1}
\draw[] (\x-0.45,\y-0.45) --++ (0.9,0.9);
\draw[] (\x+0.45,\y-0.45) --++ (-0.9,0.9);
\node[draw,shape=circle,fill=Blue, scale=0.65] at (\x,\y){};

\def\x{1.5}
\def\y{2}
\draw[] (\x-0.45,\y-0.45) --++ (0.9,0.9);
\draw[] (\x+0.45,\y-0.45) --++ (-0.9,0.9);
\node[draw,shape=circle,fill=Green, scale=0.65] at (\x,\y){};
\def\x{3.5}
\def\y{2}
\draw[] (\x-0.45,\y-0.45) --++ (0.9,0.9);
\draw[] (\x+0.45,\y-0.45) --++ (-0.9,0.9);
\node[draw,shape=circle,fill=Green, scale=0.65] at (\x,\y){};
\def\x{5.5}
\def\y{2}
\draw[] (\x-0.45,\y-0.45) --++ (0.9,0.9);
\draw[] (\x+0.45,\y-0.45) --++ (-0.9,0.9);
\node[draw,shape=circle,fill=Green, scale=0.65] at (\x,\y){};
\def\x{7.5}
\def\y{2}
\draw[] (\x-0.45,\y-0.45) --++ (0.9,0.9);
\draw[] (\x+0.45,\y-0.45) --++ (-0.9,0.9);
\node[draw,shape=circle,fill=Green, scale=0.65] at (\x,\y){};
\def\x{9}
\def\y{2}
\draw[] (\x,\y-0.45) --++ (0,0.9);
\node[draw,shape=circle,fill=Green, scale=0.65] at (\x,\y){};

\def\x{0.5}
\def\y{3}
\draw[] (\x-0.45,\y-0.45) --++ (0.9,0.9);
\draw[] (\x+0.45,\y-0.45) --++ (-0.9,0.9);
\node[draw,shape=circle,fill=Blue, scale=0.65] at (\x,\y){};
\def\x{2.5}
\def\y{3}
\draw[] (\x-0.45,\y-0.45) --++ (0.9,0.9);
\draw[] (\x+0.45,\y-0.45) --++ (-0.9,0.9);
\node[draw,shape=circle,fill=Blue, scale=0.65] at (\x,\y){};
\def\x{6.5}
\def\y{3}
\draw[] (\x-0.45,\y-0.45) --++ (0.9,0.9);
\draw[] (\x+0.45,\y-0.45) --++ (-0.9,0.9);
\node[draw,shape=circle,fill=Blue, scale=0.65] at (\x,\y){};
\def\x{8.5}
\def\y{3}
\draw[] (\x-0.45,\y-0.45) --++ (0.9,0.9);
\draw[] (\x+0.45,\y-0.45) --++ (-0.9,0.9);
\node[draw,shape=circle,fill=Blue, scale=0.65] at (\x,\y){};

\end{tikzpicture}
\caption{Strang splitting for SLIM operators for even number $N$ of sites: In each stage of the splitting scheme, only pairs of cores or single cores are altered. If no rank reduction is performed between these stages, the ranks of the resulting TT representation of $\Psi(t + \Delta t)$ are bounded by $r d^3$, where $r$ is the maximum TT rank of $\Psi(t)$. }
\label{fig: Strang}
\end{figure}
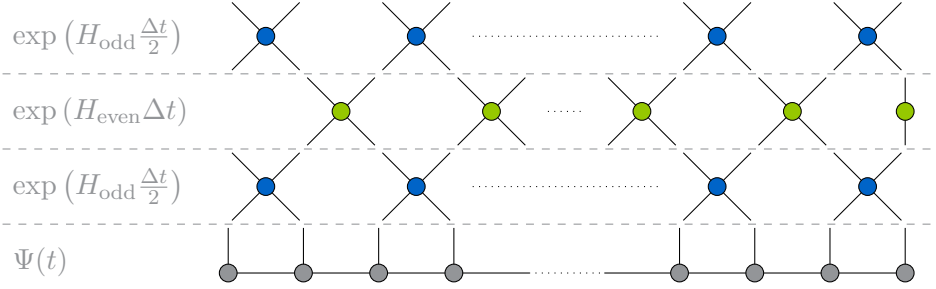

The key to splitting the exponential in the evolution~\eqref{eq:Evolution} is to separate the operators constituting the Hamiltonian~\eqref{eq:H_total} into groups that are as large as possible where all the operators within each of those groups should commute with each other  such that corresponding computations can be carried out in parallel~\citep{Volokitin2019}.
Thus, for an implementation of splitting methods for TTs using SLIM decomposition~\eqref{eq:SLIM_3}, all single site operators $S$ commute by construction, and operators $L$ and $M$ acting on odd/even NN pairs commute within their oddity groups. 
More specificially, we can split our Hamiltonian tensor operator $H \in \mathbb{R}^{(d_1 \times d_1) \times \dots \times (d_N \times d_N)}$ into two parts $H_\text{odd}$ and $H_\text{even}$ which are defined by
\begin{equation*}
\begin{split}
 H_\text{odd} &= \sum_{\substack{i=1,\\ i\,\text{odd} }}^N I^{1:i-1} \otimes S^{(i)} \otimes I^{i+1:N} + \sum_{\substack{i=1,\\ i\,\text{odd} }}^{N-1} I^{1:i-1} \otimes \left\llbracket L^{(i)} \right\rrbracket \otimes \left\llbracket M^{(i+1)} \right\rrbracket \otimes I^{i+2:N} \\
 &= \sum_{i=1}^{\ceil{\frac{N}{2}}} I^{1:2i-2} \otimes S^{(2i-1)} \otimes I^{2i:N} + \sum_{i=1}^{\ceil{\frac{N-1}{2}}} I^{1:2i-2} \otimes \left\llbracket L^{(2i-1)} \right\rrbracket \otimes \left\llbracket M^{(2i)} \right\rrbracket \otimes I^{2i+1:N} 
\end{split}
\end{equation*}
and
\begin{equation*}
\begin{split}
 H_\text{even} &= \sum_{\substack{i=1,\\ i\,\text{even} }}^N I^{1:i-1} \otimes S^{(i)} \otimes I^{i+1:N} + \sum_{\substack{i=1,\\ i\,\text{even} }}^{N-1} I^{1:i-1} \otimes \left\llbracket L^{(i)} \right\rrbracket \otimes \left\llbracket M^{(i+1)} \right\rrbracket \otimes I^{i+2:N} \\
 &= \sum_{i=1}^{\floor{\frac{N}{2}}} I^{1:2i-1} \otimes S^{(2i)} \otimes I^{2i+1:N} + \sum_{i=1}^{\floor{\frac{N-1}{2}}} I^{1:2i-1} \otimes \left\llbracket L^{(2i)} \right\rrbracket \otimes \left\llbracket M^{(2i+1)} \right\rrbracket \otimes I^{2i+2:N},
\end{split}
\end{equation*}
respectively. 
Next, we define $H_{i}$, $i=1, \dots, N-1$, to denote the components
\begin{equation*}
\begin{split}
 H_i &= I^{1:i-1} \otimes S^{(i)} \otimes I^{i+1:N} + I^{1:i-1} \otimes L^{(i)} \otimes M^{(i+1)} \otimes I^{i+2:N}\\
 &= I^{1:i-1} \otimes \left( S^{(i)} \otimes I^{(i+1)} + L^{(i)} \otimes M^{(i+1)} \right) \otimes I^{i+2:N}, 
\end{split}
\end{equation*}
whereas the last component with $i=N$ is simply given by
\begin{equation*}
 H_N = I^{1:N-1} \otimes S^{(N)} .
\end{equation*}
Note that we here only consider non-cyclic systems because defining $H_N = I^{1:N-1} \otimes S^{(N)} + M_1 \otimes J^{2:N_1} \otimes L_N$, cf.~\eqref{eq:SLIM_3}, would either lead to a significant increase of the computational complexity of the following schemes or would involve cyclic TT representations~\cite{Hackbusch2012} of  quantum states, both of which are beyond the scope of the present work.
For the use of SLIM decompositions of cyclic systems, we refer to~\cite{Gelss2017, Gelss2022a}.
With the above definition of the $H_i$, the tensors $H_\text{odd}$ and $H_\text{even}$ can then be written as
\begin{equation}\label{eq: H_odd components}
 H_\text{odd} = \sum_{i \in \mathcal{I}_\text{odd}} H_i, \quad \mathcal{I}_\text{odd} = \{1,3,\dots, 2 \ceil{N/2}-1\}
\end{equation}
and
\begin{equation}\label{eq: H_even components}
 H_\text{even} = \sum_{i \in \mathcal{I}_\text{even}} H_i, \quad \mathcal{I}_\text{even} = \{2,4,\dots, 2 \floor{N/2}\}.
\end{equation}
The components of $H_\text{odd}$ as given in \eqref{eq: H_odd components} commute pairwise. The same holds for the components of $H_\text{even}$ as given in \eqref{eq: H_even components}.
This implies that the components of the propagator in Eq.~\eqref{eq:Evolution}  can be expressed as
\begin{equation*}
\begin{split}
 \exp( -\mathrm{i}t H_\text{odd}) &= \prod_{i \in \mathcal{I}_\text{odd}} \exp(-\mathrm{i}t H_i)\\
 &= \prod_{i \in \mathcal{I}_\text{odd}} I^{1:i-1} \otimes \exp\left( -\mathrm{i}t \left( S^{(i)} \otimes I^{(i+1)} + L^{(i)} \otimes M^{(i+1)} \right) \right) \otimes I^{i+2:N}
 \end{split}
\end{equation*}
with $I_{N+1} = 1$ and $M_{N+1}=0$, see above.
An analogous statement holds for $\exp( -\mathrm{i}t H_\text{even})$.
That is, the tensor exponentials of $-\mathrm{i}t  H_\text{odd}$ and $-\mathrm{i}t  H_\text{even}$ can be written as the product of $\ceil{N/2}$ and $\floor{N/2}$, respectively, mutually commuting tensor operators acting only on two sites at most.
Thus, we can directly implement splitting schemes such as first-order Lie-Trotter (LT) or second-order Strang-Marchuk (SM), see Fig.~\ref{fig: Strang}, given by
\begin{equation*}
  \exp(-\mathrm{i} t H) \overset{\text{LT}}{=}  \exp(-\mathrm{i} t H_\text{odd}) \exp(-\mathrm{i} t H_\text{even}) + \mathcal{O}(t)
\end{equation*}
and
\begin{equation*}
  \exp(-\mathrm{i}t H) \overset{\text{SM}}{=} \exp\left(-\mathrm{i} \frac{t}{2} H_\text{odd}\right) \exp(-\mathrm{i} t H_\text{even}) \exp\left(-\mathrm{i} \frac{t}{2} H_\text{odd}\right)+ \mathcal{O}(t^2),
\end{equation*}
respectively.
Note that these two schemes are also known as TEBD1 and TEBD2 in the DMRG context of the quantum physics community~\citep{Paeckel2019}.

The SM split-operator approach offers the advantage of being unitary, symplectic, stable, symmetric, and time-reversible~\citep{Feit1982},
However, to obtain highly accurate results, this algorithm requires using  small time steps because it has only second-order accuracy. 
Hence, we consider TT-based splitting schemes with higher order by using the fact that these methods can be obtained as suitable compositions of SM steps~\citep{Roulet2019}, i.e.,
\begin{equation*}
 \exp(-\mathrm{i}tH) \approx \text{SM}(-\mathrm{i} \gamma_s t H) \cdots \text{SM}(-\mathrm{i} \gamma_1 t H)
\end{equation*}
with
\begin{equation*}
 \text{SM}(-\mathrm{i}\gamma t H) := \exp\left(-\mathrm{i} \frac{\gamma t}{2} H_\text{odd}\right) \exp(-\mathrm{i} \gamma t H_\text{even}) \exp\left(-\mathrm{i} \frac{\gamma t}{2} H_\text{odd}\right),
\end{equation*}
where $s$ is the number of stages and where consistency implies that $\sum_{j=1}^s\gamma_j=1$.
For more information see Ref.~\cite{Lubich2008}, where an overview of splitting methods with different order is given.
In the present work, we consider the palindromic Yoshida-Neri (YN, order $4$ with 3 stages)~\cite{Yoshida1990,Blanes2006} and Kahan-Li (KL, order $8$ with 17 stages) compositions with coefficients~\cite{Kahan1997}
\begin{equation*}
 \gamma_1=\gamma_3=\frac{1}{2-\sqrt[3]{2}}\approx 1.3512, \quad \gamma_2 = 1-2\gamma_1 =\frac{1}{1-\sqrt[3]{4}} \approx-1.7024,
\end{equation*}
and
\begin{gather*}
 \gamma_1=\gamma_{17}= \phantom{-}0.1302, \quad \gamma_2 = \gamma_{16} = 0.5612, \quad \gamma_3 = \gamma_{15} = -0.3895, \quad \gamma_4 = \gamma_{14} = 0.1588,\\
 \gamma_5 = \gamma_{13} = -0.3959, \quad \gamma_6 = \gamma_{12} = 0.1845, \quad \gamma_7 = \gamma_{11} = \phantom{-}0.2584, \quad \gamma_8 = \gamma_{10} = 0.2950,\\
 \gamma_9 = -0.6055,
\end{gather*}
respectively, which are largely based on a stability and error analysis~\citep{Blanes2015}. 
While the 4th order YN scheme is very similar to the TEBD4 scheme known in the quantum physics community~\citep{Paeckel2019}, the 8th-order KL scheme has not been utilized in the context of TT/MPS representations to the best of our knowledge.
Because some of the coefficients are negative, both the YN and the KL scheme can be considered as combinations of backward and forward propagations in time which may generate non-negligible error terms~\citep{Blanes2006}.

Note that the composition schemes described above are time-reversible, symplectic and unitary, i.e., norm-preserving.
In analogy to the higher-order differencing schemes, see Sec.~\ref{sec:differencing}, the advantages of TT-based splitting methods are that, in contrast to implicit time-integration methods, we do not have to rely on iterative solvers for linear systems like the \emph{alternating linear scheme} (ALS)~\cite{Holtz2012} and we can, in principle, parallelize the tensor contractions and rank reductions in each stage.
In particular, homogeneous systems can be efficiently simulated by splitting methods since the two-site tensor exponentials of $H_\text{odd}$ and $H_\text{even}$ are identical for all sites.
In general, each two-site operator $\exp(\mathrm{i}t H_i)$ has a TT representation with rank bounded by $d^2$. 
Thus, by orthonormalizing (without truncation, see Sec.~\ref{sec:differencing}) between the stages, the ranks of the resulting tensor after one SM step are alternately bounded by $r d^2$ and $r d^3$. 
In this case, the computational complexity for sufficiently large $N$ can be estimated by $\mathcal{O}(N r^3 d^8)$, where we assume that in each stage the matricizations of the two-site operators are multiplied by the contractions of the corresponding cores and the resulting matrix is reshaped and decomposed by using an SVD.
Note that a subsequent truncation for reducing the ranks of $\Psi(t+\Delta t)$ to $r$ would take $O(N r^3 d^5)$ floating-point operations.
However, if we only allow TT ranks bounded by some number $s < r d$ between the SM stages, we can reduce the computational costs to $O(N s^3 d^4)$. 
Truncating the ranks of the resulting tensor train to $r$ then only takes $O(N r s^2 d)$ operations.
In this way, we can significantly reduce the computational complexity but we have to accept losses in accuracy since our experiments showed that higher TT ranks are needed particularly in between the SM stages.
Choosing $s=2r$, however, turned out to yield nearly the same numerical results in our test cases as when allowing arbitrarily large TT ranks in between the SM stages.

\subsection{Time-dependent variational principle}
\label{sec:tdvp}
	
Another approach to integrate the TDSE \eqref{eq:TDSE} is through the Dirac-Frenkel time-dependent variational principle (TDVP). 
For a detailed explanation, we refer to \citep{Lubich2014,Lubich2015}. 
In this method, the time evolution is constrained to the manifold of tensor trains with predefined rank dimensions. 
The action of the Hamiltonian is projected onto the tangent space of this manifold, and the corresponding TDSE is solved exclusively within the manifold. 
The resulting algorithm closely resembles ALS/DMRG. 
That is, we construct the same effective Hamiltonians as we did for the time-independent case in~\citep{Gelss2022a}, but instead of solving an eigenvalue problem, two time-evolution steps are performed. 
Given the core $\Psi^{(i)}$ of a properly orthogonalized TT representation of a quantum state, a TDVP step can be described in terms of local Schrödinger equations in simplified notation as
\begin{equation*}
	\begin{split}
		\frac{\mathrm{d}}{\mathrm{d}t} \Psi^{(i)} &= - \mathrm{i} H^{(\text{eff})}_i \Psi^{(i)}, \\
		\frac{\mathrm{d}}{\mathrm{d}t} R^{(i)} &= + \mathrm{i} \left( H^{(\text{eff})}_i \cdot Q^{(i)}\right) R^{(i)}, 
	\end{split}
\end{equation*}
where $Q^{(i)}$ and $R^{(i)}$ are given by a QR decomposition of $\Psi^{(i)}$, i.e., $\Psi^{(i)} = Q^{(i)} R^{(i)}$, and $H^{(\text{eff})}_i$ is constructed by contracting the given Hamiltonian with the orthonormalized cores of $\Psi$.
The tensor cores are iteratively optimized for representing the quantum state at $t + \Delta t$ while the global state is kept at the original time by backward evolution.
When we finally reach the last core, we keep the local state tensor at the new time and hence obtain a TT representation of the global state $\Psi(t + \Delta t)$.
The single-site variant of TDVP, described in \citep{Haegeman2016, Paeckel2019}, exhibits exact norm and energy conservation when the Hamiltonian is time-independent, a property that will be evident in our later experiments. 
By utilizing iterative Lanczos schemes \citep{Hochbruck1997} or scaling and squaring methods \citep{AlMohy2011}, the total cost can be estimated as $\mathcal{O}(N R^2 r^3 d^2)$.

\subsection{Krylov subspace methods}
\label{sec:krylov}

The iterative Lanczos propagation~\cite{Lanczos1950} based on Krylov subspaces is a widely used approach for approximating time evolution, particularly in the context of full grid representations of tensors.
However, when applied to low-rank tensor train (TT) representations, challenges arise due to the reliance on repeated applications of the Hamiltonian operator.
These operations, similar to those in the Chebyshev scheme (see Sec.~\ref{sec:other}), can lead to significant computational overhead and numerical instability, especially in high-dimensional systems, see also Sec.~\ref{sec:differencing}.
Despite these challenges, Krylov subspace methods remain an attractive option because they allow the projection of the time evolution problem onto a reduced subspace, making them a potentially highly efficient tool for propagating quantum state vectors in the context of Schrödinger equations, particularly when the Hamiltonian operator $H$ is high-dimensional.

The key idea behind the Krylov method is to approximate the system's dynamics in a low-dimensional subspace that captures the essential features of the time evolution. 
To construct a Krylov subspace for solving the Schr\"{o}dinger equation~\eqref{eq:TDSE}, we start with an initial state $\Psi_0$ and generate a sequence of tensors by repeated application of the Hamiltonian operator. 
The Krylov subspace of dimension $m$ is defined as
\begin{equation*}
K_m = \textrm{span} \{ \Psi_0, H \Psi_0, H^2 \Psi_0, \dots, {H}^{m-1} \Psi_0 \}.
\end{equation*}
where the construction of the Krylov subspace is performed iteratively using the tensor-based Lanczos algorithm described in \cite{Paeckel2019}. 
In each iteration step, the new tensor $H^k \Psi_0$ is orthogonalized against the previously generated tensors to build an orthogonal basis.
If necessary, truncation of the tensor train representations can be applied to maintain computational efficiency.
Within this subspace, the time evolution $\Psi(t) = \textrm{exp}(-\mathrm{i} t H ) \Psi_0$ can be approximated using the effective Hamiltonian $H_\textrm{eff} = V \, H \, V^\dagger$, where $V \in \mathbb{C}^{m \times (d_1 \times \dots \times d_N)}$ is the tensor that contains the orthonormal basis tensors of the Krylov subspace such that $V_{k, :, \dots, :}$, $k = 1, \dots, m$, is the orthonormalized version of the tensor $H^{k-1} \Psi_0$. 

Under exact arithmetic, $H_\textrm{eff}$ has a tridiagonal form, and orthogonality among the Krylov tensors is preserved.
This property significantly reduces the computational effort, as each new Krylov vector only needs to be orthogonalized against the last two basis vectors. 
While the enforced tridiagonal structure of the effective Hamiltonian and the numerical implementation of the Lanczos algorithm can, in principle, lead to the loss of orthogonality and numerical instabilities, we did not observe such issues in our experiments with relatively small Krylov subspaces. 
This simplification renders the method not only computationally efficient but also well-suited for tackling large-scale quantum systems.
The TDSE~\eqref{eq:TDSE} can then be efficiently solved in the Krylov subspace by computing the exponential of $H_\textrm{eff}$. 
This method provides a very accurate approximation of the time evolution $\Psi(t)$ without requiring direct access to the full tensor $\textrm{exp}(-\mathrm{i} t H)$, i.e., we obtain the next time step by $\Psi(t) \approx  V^\dagger \textrm{exp}(-\mathrm{i} t H_\textrm{eff}) V \Psi_0$. 
For more information, we refer to \cite{Paeckel2019}. 

The computational costs of the global Krylov method for each time step can be estimated as ${\mathcal{O}(m N d^2 R^2 r^2 + m N d R^3 r^3 + m N d R^2 r ^4)}$. 
This expression combines contributions from three main operations: the construction of Krylov tensors through tensor contractions, the orthonormalization and truncation of these tensors, and the computation of the effective Hamiltonian. 
Despite these costs, the approach remains efficient, as the effective Hamiltonian $H_\textrm{eff}$ can be exponentiated using standard diagonalization routines. 
Here, $R$ is the rank of the Hamiltonian operator $H$, and $r$ is the maximum rank of the Krylov tensors, ensuring the method's scalability for low-rank tensor representations in high-dimensional systems.

In addition to the global Krylov method above, Ref.~\citep{Paeckel2019} also describes a local Krylov method that exploits a Lie--Trotter decomposition of the Hamiltonian to approximate the action of the propagator on a given quantum state in TT/MPS format. 
In this approach, similar to the Time-Dependent Variational Principle (TDVP), all tensor cores are iteratively optimized to represent the next time step while the global state remains fixed at the original time until the final core is updated, yielding the global state at the new time point.

However, the local Krylov method is not without its drawbacks. 
Due to the specific basis transformations involved, it introduces an additional error that accumulates over time, leading to large quantitative deviations at longer time scales. The method exhibits a strong dependence on the step size, with smaller steps required for accurate results, which in turn increases computational costs. 
Furthermore, the runtime of the method scales unfavorably compared to other approaches, depending on the bond dimension and the number of Krylov tensors involved. 
As a result, this method may fail to adequately reproduce the desired dynamics, particularly in cases where precision and efficiency are critical, cf.~\citep{Paeckel2019}.
Hence, we resort to the use of the global Krylov method (termed  K2, K4, K6, K8 with $m=2,4,6,8$) throughout the remainder of this work.

\subsection{Other integration schemes}
\label{sec:other}
Finally, we mention a few other integrator schemes from the classical (1991) review \citep{Leforestier1991} of propagation schemes for solving the TDSE in situations where the involved tensors can still be stored in full format, e.~g., for conventional grid representations. 

First, the Chebychev global propagation scheme builds on an expansion of the time-evolution operator, see \eqref{eq:Evolution}, in terms of (imaginary) Chebychev polynomials.
For the case of time-independent Hamiltonians it has become a quasi-standard because it is known to allow very large time steps.
However, this integrator is not suitable for low-rank tensor train representations investigated in the present work.
In general, the ranks of a tensor-train operator $A$ resulting from the multiplication of two operators, say $A_1$ and $A_2$, equal the products of the corresponding ranks of $A_1$ and $A_2$. 
Thus, the ranks of $A^n$ are only bounded by $r^n$ where $r$ is the maximum rank of $A$. In practice, this often leads to extremely high ranks, resulting in disproportionately high memory requirements and computational time. 
Only by truncating the tensor ranks appropriately the numerical cost could be kept under control, but this, in turn, leads to imprecise approximations of the propagator.

Finally, we also inspected the advanced techniques to design efficient symplectic  splitting integrators from Refs.~\citep{Blanes2006,Blanes2011,Blanes2015}.
While those strategies can lead to highly efficient methods for cases where quantum operators and state vectors can be represented in terms of conventional spatial discretization schemes, they are again not suitable for use with low-rank tensor decompositions studied in the present work.
This is because they also involve polynomials in $\mathbf{H}$ acting on wave functions which is problematic because of the rapidly increasing ranks, see above.

\subsection{Software}
\label{sec:numerics_soft}

All simulations presented in this work are carried out using our recently developed Python package \textsc{WaveTrain} which is publicly available via the GitHub platform~\cite{Riedel2023}. 
It encompasses various numerical solvers for TISE and  TDSE for Hamiltonians in TT form using a SLIM representation, among them all the integrators mentioned in the present work.
\textsc{WaveTrain} builds on \textsc{Scikit-TT}~\cite{Gelss2021}, an open-source tensor train toolbox for Python based on NumPy and SciPy.
For short chains, $N \leq 5$, our results obtained with \textsc{WaveTrain} are in excellent agreement with those obtained from the conventional, grid-based solvers for general Hamiltonians available within the \textsc{WavePacket} software package ~\cite{Schmidt2017,Schmidt2018,Schmidt2019}.

\section{Results and Discussion}

\label{sec:results}
In this Section, we aim at assessing the performance and accuracy of the TT techniques, in particular the various propagators to numerically solve the TDSE for excitons, for phonons, as well as for the coupled systems.
First of all, our comparisons are based on the conservation of norm and energy over time.
Where possible, we also compare our numerical results with semi-analytical solutions which are available for short chains, see Sec.~\ref{sec:results_comp_qd}, and/or for harmonic phonons, see Sec.~\ref{sec:results_comp_cd}.

\subsection{Model parameters}
\label{sec:results_params}

In our computer experiments, we restrict ourselves to the case of non-periodic, homogeneous chains, with the model parameters $\alpha, \beta, m, \nu, \omega, \sigma$ being equal for all sites, with their numerical values adapted from our previous work~\cite{Gelss2022a}.
For the excitonic Hamiltonian of~\eqref{eq:H_ex}, we choose a local excitation energy $\alpha=0.1 E_h$ ($\approx 2.7 $ eV) which is of the order of typical band gaps in organic semiconductors, whereas the NN coupling energy is chosen one order of magnitude smaller, $\beta=-0.01 E_h$.
The phononic Hamiltonian of~\eqref{eq:H_ph0} is parametrized in terms of mass-weighted displacement coordinates, $\tilde{R}=\sqrt{m}R$, here with unit masses, $m=1$, and harmonic frequencies $\nu=10^{-3} E_h/\hbar$ ($\approx 220$ cm$^{-1}$) and $\omega=\sqrt{2}\nu$ for the restraining and for the NN oscillators, respectively. 
The coupling constant occurring in the (symmetric) EPC mechanism of Eq.~\eqref{eq:H_couple} is chosen even smaller, $\sigma=2 \times 10^{-4} E_h/a_0$.
Note that for this value of the EPC constant, the stationary states obtained as solutions of the TISE display mutual trapping of excitons and phonons, the former ones localized with a full width at half maximum of about 7 sites, see our previous work ~\cite{Gelss2022a}.
Note that the use of mass-weighted coordinates implies a scaling of the EPC
constant with $\tilde{\sigma}=\sigma/\sqrt{m}$.
Thus, it is straight-forward to transfer all our results described below to arbitrary values of the particle masses involved.
Except where stated otherwise, we employ a basis set consisting of $d^{\mathrm{(ex)}}=2$ (electronic two-state model) and $d^{\mathrm{(ph)}}=8$ vibrational basis functions (from the second quantization introduced in Eq.~\eqref{eq:H_ph2} in Sec.~\ref{sec:model}).

\subsection{Time scales}
\label{sec:results_ts}

For a homogeneous, non-periodic chain of infinite length, the TDSE for the purely excitonic two-state Hamiltonian of Eq.~\eqref{eq:H_ex} can be solved analytically: 
For the case of an initial excitation localized at site $i_0$, the probability of finding the exciton at time $t$ on site $i$ is given by
\begin{equation}
\mathcal{P}_{i}(t) = J_{i-i_0}^2 (2|\beta| t)
\end{equation}
where the $J_n$ are $n$-th order Bessel functions of the first kind~\cite{Kenkre1984,Scott1992}.
In practice, these solutions are of limited use as a benchmark for numerical quantum dynamics because typically the excitations quickly reach the end of a finite length chain.
However, the above solution is instrumental in defining a time scale for the excitonic energy transport as the time lag between the first maxima of $J_0$ (at $t=0$) and $J_1$ (at $2|\beta|t/ \approx  1.8412$). 
Hence, for our example we suggest a rounded value for the time scale, $\tau^{\mathrm{(ex)}} \equiv 1/|\beta| = 100$ in atomic units, or approximately 2.4 fs.
In the absence of analytical results for phonon dynamics given by the Hamiltonian of Eq.~\eqref{eq:H_ph2}, we simply define an analogous time scale for phononic energy transport, $\tau^{\mathrm{(ph)}} \equiv 1/\nu = 1000$, or approximately 24 fs.
This value qualitatively coincides with our empirical observation of vibrational excitations moving from one site to its nearest neighbors.
Similarly, the time scale of the exciton-phonon coupling of \eqref{eq:H_couple} is defined as $\tau^{\mathrm{(epc)}} \equiv 1/|\sigma|=5000$, or approximately 120 fs, characterizing the redistribution of energy between excitonic and phononic subsystems. 

\subsection{Comparison with quantum dynamics}
\label{sec:results_comp_qd}

For short chains, it is straight-forward to obtain semi-analytical solutions to the TDSE~\eqref{eq:TDSE} by calculating the matrix exponential in the formal solution~\eqref{eq:Evolution} directly.
Assuming a maximum matrix size of $4096 \times 4096$, the semi-analytical approach is limited to $N=12$, $N=4$, or $N=3$ for excitons with $d^{\mathrm(ex)}=2$, phonons with $d^{\mathrm(ph)}=8$, or coupled systems with $d=d^{\mathrm(ex)}d^{\mathrm(ph)}=16$, respectively.

First, we run semi-analytical propagations  with 100 main time steps of length $\tau^{\mathrm{(ex)}}/2 = 50$ for excitons with and without EPC and $\tau^{\mathrm{(ph)}}/2 = 500$ for phonons only. 
As expected, these simulations preserve norm and energy to machine precision.
Subsequently, we use the resulting state vectors as a reference in assessing the quality of TT-based propagations of the same duration where we split the main time steps into 1, 2, 5, 10, 20, 50, 100, 200, 500, or 1000 sub-steps each.
Propagations are carried out using the second order symmetric Euler (S2) method, see Sec.~\ref{sec:differencing}, and four different splitting schemes, i.e., Lie-Trotter (LT), Strang-Marchuk (SM), Yoshida-Neri (YN), and Kahan-Li (KL), see Sec.~\ref{sec:splitting}, as well as the variational principle (VP) approach of Sec.~\ref{sec:tdvp} and the global Krylov schemes K2, K4, K6, K8 of Sec.~\ref{sec:krylov}.
In all simulations of exciton dynamics with or without EPC, the initial state corresponds to a single site (near the center of the chain) being excited.
Our simulations of phonon dynamics are initialized with the vibration of that site being in a coherent state where a mean displacement $\langle \Delta R \rangle = 1$ has been chosen~\cite{Schleich2001}.  
Note that this value is small enough such that the truncation of the vibrational basis (here $d^{\mathrm{(ph)}}=8$) does not have a notable influence on the precision of the results.
All other sites were prepared in their excitonic and/or vibrational ground states.

\begin{figure}[htbp]
\centering
\includegraphics[width=1.0\textwidth]{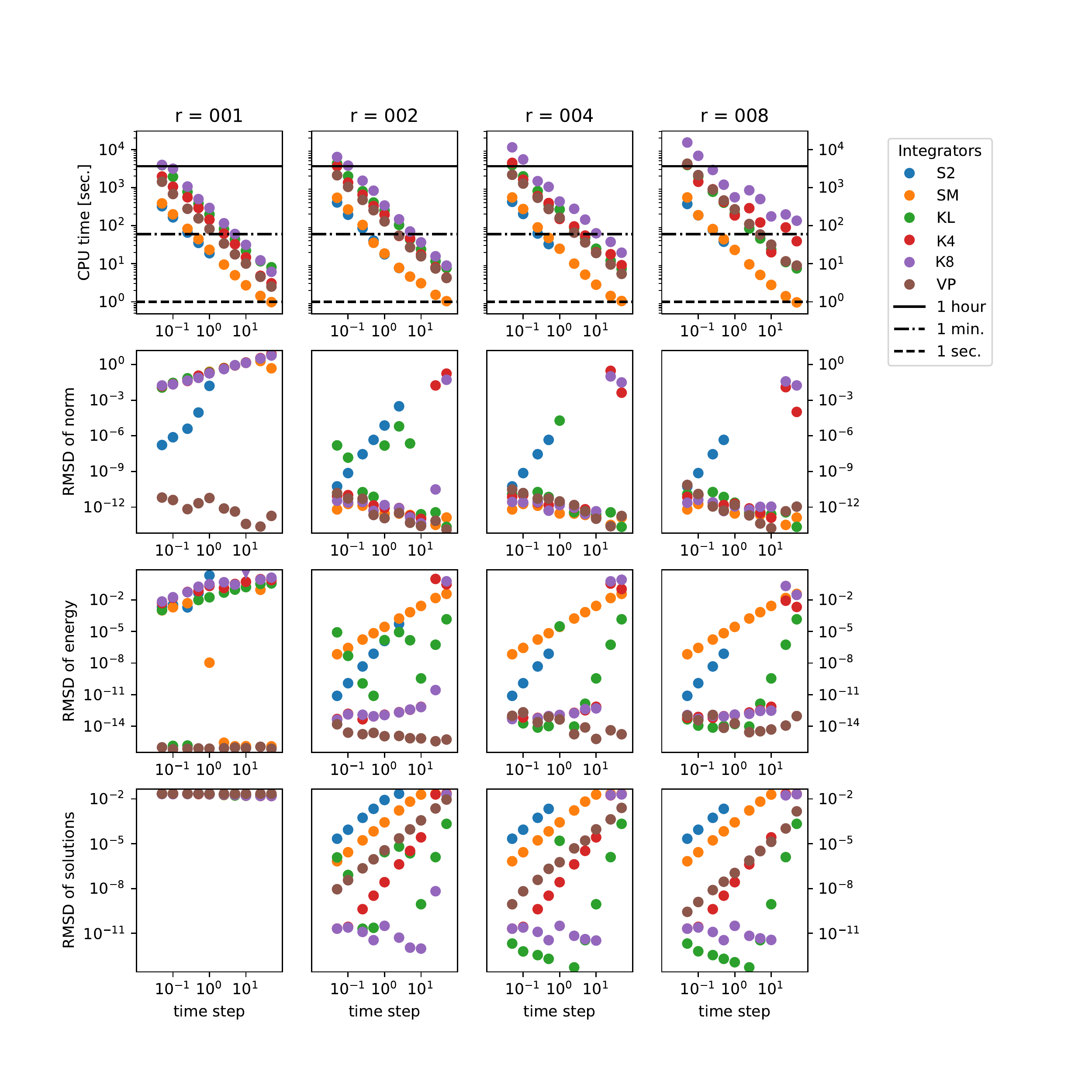} 
\caption{Quantum dynamics of purely excitonic chains with $d=2$ for  $N=12$ sites. 
From left to right: Maximum number of ranks, $r$, of state vectors increasing. 
From top to bottom: CPU-time versus size of temporal sub-steps, deviation of norm from unity, relative deviation of energy from initial value, deviation of state vectors from semi-analytical reference data. 
For second order symmetric Euler (S2), Strang-Marchuk (SM), Kahan-Li (KL), 4-th and 8-th order global Krylov (K4, K8) and time-dependent variational principle (VP) integration methods}
\label{fig:exci_12}
\end{figure}

\begin{figure}[htbp]
\centering
\includegraphics[width=1.0\textwidth]{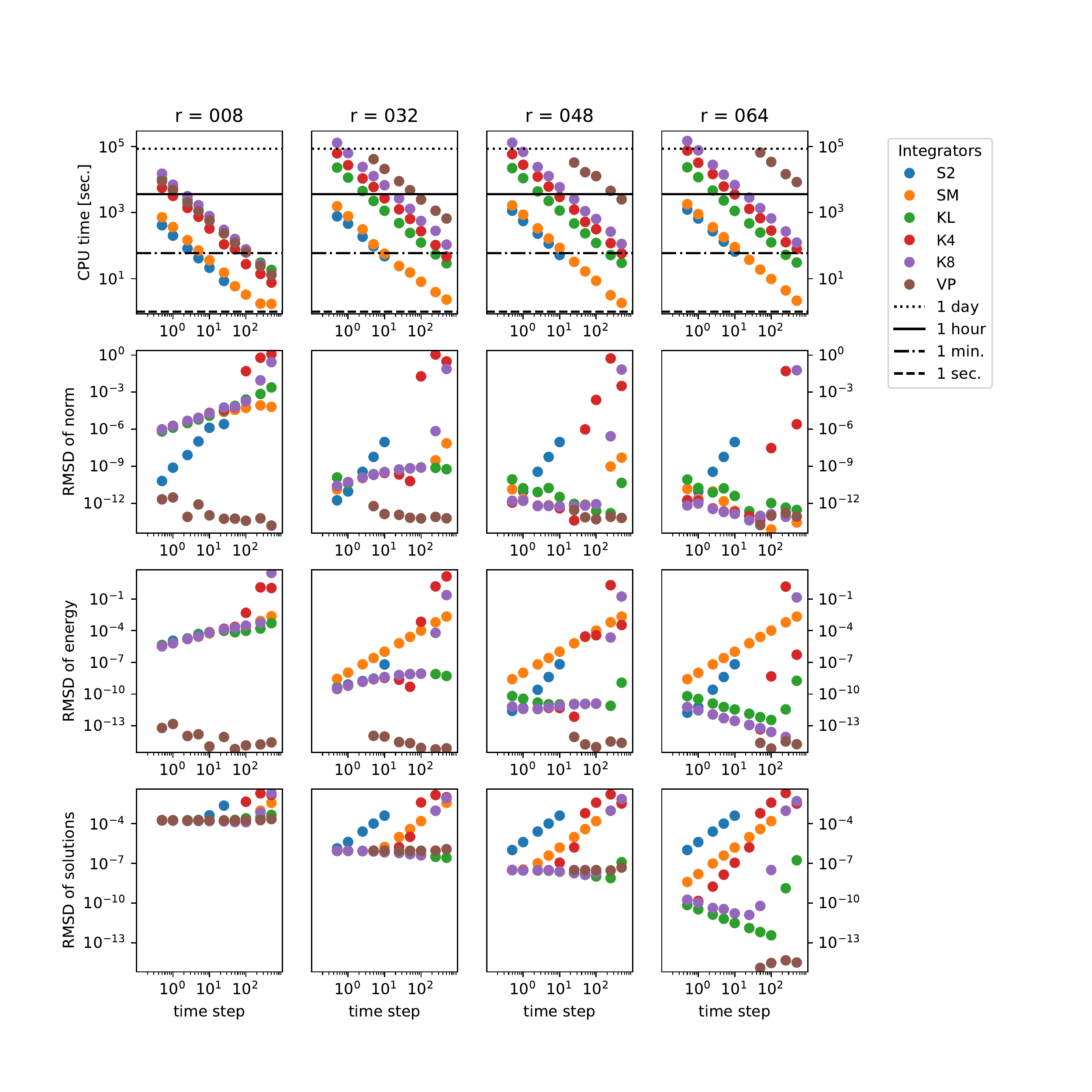} 
\caption{Quantum dynamics of purely phononic chains with $d=8$ for $N=4$ sites. 
For more details, see caption of Fig.~\ref{fig:exci_12}}
\label{fig:phon_04}
\end{figure}

\begin{figure}[htbp]
\centering
\includegraphics[width=1.0\textwidth]{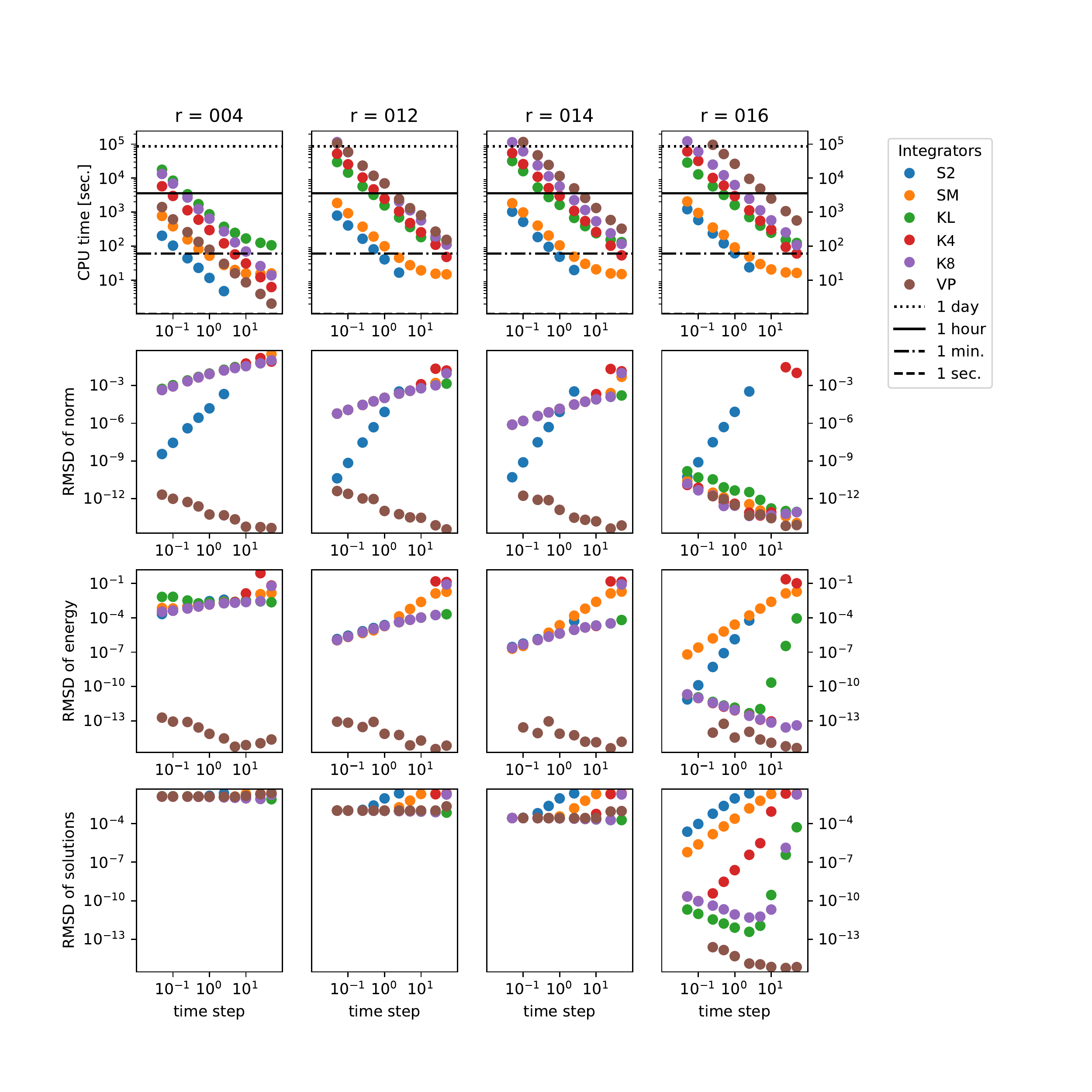} 
\caption{Quantum dynamics of coupled excitons and phonons for $d=16$ for $N=3$ sites.
For more details, see caption of Fig.~\ref{fig:exci_12}}
\label{fig:coup_03}
\end{figure}

Our results for the three classes of systems investigated are displayed in Figs.~\ref{fig:exci_12}, \ref{fig:phon_04}, and \ref{fig:coup_03}. 
Note that results for LT, YN, K2, and K6 are not shown there for reasons of clarity, but they are given in additional figures included in the Supporting Information. 
While typically the results for LT and K2 are of poor quality, the results for YN lie in between those for SM and KL and the results for K6 are found between those for K4 and K8, as expected.
The columns of Figs.~\ref{fig:exci_12}--\ref{fig:coup_03} correspond to different values of $r$, i.~e., the maximum ranks of the solutions, characterizing the flexibility of the solutions.
Note that for given dimension $d$ of the Hilbert spaces for each of the $N$ sites, the maximum rank possible, $r_\mathrm{max}$, of a tensor train (after left- and right-orthonormalization) is given by $d^{\left\lfloor N/2 \right\rfloor}$ where the values of $d$ are 2, 8, or 16 for excitons, phonons, or coupled systems, respectively.
Hence, we have $r_\mathrm{max}=64$ both for pure excitons ($N=12$) and pure phonons ($N=4$) whereas we have $r_\mathrm{max}=16$ for coupled systems ($N=3$).
The CPU times for a single thread of a Xeon Skylake 6130 processor, shown in the upper rows of the figures, increase linearly with the number of time sub-steps.
Owing to the simplicity of the integrator~\eqref{eq:symm_Euler}, the S2 is the fastest of the six integrators compared here.
It is followed by the second order (SM) splitting schemes the computational effort of which is only slightly higher.
Note that results for S2 are partly not available for the larger sub-step sizes where that propagator becomes unstable.
The elapsed CPU times are considerable higher for the KL splitting scheme and often slightly higher for the global Krylov schemes K4 and K8.
For the cases of phonons (Fig.~\ref{fig:phon_04}) and coupled systems (Fig.~\ref{fig:coup_03}), the numerical effort for these three integrators exceeds that for S2 and SM by about one order of magnitude.
This discrepancy is attributed to the cubic dependence of the effort on the ranks, see Sec.~\ref{sec:tdvp}. 
Finally, the VP integrator is the slowest one for simulations of phonons and coupled systems, whereas for the excitons it is found to be comparable to K4.

The panels in the second rows of Figs.~\ref{fig:exci_12}--\ref{fig:coup_03} show the root mean squared deviation (RMSD, averaged over the 100 time steps considered) of the norm of the quantum-mechanical states from unity.
The S2 method reproduces the preservation of norm moderately well with RMSD values typically between $10^{-2}$ and  $10^{-11}$, steeply descending with decreasing step size. 
A similar behavior is often found for the Krylov schemes K4 and K8, but there the decrease is much less steep.
RMSD values between $10^{-9}$ and $10^{-13}$ are often found for the  splitting methods (SM and KL) which is expected because of the unitarity of those schemes.
However, this favorable behavior is only found where the ranks of the solutions, $r$, are chosen high enough.
While this is already the case for $r=2$ for the exciton results shown in Fig.~\ref{fig:exci_12}, one has to go close to the maximum ranks for the other two systems, i.e., $r=r_\mathrm{max}=64$ for the phonons shown in Fig.~\ref{fig:phon_04} or $r=r_\mathrm{max}=16$ for the coupled systems, see Fig.~\ref{fig:coup_03}. 
Finally, we note that the VP integrator displays the smallest deviations of the norm of the quantum states from unity, with all values below $10^{-11}$, essentially irrespective of the ranks of the solutions and/or the size of the time steps used.

A similar behavior is found in our (relative) RMSDs of energies w.r.t.~their initial values and for the (absolute) RMSDs of the TT-based solutions from the semi-analytical solutions, as can be seen in the third and fourth rows of Figs.~\ref{fig:exci_12}--\ref{fig:coup_03}.
Once the ranks of the solutions, $r$, exceed the above-mentioned thresholds, these two quantities can be brought down to $10^{-11}$ or even lower for small enough time steps.
While such a high quality of the numerical solutions requires rather small time steps for the lower order propagators, it is reached already for much larger time steps when using the higher order schemes. 
As expected, our data in the last rows (RMSDs of the state vectors) of the right columns of Figs.~\ref{fig:exci_12}--\ref{fig:coup_03} are well approximated by power laws with exponents of approximately 2 (S2, SM), 4 (K4), and 8 (K8, KL).
The deviating behavior of the RMSDs when applying K8 and KL for very small time steps can be explained by the high number of SVDs used for truncating/orthonormalizing the intermediate solutions, see Sec.~\ref{sec:splitting} and \ref{sec:krylov}. 
In those cases, the (numerical) errors induced by the core decompositions appear to outweigh the time-discretization error.
Finally, the VP integrator displays by far the smallest variations of the energy, with all values below $10^{-12}$, irrespective of the ranks of the solutions and/or the size of the time steps used, as was also observed for the norm of the quantum states.
For the RMSDs of the solutions from the semi-analytic solutions, the performance of the VP scheme appears inconsistent: For the pure excitons, the approximation quality lies between that of the SM and the K4 scheme, whereas for pure phonons and coupled systems it exceeds even that of the KL scheme, even for rather long time steps.

\begin{figure}[htbp]
\centering
\includegraphics[width=0.3\textwidth]{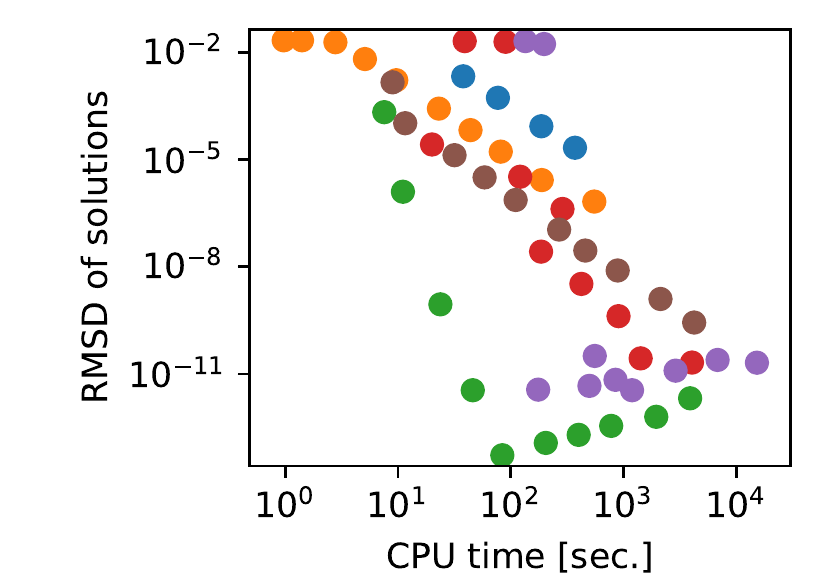} \hspace{-0.03\textwidth} 
\includegraphics[width=0.3\textwidth]{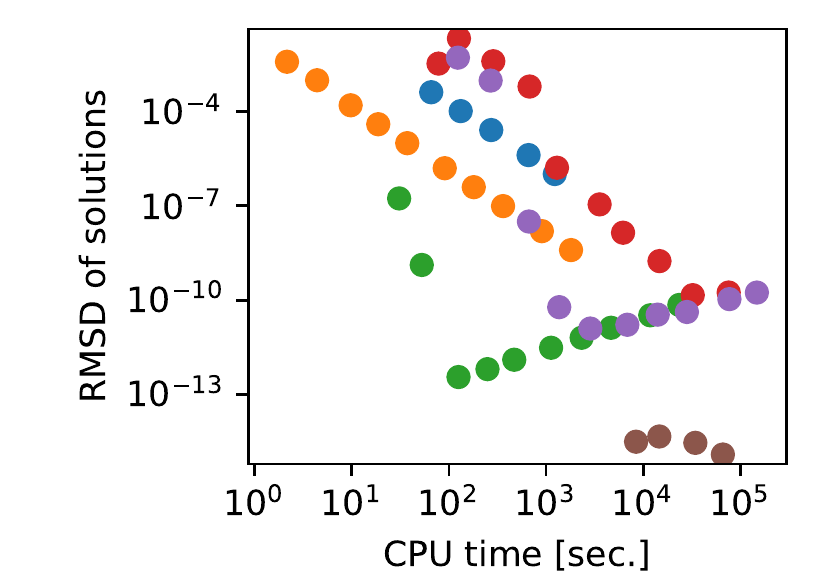} \hspace{-0.02\textwidth}
\includegraphics[width=0.3\textwidth]{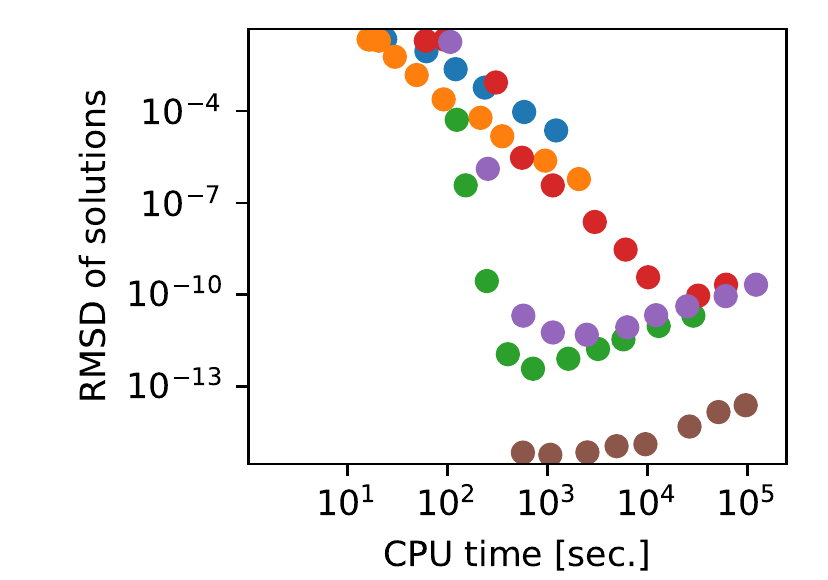} 
\includegraphics[width=0.1\textwidth]{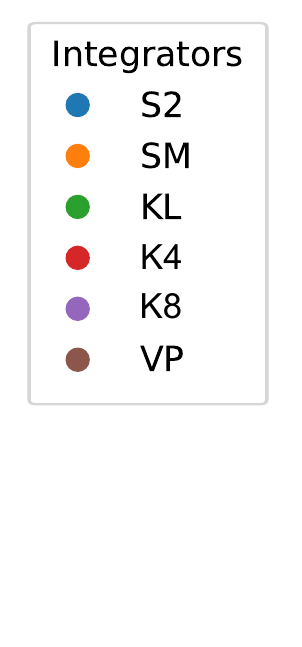} 
\caption{Accuracy of solutions versus computational effort. 
From left to right: excitons ($N=12$, $r=8$, $d=2$), phonons ($N=4$, $r=64$, $d=8$), and coupled systems ($N=3$, $r=16$, $d=16$).
For more details, see caption of Fig.~\ref{fig:exci_12}}
\label{fig:ex_ph_co}
\end{figure}

To summarize, Fig.~\ref{fig:ex_ph_co} shows a synopsis of our results for excitons, phonons, and coupled systems (left to right). 
Again, the respective chain lengths of $N=12$, $N=4$, and $N=3$ are equivalent in the sense that the dimensions of the underlying Hilbert spaces, $(d^{\mathrm(ex)})^{12}=(d^{\mathrm(ph)})^4=(d^{\mathrm(ex)}d^{\mathrm(ph)})^3=4096$, are equal.
In the figure, the accuracy, calculated as RMSD between semi-analytical and TT-based numerical solutions, is shown as a function of the total CPU time elapsed.
In essence, the comparison shows the superiority of the higher-order splitting integrators as well as the variational integrator when high precision is required.
For pure excitons, only the 4-th order (K4) and 8-th order schemes (KL, K8) are able to reach a precision of $10^{-10}$ and below, where the most precise results are obtained for KL for suitable choice of the time step size.
Essentially, the same holds for phonons and coupled systems, but here the variational (VP) integrator yields even more accurate results with errors below $10^{-13}$ which, however, requires rather long CPU times between $10^3$ and $10^5$ s.
In contrast, the CPU times required to reach an excellent precision using the KL scheme are below $10^2$ s for excitons and phonons, whereas about $10^3$ s are found for the coupled systems.
Since we here consider rather short chains, especially for phonons and coupled systems, we observe only a weak influence of the maximum TT rank, $r$, on the computational costs.
But the complexity of the simulations employing the splitting schemes still depends on $d^4$, see Sec.~\ref{sec:splitting}, which we identify as the reason for the higher CPU times required for computing accurate solutions for the coupled exciton-phonon systems.

\clearpage

\subsection{Comparison with classical dynamics}
\label{sec:results_comp_cd}

For the dynamics of phonons, reference data for assessing our TT-based results can also be generated on the basis of the quantum-classical correspondence.
According to the Ehrenfest theorem, the quantum-mechanical expectation values of positions and momenta coincide with results from classical trajectories, as long as the vibrational Hamiltonian is a polynomial of order not higher than two~\cite{Cohen1977}.
This is indeed the case for our vibrational model Hamiltonian~\eqref{eq:H_ph0} which can be re-formulated as $H^{\mathrm{(ph)}} = R^T \mathbf{F} R + P^T \mathbf{G} P$
where $\mathbf{F}$ and $\mathbf{G}$ stand for the Hessians of the Hamiltonian function with respect to positions, $R$, and momenta, $P$, respectively. 
Under these assumptions, Hamilton's classical equations of motion can be solved in a semi-analytical manner using a matrix exponential
\begin{equation}
	\left( \begin{array}{c}R(t) \\P(t)\end{array}	\right) = 
	\exp \left[
	\left( \begin{array}{cc} \mathbf{0} & \mathbf{G} \\ -\mathbf{F} &\mathbf{0} \end{array}	\right)  
	t \right]  
	\left( \begin{array}{c}R(0) \\P(0)\end{array} \right)  
	\label{eq:ph_semi}
\end{equation}
where a symplectic phase space formulation~\cite{Hairer2006} is used and where $\mathbf{0}$ stands for a zero matrix of size $N\times N$.
Again, our classical simulations are initialized with a single site (at the center of the chain) displaced by $\Delta R=1$ while all other sites are at their equilibrium positions.
We generate a set of classical simulations with a duration of 100 time steps of size $\tau^{\mathrm{(ph)}}/2 = 500$ for $N\in\{4,8,16\}$.
As expected for these system sizes, the semi-analytical solutions display excellent preservation of energy within machine accuracy.

\begin{figure}[htbp]
\centering
\includegraphics[width=1.0\textwidth]{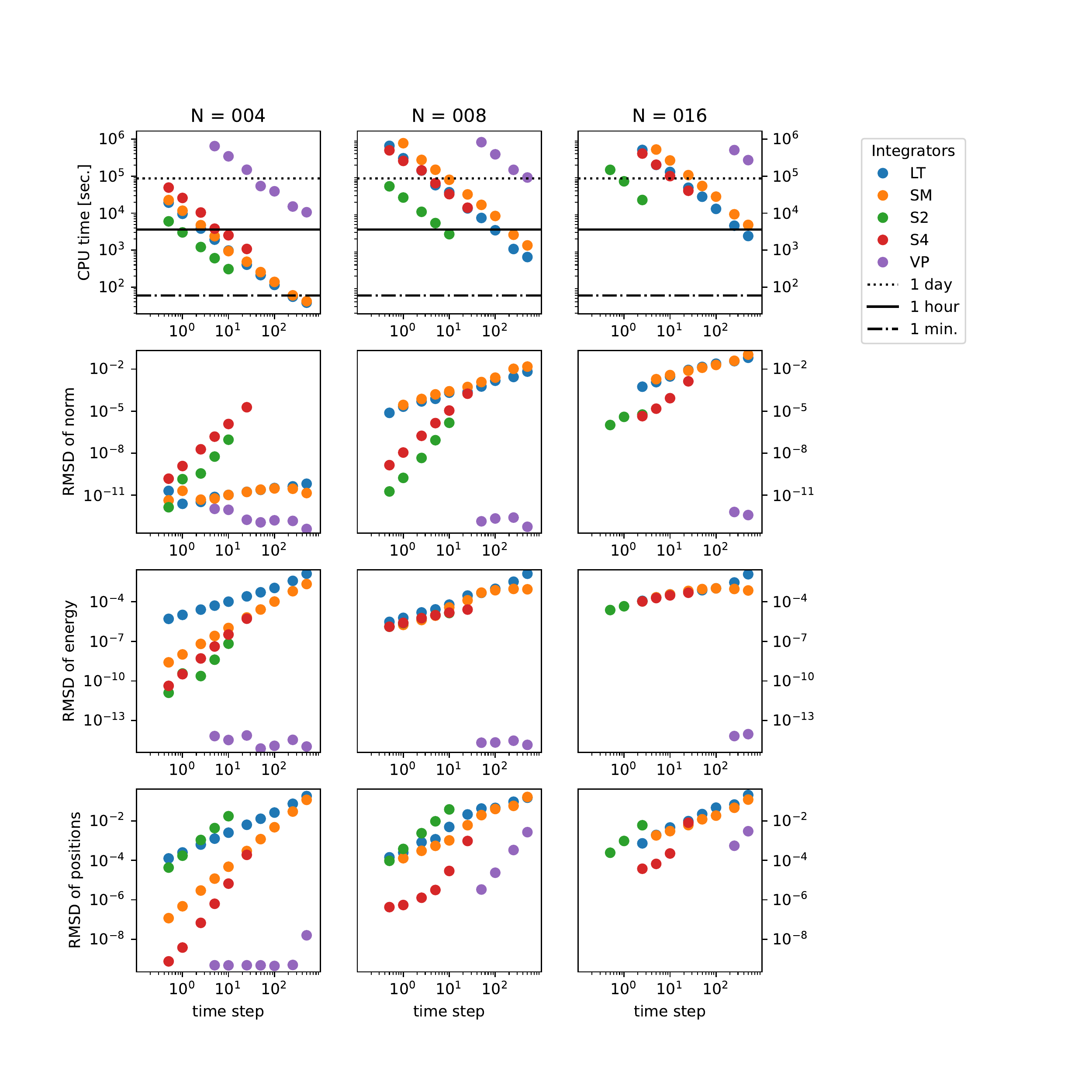} 
\caption{Quantum dynamics of phonons for $N=4$, $N=8$, and $N=16$ sites (from left to right) for $d=16$ and $r=32$.
Note that the last row shows the RMSDs of the displacements (position coordinates) of the oscillators}
\label{fig:phon_16}
\end{figure}

Next, we conduct  TT-based quantum propagations of the same duration ($50\tau^{\mathrm{(ph)}}=50000$,  approximately 1.2 ps) using splitting (LT, SM) and differencing (S2, S4), as well as the variational principle (VP) propagators.
In analogy to our classical simulations of phonon dynamics, we initialize our quantum simulations by using a coherent state  for a single site with the mean position matching the displacement given above.
In contrast to our approach in the previous subsection~\ref{sec:results_comp_qd}, we do not conduct our comparison at the level of quantum state vectors but only at the level of expectation values of vibrational displacement coordinates.
Again, we sub-divide the main time steps into 1, 2, 5, 10, 20, 50, 100, 200, 500, 1000 sub-steps each.
Moreover, we restrict our analysis to the results for $r=32$.
For shorter chains with $N=4$, this value has already been shown to yield reasonable results for the RMSDs of the state vectors, see Fig.~\ref{fig:phon_04}.
Note that we here exclude the higher-order splitting and Krylov schemes (YN, KL, K4, K8) due to the exceedingly high computational effort for longer chains.
Furthermore, we find that the S6 as well as the S8 splitting methods yield results that are almost identical to those obtained for the S4 method which is the reason why we do not show them anywhere in Fig.~\ref{fig:phon_16}.
We explain this effect as a consequence of the used orthonormalizations and rank reductions which eliminate the information stored in the terms corresponding to small singular values and expect S6 or S8 to show better results than S4 only when allowing for higher TT ranks.

Our results are shown in Fig.~\ref{fig:phon_16} where RMSD comparisons of quantum versus classical position values of the vibrational coordinates are shown in the last row.
In contrast to our simulations of phonon dynamics presented in Sec.~\ref{sec:results_comp_qd}, we use a larger basis set with  $d^{\mathrm{(ph)}}=16$ here which has turned out to be mandatory when aiming at high accuracy for the comparison of vibrational coordinates.

The CPU times for the five integrators used here are shown in the top panels of Fig.~\ref{fig:phon_16} for chains of 4, 8, and 16 repeat units.
We see an increase in the CPU times of roughly one order of magnitude upon doubling the chain length for all propagators considered, with the VP integrator being far slower than the other schemes such that only simulations employing the longest time steps could be considered for the longer chains.
As stated in Sec.~\ref{sec:differencing},  Sec.~\ref{sec:splitting}, and Sec.~\ref{sec:tdvp}, the computational complexities of the differencing, splitting, and variational schemes, respectively, should scale linearly with the chain length.
The reason why this behavior cannot be observed in Fig.~\ref{fig:phon_16} is the comparatively small number of sites and the fact that the ranks $r_1$ and $r_{N-1}$ of a TT representation of a quantum state vector with dimensions $d$ are always bounded by $d$.
Thus, in this experiment, every tensor train with $N=4$ has only one rank bounded by $32$, whereas it has five when $N=8$ and $13$ when $N=16$.
Hence, due to the dominant computational costs of contractions and orthonormalizations of TT cores with maximum ranks, the CPU times in Fig.~\ref{fig:phon_16} are not expected to show a linear scaling when doubling the chain length.
However, by bounding the TT ranks of the intermediate solutions in between the SM stages to $2r$, cf.~Section~\ref{sec:splitting}, we are able to reduce the increase of the CPU times for splitting methods significantly while observing almost no loss in accuracy.

The behavior of the norm of the state vectors is shown in the second row of Fig.~\ref{fig:phon_16}. 
The norm is preserved up to machine precision for the unitary LT and SM splitting schemes only for $N=4$ whereas notable deviations from unity are found for $N=8$ and $N=16$.
This behavior is attributed to the finite ranks, $r=32$, used here, see also Fig.~\ref{fig:phon_04} and our discussion of rank effects in Sec.~\ref{sec:results_comp_qd}.
In contrast, for the S2 and S4 differencing schemes, the norm is approximately conserved only for short time steps. 
For $N=4$ and $N=8$, the deviation from unity scales approximately proportional to $\tau^{-4}$ for the S2 scheme as expected~\citep{Askar1978}, whereas it scales as $\tau^{-3}$ for the S4 scheme.
Note that the RMSDs of the norm for longer chains ($N=16$) are much larger due to finite-rank effects both for differencing and splitting schemes.
In marked contrast, the preservation of the norm is extremely good for the variational integrator, irrespective of the chain length and of the time step, as was already found in Sec.~\ref{sec:results_comp_qd}.

The third row of Fig.~\ref{fig:phon_16} shows that the energies are not preserved exactly for the differencing or the splitting schemes, as expected for these integrators which are known not to conserve quadratic invariants rigorously. 
The comparison of our results does not show any characteristic differences between the different integrators for longer chains ($N=8$ and $N=16$).
Only for short chains ($N=4$), the double-logarithmic representations show an approximately linear behavior with slopes of 1.0 (LT), 2.0 (SM), 2.5 (S2) and 3.0 (S4).
As shown already in Sec.~\ref{sec:results_comp_qd}, for the VP integrator the preservation of the energy is reproduced practically to machine precision, again without notable dependence on the time step or the chain length.

Finally, we consider the deviations (RMSD) of our numerical TT quantum results for the calculated positions from our (semi-analytical) classical results described at the beginning of this section.
Again, only for short chains we find straight lines in the bottom panels of Fig.~\ref{fig:phon_16}, this time for $N=4$ with slopes of 1.0 (LT), 2.0 (SM and S2) and 4.0 (S4) as expected.
For that chain length, the VP integrator yields far better results, especially for the longest time steps considered, where the RMSDs are more than 6 orders of magnitude below those for any other integrator.
Also for the longer chains with $N=8$ and $N=16$ we find that the higher order differencing schemes are superior to the lower order ones.
Except for the largest time steps, the SM scheme clearly beats the LT scheme, and the S4 scheme is often several orders of magnitude more precise than the S2 scheme which, however, is the fastest.
For the chains of length $N=8$ and $N=16$, the variational (VP) scheme yields similarly good results as the S4 scheme, but partly using much longer time steps. 
However, this advantage is essentially outweighed by the largely increased CPU time per time step for the VP scheme.

\clearpage

\section{Conclusions and Prospects}
\label{sec:conclude}
In the present work, we have shown how to apply tensor-trains (TT) to solve the time-dependent Schr\"{o}dinger equation for chain-like systems with nearest neighbor (NN) interactions only. 
The aim of this approach is to mitigate the curse of dimensionality, i.~e., to reduce the memory consumption as well as the computational costs, as much as possible.
Throughout this study, we have employed memory-efficient SLIM representations to construct TT representations of the quantum-mechanical Hamiltonians and state vectors.


Four classes of explicit integrators to solve the Schr\"{o}dinger equation were investigated here. 
First, we considered time-symmetrized Euler integrators, see Sec.~\ref{sec:differencing}, based on the second order differencing method regularly used since the early days of numerical quantum dynamics.
Second, we studied splitting propagators such as the well-known first order Lie-Trotter and the second-order Strang-Marchuk schemes, see Sec.~\ref{sec:splitting}, originally based on a splitting of the Hamiltonian into kinetic and potential energy terms.
However, in the present context of chain-like systems we built on a splitting of the Hamiltonian into groups of interleaved NN interactions commutating within each group.
Both for the differencing and splitting methods, we have extended the traditional schemes to higher order ones.
Third, we also investigated a propagator based on the time-dependent variational principle, see Sec.~\ref{sec:tdvp}, which is commonly used in the context of matrix product states.
Finally, also global Krylov schemes of various orders were considered, see Sec.~\ref{sec:krylov}. 

This work presents rigorous tests of these approaches for three different classes of physical systems, i.e., for two-state excitons, for harmonic acoustic phonons, and for coupled exciton-phonon systems modeled in terms of Frenkel-Fr\"{o}hlich-Holstein type Hamiltonians.
In particular, we conducted extensive investigations of the influence of the TT ranks, the time step size, and the orders of the individual propagators for homogeneous and non-periodic chains of various lengths.

The focus of our studies in Sec.~\ref{sec:results_comp_qd} is on the dependence of the conservation properties as well as the approximation quality on the maximum number of ranks, $r$, of the solutions.
Typically we find a threshold behavior, i.e., once the number $r$ exceeds a certain threshold, the quality of the solutions increases drastically.
For the three classes of physical systems investigated here, this threshold is found at very different values of $r$.
For example, for an excitonic chain of length $N=12$, we find excellent agreement with semi-analytic results already for $r=2$.
This may be due to the fact that the exciton dynamics is confined to the Fock space of states with one excitation which also contains our initial state.
(Note that the number of excitons is a conserved quantity because it commutes with the Hamiltonian~\eqref{eq:H_ex}).
In marked contrast, for a phononic chain of length $N=4$ and for our (relatively large) value of the initial displacements of the oscillators, one has to go beyond $r=50$ to find good approximations.
Nevertheless, we summarize that the TT-based propagators provide excellent solutions once the ranks are chosen large enough.
In particular, the higher order (here 8-th order) splitting method KL and the global Krylov scheme K8 represent good choices. Where very high precision (close to machine precision) is required, the variational VP integrator is recommended which, however, requires rather long CPU times.

The main objective of our studies presented in Sec.~\ref{sec:results_comp_cd} is the dependence of the approximation quality on the number of sites, $N$, for the example of phonons.
For larger values of $N$, neither splitting nor Krylov methods are preferable.
Instead, differencing methods turn out to be more efficient, i.e. the S2 scheme is the fastest one whereas the S4 scheme is more accurate than LT or SM splitting. 
However, one has to be aware that even S4 achieves only moderate precision. 
For longer chains, the performance of the variational VP scheme is similar that of the S4 scheme w.r.t. the accuracy versus CPU time required.
However, the strong point of the VP integrator is the excellent conservation of norm and energy, which we found to be independent of ranks of the solutions, time step size, and chain length, for each of the three Hamiltonians investigated.

As an outlook toward future work, we expect our approach to be extended to more complex scenarios than the ones investigated in the present work. 
For example, considering more than one vibrational degree of freedom per site becomes necessary, e.g., for realistically modeling the conformational dynamics of $\pi$-conjugated polymer chains \cite{Binder2018,DiMaiolo2020}.
Obviously, this requires larger dimensions, $d$, of the local Hilbert spaces.
In the practice of our approaches using TT based propagators, in that case differencing methods are expected to be superior to splitting methods because of the more favorable scaling of the computational costs with $d$ for the former ones, see our estimates of the complexities at the end of Secs.~\ref{sec:differencing} and~\ref{sec:splitting}.

Finally we remark that we have not discussed implicit integrators throughout the present work.
For example, the implicit trapezoidal scheme~\citep{Hairer2006} offers the promising advantage of conserving quadratic invariants such as the energy associated with the Hamiltonians proposed in Sec.~\ref{sec:model}. 
Unfortunately, in our test calculations this integrator was found  to be considerably slower than any of the explicit schemes considered.
This is because it involves the solution of a large set of equations for every time step.
In the context of TT techniques, this is typically done by means of the alternating linear scheme~\cite{Holtz2012} which is rather time-consuming, see also our previous TISE work~\cite{Gelss2022a}.
However, implicit schemes may become more attractive when combined with spatially and/or temporally adaptive schemes, e.~g., similar to the trapezoidal rule for adaptive integration of Liouville equations (TRAIL)~\cite{Horenko2003,Horenko2004}.

\begin{acknowledgments}
This work was funded by the Deutsche Forschungsgemeinschaft (DFG, German Research Foundation) through Germany's Excellence Strategy via the Berlin Mathematics Research Center MATH+ (EXC-2046/1, project ID: 390685689) 
and the CRC 1114 ``Scaling Cascades in Complex Systems'' (project ID: 235221301, project B06), as well as by the QuantERA II programme funded by the European Union's Horizon 2020 research and innovation programme 
under Grant Agreement No. 101017733.
Felix Henneke is acknowledged for insightful discussions and Jerome Riedel (FU Berlin) for valuable help with the implementation of the WaveTrain Python codes. 
Moreover, the authors would like to thank the HPC Service of ZEDAT, FU Berlin, for generous allocation of computing resources.
\end{acknowledgments}

\section*{Author declarations}
\subsection*{Conflict of Interest}
No potential conflict of interest was reported by the authors.

\section*{Data Availability}
The Python and Bash scripts used to generate the simulation results shown in Figs.~\ref{fig:exci_12}--\ref{fig:phon_16} are openly available in the \textsc{Zenodo} repository at https://doi.org/10.5281/zenodo.7351812.
Moreover, we used version 1.0 of our open-source \textsc{WaveTrain} software~\citep{Riedel2023}, based on version 1.2 of the \textit{scikit\_tt} tensor train package, both of which are freely available from the GitHub platform at https://github.com/PGelss/.

\newpage

\bibliography{TDSE_TT_1} 

\begin{thebibliography}{86}%
\makeatletter
\providecommand \@ifxundefined [1]{%
 \@ifx{#1\undefined}
}%
\providecommand \@ifnum [1]{%
 \ifnum #1\expandafter \@firstoftwo
 \else \expandafter \@secondoftwo
 \fi
}%
\providecommand \@ifx [1]{%
 \ifx #1\expandafter \@firstoftwo
 \else \expandafter \@secondoftwo
 \fi
}%
\providecommand \natexlab [1]{#1}%
\providecommand \enquote  [1]{``#1''}%
\providecommand \bibnamefont  [1]{#1}%
\providecommand \bibfnamefont [1]{#1}%
\providecommand \citenamefont [1]{#1}%
\providecommand \href@noop [0]{\@secondoftwo}%
\providecommand \href [0]{\begingroup \@sanitize@url \@href}%
\providecommand \@href[1]{\@@startlink{#1}\@@href}%
\providecommand \@@href[1]{\endgroup#1\@@endlink}%
\providecommand \@sanitize@url [0]{\catcode `\\12\catcode `\$12\catcode
  `\&12\catcode `\#12\catcode `\^12\catcode `\_12\catcode `\%12\relax}%
\providecommand \@@startlink[1]{}%
\providecommand \@@endlink[0]{}%
\providecommand \url  [0]{\begingroup\@sanitize@url \@url }%
\providecommand \@url [1]{\endgroup\@href {#1}{\urlprefix }}%
\providecommand \urlprefix  [0]{URL }%
\providecommand \Eprint [0]{\href }%
\providecommand \doibase [0]{https://doi.org/}%
\providecommand \selectlanguage [0]{\@gobble}%
\providecommand \bibinfo  [0]{\@secondoftwo}%
\providecommand \bibfield  [0]{\@secondoftwo}%
\providecommand \translation [1]{[#1]}%
\providecommand \BibitemOpen [0]{}%
\providecommand \bibitemStop [0]{}%
\providecommand \bibitemNoStop [0]{.\EOS\space}%
\providecommand \EOS [0]{\spacefactor3000\relax}%
\providecommand \BibitemShut  [1]{\csname bibitem#1\endcsname}%
\let\auto@bib@innerbib\@empty
\bibitem [{\citenamefont {Mikhnenko}\ \emph {et~al.}(2015)\citenamefont
  {Mikhnenko}, \citenamefont {Blom},\ and\ \citenamefont
  {Nguyen}}]{Mikhnenko2015}%
  \BibitemOpen
  \bibfield  {author} {\bibinfo {author} {\bibfnamefont {O.~V.}\ \bibnamefont
  {Mikhnenko}}, \bibinfo {author} {\bibfnamefont {P.~W.~M.}\ \bibnamefont
  {Blom}},\ and\ \bibinfo {author} {\bibfnamefont {T.-Q.}\ \bibnamefont
  {Nguyen}},\ }\bibfield  {title} {\bibinfo {title} {{Exciton diffusion in
  organic semiconductors}},\ }\href {https://doi.org/10.1039/C5EE00925A}
  {\bibfield  {journal} {\bibinfo  {journal} {Energy Environ. Sci.}\ }\textbf
  {\bibinfo {volume} {8}},\ \bibinfo {pages} {1867} (\bibinfo {year}
  {2015})}\BibitemShut {NoStop}%
\bibitem [{\citenamefont {Kranz}\ and\ \citenamefont
  {Elstner}(2016)}]{Kranz2016}%
  \BibitemOpen
  \bibfield  {author} {\bibinfo {author} {\bibfnamefont {J.~J.}\ \bibnamefont
  {Kranz}}\ and\ \bibinfo {author} {\bibfnamefont {M.}~\bibnamefont
  {Elstner}},\ }\bibfield  {title} {\bibinfo {title} {{Simulation of Singlet
  Exciton Diffusion in Bulk Organic Materials}},\ }\href
  {https://doi.org/10.1021/acs.jctc.6b00235} {\bibfield  {journal} {\bibinfo
  {journal} {Journal of Chemical Theory and Computation}\ }\textbf {\bibinfo
  {volume} {12}},\ \bibinfo {pages} {4209} (\bibinfo {year}
  {2016})}\BibitemShut {NoStop}%
\bibitem [{\citenamefont {Schr{\"{o}}ter}\ \emph {et~al.}(2015)\citenamefont
  {Schr{\"{o}}ter}, \citenamefont {Ivanov}, \citenamefont {Schulze},
  \citenamefont {Polyutov}, \citenamefont {Yan}, \citenamefont {Pullerits},\
  and\ \citenamefont {K{\"{u}}hn}}]{Schroter2015}%
  \BibitemOpen
  \bibfield  {author} {\bibinfo {author} {\bibfnamefont {M.}~\bibnamefont
  {Schr{\"{o}}ter}}, \bibinfo {author} {\bibfnamefont {S.~D.}\ \bibnamefont
  {Ivanov}}, \bibinfo {author} {\bibfnamefont {J.}~\bibnamefont {Schulze}},
  \bibinfo {author} {\bibfnamefont {S.~P.}\ \bibnamefont {Polyutov}}, \bibinfo
  {author} {\bibfnamefont {Y.}~\bibnamefont {Yan}}, \bibinfo {author}
  {\bibfnamefont {T.}~\bibnamefont {Pullerits}},\ and\ \bibinfo {author}
  {\bibfnamefont {O.}~\bibnamefont {K{\"{u}}hn}},\ }\bibfield  {title}
  {\bibinfo {title} {{Exciton-vibrational coupling in the dynamics and
  spectroscopy of Frenkel excitons in molecular aggregates}},\ }\href
  {https://doi.org/10.1016/j.physrep.2014.12.001} {\bibfield  {journal}
  {\bibinfo  {journal} {Physics Reports}\ }\textbf {\bibinfo {volume} {567}},\
  \bibinfo {pages} {1} (\bibinfo {year} {2015})}\BibitemShut {NoStop}%
\bibitem [{\citenamefont {Zhugayevych}\ and\ \citenamefont
  {Tretiak}(2015)}]{Zhugayevych2015}%
  \BibitemOpen
  \bibfield  {author} {\bibinfo {author} {\bibfnamefont {A.}~\bibnamefont
  {Zhugayevych}}\ and\ \bibinfo {author} {\bibfnamefont {S.}~\bibnamefont
  {Tretiak}},\ }\bibfield  {title} {\bibinfo {title} {{Theoretical Description
  of Structural and Electronic Properties of Organic Photovoltaic Materials}},\
  }\href {https://doi.org/10.1146/annurev-physchem-040214-121440} {\bibfield
  {journal} {\bibinfo  {journal} {Annual Review of Physical Chemistry}\
  }\textbf {\bibinfo {volume} {66}},\ \bibinfo {pages} {305} (\bibinfo {year}
  {2015})}\BibitemShut {NoStop}%
\bibitem [{\citenamefont {Devreese}\ and\ \citenamefont
  {Alexandrov}(2009)}]{Devreese2009}%
  \BibitemOpen
  \bibfield  {author} {\bibinfo {author} {\bibfnamefont {J.~T.}\ \bibnamefont
  {Devreese}}\ and\ \bibinfo {author} {\bibfnamefont {A.~S.}\ \bibnamefont
  {Alexandrov}},\ }\bibfield  {title} {\bibinfo {title} {{Fr{\"{o}}hlich
  polaron and bipolaron: Recent developments}},\ }\href
  {https://doi.org/10.1088/0034-4885/72/6/066501} {\bibfield  {journal}
  {\bibinfo  {journal} {Reports on Progress in Physics}\ }\textbf {\bibinfo
  {volume} {72}},\ \bibinfo {pages} {066501} (\bibinfo {year}
  {2009})}\BibitemShut {NoStop}%
\bibitem [{\citenamefont {Davydov}(1985)}]{Davydov1985}%
  \BibitemOpen
  \bibfield  {author} {\bibinfo {author} {\bibfnamefont {A.~S.}\ \bibnamefont
  {Davydov}},\ }\href@noop {} {\emph {\bibinfo {title} {{Solitons in Molecular
  Systems}}}}\ (\bibinfo  {publisher} {Reidel},\ \bibinfo {year}
  {1985})\BibitemShut {NoStop}%
\bibitem [{\citenamefont {Scott}(1992)}]{Scott1992}%
  \BibitemOpen
  \bibfield  {author} {\bibinfo {author} {\bibfnamefont {A.~C.}\ \bibnamefont
  {Scott}},\ }\bibfield  {title} {\bibinfo {title} {{Davydov's soliton}},\
  }\href {https://doi.org/10.1016/0370-1573(92)90093-F} {\bibfield  {journal}
  {\bibinfo  {journal} {Physics Reports}\ }\textbf {\bibinfo {volume} {217}},\
  \bibinfo {pages} {1} (\bibinfo {year} {1992})}\BibitemShut {NoStop}%
\bibitem [{\citenamefont {Georgiev}\ and\ \citenamefont
  {Glazebrook}(2019)}]{Georgiev2019}%
  \BibitemOpen
  \bibfield  {author} {\bibinfo {author} {\bibfnamefont {D.~D.}\ \bibnamefont
  {Georgiev}}\ and\ \bibinfo {author} {\bibfnamefont {J.~F.}\ \bibnamefont
  {Glazebrook}},\ }\bibfield  {title} {\bibinfo {title} {{On the quantum
  dynamics of Davydov solitons in protein $\alpha$-helices}},\ }\href
  {https://doi.org/10.1016/j.physa.2018.11.026} {\bibfield  {journal} {\bibinfo
   {journal} {Physica A: Statistical Mechanics and its Applications}\ }\textbf
  {\bibinfo {volume} {517}},\ \bibinfo {pages} {257} (\bibinfo {year}
  {2019})}\BibitemShut {NoStop}%
\bibitem [{\citenamefont {Binder}\ \emph {et~al.}(2018)\citenamefont {Binder},
  \citenamefont {Lauvergnat},\ and\ \citenamefont {Burghardt}}]{Binder2018}%
  \BibitemOpen
  \bibfield  {author} {\bibinfo {author} {\bibfnamefont {R.}~\bibnamefont
  {Binder}}, \bibinfo {author} {\bibfnamefont {D.}~\bibnamefont {Lauvergnat}},\
  and\ \bibinfo {author} {\bibfnamefont {I.}~\bibnamefont {Burghardt}},\
  }\bibfield  {title} {\bibinfo {title} {{Conformational Dynamics Guides
  Coherent Exciton Migration in Conjugated Polymer Materials: First-Principles
  Quantum Dynamical Study}},\ }\href
  {https://doi.org/10.1103/PhysRevLett.120.227401} {\bibfield  {journal}
  {\bibinfo  {journal} {Physical Review Letters}\ }\textbf {\bibinfo {volume}
  {120}},\ \bibinfo {pages} {227401} (\bibinfo {year} {2018})}\BibitemShut
  {NoStop}%
\bibitem [{\citenamefont {{Di Maiolo}}\ \emph {et~al.}(2020)\citenamefont {{Di
  Maiolo}}, \citenamefont {Brey}, \citenamefont {Binder},\ and\ \citenamefont
  {Burghardt}}]{DiMaiolo2020}%
  \BibitemOpen
  \bibfield  {author} {\bibinfo {author} {\bibfnamefont {F.}~\bibnamefont {{Di
  Maiolo}}}, \bibinfo {author} {\bibfnamefont {D.}~\bibnamefont {Brey}},
  \bibinfo {author} {\bibfnamefont {R.}~\bibnamefont {Binder}},\ and\ \bibinfo
  {author} {\bibfnamefont {I.}~\bibnamefont {Burghardt}},\ }\bibfield  {title}
  {\bibinfo {title} {{Quantum dynamical simulations of intra-chain exciton
  diffusion in an oligo ( para -phenylene vinylene) chain at finite
  temperature}},\ }\href@noop {} {\bibfield  {journal} {\bibinfo  {journal}
  {The Journal of Chemical Physics}\ }\textbf {\bibinfo {volume} {153}},\
  \bibinfo {pages} {184107} (\bibinfo {year} {2020})}\BibitemShut {NoStop}%
\bibitem [{\citenamefont {Kosloff}(1988)}]{Kosloff1988}%
  \BibitemOpen
  \bibfield  {author} {\bibinfo {author} {\bibfnamefont {R.}~\bibnamefont
  {Kosloff}},\ }\bibfield  {title} {\bibinfo {title} {{Time-dependent
  quantum-mechanical methods for molecular dynamics}},\ }\href
  {https://doi.org/10.1021/j100319a003} {\bibfield  {journal} {\bibinfo
  {journal} {The Journal of Physical Chemistry}\ }\textbf {\bibinfo {volume}
  {92}},\ \bibinfo {pages} {2087} (\bibinfo {year} {1988})}\BibitemShut
  {NoStop}%
\bibitem [{\citenamefont {Lenz}(1951)}]{Lenz1951}%
  \BibitemOpen
  \bibfield  {author} {\bibinfo {author} {\bibfnamefont {F.}~\bibnamefont
  {Lenz}},\ }\bibfield  {title} {\bibinfo {title} {{The Ratio of Proton and
  Electron Masses}},\ }\href {https://doi.org/10.1103/PhysRev.82.554.2}
  {\bibfield  {journal} {\bibinfo  {journal} {Physical Review}\ }\textbf
  {\bibinfo {volume} {82}},\ \bibinfo {pages} {554} (\bibinfo {year}
  {1951})}\BibitemShut {NoStop}%
\bibitem [{\citenamefont {Bornemann}\ \emph {et~al.}(1996)\citenamefont
  {Bornemann}, \citenamefont {Nettesheim},\ and\ \citenamefont
  {Sch{\"{u}}tte}}]{Bornemann1996}%
  \BibitemOpen
  \bibfield  {author} {\bibinfo {author} {\bibfnamefont {F.~A.}\ \bibnamefont
  {Bornemann}}, \bibinfo {author} {\bibfnamefont {P.}~\bibnamefont
  {Nettesheim}},\ and\ \bibinfo {author} {\bibfnamefont {C.}~\bibnamefont
  {Sch{\"{u}}tte}},\ }\bibfield  {title} {\bibinfo {title} {{Quantum-classical
  molecular dynamics as an approximation to full quantum dynamics}},\ }\href
  {https://doi.org/10.1063/1.471952} {\bibfield  {journal} {\bibinfo  {journal}
  {The Journal of Chemical Physics}\ }\textbf {\bibinfo {volume} {105}},\
  \bibinfo {pages} {1074} (\bibinfo {year} {1996})}\BibitemShut {NoStop}%
\bibitem [{\citenamefont {Nettesheim}\ \emph {et~al.}(1996)\citenamefont
  {Nettesheim}, \citenamefont {Bornemann}, \citenamefont {Schmidt},\ and\
  \citenamefont {Sch{\"{u}}tte}}]{Nettesheim1996}%
  \BibitemOpen
  \bibfield  {author} {\bibinfo {author} {\bibfnamefont {P.}~\bibnamefont
  {Nettesheim}}, \bibinfo {author} {\bibfnamefont {F.~A.}\ \bibnamefont
  {Bornemann}}, \bibinfo {author} {\bibfnamefont {B.}~\bibnamefont {Schmidt}},\
  and\ \bibinfo {author} {\bibfnamefont {C.}~\bibnamefont {Sch{\"{u}}tte}},\
  }\bibfield  {title} {\bibinfo {title} {{An explicit and symplectic integrator
  for quantum-classical molecular dynamics}},\ }\href
  {https://doi.org/10.1016/0009-2614(96)00471-X} {\bibfield  {journal}
  {\bibinfo  {journal} {Chemical Physics Letters}\ }\textbf {\bibinfo {volume}
  {256}},\ \bibinfo {pages} {581} (\bibinfo {year} {1996})}\BibitemShut
  {NoStop}%
\bibitem [{\citenamefont {Choi}\ and\ \citenamefont
  {Van{\'{i}}{\v{c}}ek}(2021)}]{Choi2021}%
  \BibitemOpen
  \bibfield  {author} {\bibinfo {author} {\bibfnamefont {S.}~\bibnamefont
  {Choi}}\ and\ \bibinfo {author} {\bibfnamefont {J.}~\bibnamefont
  {Van{\'{i}}{\v{c}}ek}},\ }\bibfield  {title} {\bibinfo {title} {{High-order
  geometric integrators for representation-free Ehrenfest dynamics}},\ }\href
  {https://doi.org/10.1063/5.0061878} {\bibfield  {journal} {\bibinfo
  {journal} {The Journal of Chemical Physics}\ }\textbf {\bibinfo {volume}
  {155}},\ \bibinfo {pages} {124104} (\bibinfo {year} {2021})}\BibitemShut
  {NoStop}%
\bibitem [{\citenamefont {Tully}(1990)}]{Tully1990}%
  \BibitemOpen
  \bibfield  {author} {\bibinfo {author} {\bibfnamefont {J.~C.}\ \bibnamefont
  {Tully}},\ }\bibfield  {title} {\bibinfo {title} {{Molecular dynamics with
  electronic transitions}},\ }\href {https://doi.org/10.1063/1.459170}
  {\bibfield  {journal} {\bibinfo  {journal} {The Journal of Chemical Physics}\
  }\textbf {\bibinfo {volume} {93}},\ \bibinfo {pages} {1061} (\bibinfo {year}
  {1990})}\BibitemShut {NoStop}%
\bibitem [{\citenamefont {Xie}\ \emph {et~al.}(2020)\citenamefont {Xie},
  \citenamefont {Holub}, \citenamefont {Kuba{\v{s}}},\ and\ \citenamefont
  {Elstner}}]{Xie2020}%
  \BibitemOpen
  \bibfield  {author} {\bibinfo {author} {\bibfnamefont {W.}~\bibnamefont
  {Xie}}, \bibinfo {author} {\bibfnamefont {D.}~\bibnamefont {Holub}}, \bibinfo
  {author} {\bibfnamefont {T.}~\bibnamefont {Kuba{\v{s}}}},\ and\ \bibinfo
  {author} {\bibfnamefont {M.}~\bibnamefont {Elstner}},\ }\bibfield  {title}
  {\bibinfo {title} {{Performance of Mixed Quantum-Classical Approaches on
  Modeling the Crossover from Hopping to Bandlike Charge Transport in Organic
  Semiconductors}},\ }\href {https://doi.org/10.1021/acs.jctc.9b01271}
  {\bibfield  {journal} {\bibinfo  {journal} {Journal of Chemical Theory and
  Computation}\ }\textbf {\bibinfo {volume} {16}},\ \bibinfo {pages} {2071}
  (\bibinfo {year} {2020})}\BibitemShut {NoStop}%
\bibitem [{\citenamefont {Nelson}\ \emph {et~al.}(2020)\citenamefont {Nelson},
  \citenamefont {White}, \citenamefont {Bjorgaard}, \citenamefont {Sifain},
  \citenamefont {Zhang}, \citenamefont {Nebgen}, \citenamefont
  {Fernandez-Alberti}, \citenamefont {Mozyrsky}, \citenamefont {Roitberg},\
  and\ \citenamefont {Tretiak}}]{Nelson2020}%
  \BibitemOpen
  \bibfield  {author} {\bibinfo {author} {\bibfnamefont {T.~R.}\ \bibnamefont
  {Nelson}}, \bibinfo {author} {\bibfnamefont {A.~J.}\ \bibnamefont {White}},
  \bibinfo {author} {\bibfnamefont {J.~A.}\ \bibnamefont {Bjorgaard}}, \bibinfo
  {author} {\bibfnamefont {A.~E.}\ \bibnamefont {Sifain}}, \bibinfo {author}
  {\bibfnamefont {Y.}~\bibnamefont {Zhang}}, \bibinfo {author} {\bibfnamefont
  {B.}~\bibnamefont {Nebgen}}, \bibinfo {author} {\bibfnamefont
  {S.}~\bibnamefont {Fernandez-Alberti}}, \bibinfo {author} {\bibfnamefont
  {D.}~\bibnamefont {Mozyrsky}}, \bibinfo {author} {\bibfnamefont {A.~E.}\
  \bibnamefont {Roitberg}},\ and\ \bibinfo {author} {\bibfnamefont
  {S.}~\bibnamefont {Tretiak}},\ }\bibfield  {title} {\bibinfo {title}
  {{Non-adiabatic Excited-State Molecular Dynamics: Theory and Applications for
  Modeling Photophysics in Extended Molecular Materials}},\ }\href
  {https://doi.org/10.1021/acs.chemrev.9b00447} {\bibfield  {journal} {\bibinfo
   {journal} {Chemical Reviews}\ }\textbf {\bibinfo {volume} {120}},\ \bibinfo
  {pages} {2215} (\bibinfo {year} {2020})}\BibitemShut {NoStop}%
\bibitem [{\citenamefont {Freixas}\ \emph {et~al.}(2021)\citenamefont
  {Freixas}, \citenamefont {White}, \citenamefont {Nelson}, \citenamefont
  {Song}, \citenamefont {Makhov}, \citenamefont {Shalashilin}, \citenamefont
  {Fernandez-Alberti},\ and\ \citenamefont {Tretiak}}]{Freixas2021}%
  \BibitemOpen
  \bibfield  {author} {\bibinfo {author} {\bibfnamefont {V.~M.}\ \bibnamefont
  {Freixas}}, \bibinfo {author} {\bibfnamefont {A.~J.}\ \bibnamefont {White}},
  \bibinfo {author} {\bibfnamefont {T.}~\bibnamefont {Nelson}}, \bibinfo
  {author} {\bibfnamefont {H.}~\bibnamefont {Song}}, \bibinfo {author}
  {\bibfnamefont {D.~V.}\ \bibnamefont {Makhov}}, \bibinfo {author}
  {\bibfnamefont {D.}~\bibnamefont {Shalashilin}}, \bibinfo {author}
  {\bibfnamefont {S.}~\bibnamefont {Fernandez-Alberti}},\ and\ \bibinfo
  {author} {\bibfnamefont {S.}~\bibnamefont {Tretiak}},\ }\bibfield  {title}
  {\bibinfo {title} {{Nonadiabatic Excited-State Molecular Dynamics
  Methodologies: Comparison and Convergence}},\ }\href
  {https://doi.org/10.1021/acs.jpclett.1c00266} {\bibfield  {journal} {\bibinfo
   {journal} {Journal of Physical Chemistry Letters}\ }\textbf {\bibinfo
  {volume} {12}},\ \bibinfo {pages} {2970} (\bibinfo {year}
  {2021})}\BibitemShut {NoStop}%
\bibitem [{\citenamefont {Trugman}\ \emph {et~al.}(2004)\citenamefont
  {Trugman}, \citenamefont {Ku},\ and\ \citenamefont
  {Bon{\v{c}}a}}]{Trugman2004}%
  \BibitemOpen
  \bibfield  {author} {\bibinfo {author} {\bibfnamefont {S.~A.}\ \bibnamefont
  {Trugman}}, \bibinfo {author} {\bibfnamefont {L.-C.}\ \bibnamefont {Ku}},\
  and\ \bibinfo {author} {\bibfnamefont {J.}~\bibnamefont {Bon{\v{c}}a}},\
  }\bibfield  {title} {\bibinfo {title} {{Jahn–Teller and the Dynamics of
  Polaron Formation}},\ }\href
  {https://doi.org/10.1023/B:JOSC.0000021212.94285.d5} {\bibfield  {journal}
  {\bibinfo  {journal} {Journal of Superconductivity}\ }\textbf {\bibinfo
  {volume} {17}},\ \bibinfo {pages} {193} (\bibinfo {year} {2004})}\BibitemShut
  {NoStop}%
\bibitem [{\citenamefont {Ku}\ and\ \citenamefont {Trugman}(2007)}]{Ku2007}%
  \BibitemOpen
  \bibfield  {author} {\bibinfo {author} {\bibfnamefont {L.-C.}\ \bibnamefont
  {Ku}}\ and\ \bibinfo {author} {\bibfnamefont {S.~A.}\ \bibnamefont
  {Trugman}},\ }\bibfield  {title} {\bibinfo {title} {{Quantum dynamics of
  polaron formation}},\ }\href {https://doi.org/10.1103/PhysRevB.75.014307}
  {\bibfield  {journal} {\bibinfo  {journal} {Physical Review B}\ }\textbf
  {\bibinfo {volume} {75}},\ \bibinfo {pages} {014307} (\bibinfo {year}
  {2007})}\BibitemShut {NoStop}%
\bibitem [{\citenamefont {Beck}\ \emph {et~al.}(2000)\citenamefont {Beck},
  \citenamefont {J{\"{a}}ckle}, \citenamefont {Worth},\ and\ \citenamefont
  {Meyer}}]{Beck2000}%
  \BibitemOpen
  \bibfield  {author} {\bibinfo {author} {\bibfnamefont {M.~H.}\ \bibnamefont
  {Beck}}, \bibinfo {author} {\bibfnamefont {A.}~\bibnamefont {J{\"{a}}ckle}},
  \bibinfo {author} {\bibfnamefont {G.~A.}\ \bibnamefont {Worth}},\ and\
  \bibinfo {author} {\bibfnamefont {H.-D.}\ \bibnamefont {Meyer}},\ }\bibfield
  {title} {\bibinfo {title} {{The multiconfiguration time-dependent Hartree
  (MCTDH) method: a highly efficient algorithm for propagating wavepackets}},\
  }\href {https://doi.org/10.1016/S0370-1573(99)00047-2} {\bibfield  {journal}
  {\bibinfo  {journal} {Physics Reports}\ }\textbf {\bibinfo {volume} {324}},\
  \bibinfo {pages} {1} (\bibinfo {year} {2000})}\BibitemShut {NoStop}%
\bibitem [{\citenamefont {Meyer}\ \emph {et~al.}(2009)\citenamefont {Meyer},
  \citenamefont {Gatti},\ and\ \citenamefont {Worth}}]{Meyer2009}%
  \BibitemOpen
  \bibinfo {editor} {\bibfnamefont {H.~D.}\ \bibnamefont {Meyer}}, \bibinfo
  {editor} {\bibfnamefont {F.}~\bibnamefont {Gatti}},\ and\ \bibinfo {editor}
  {\bibfnamefont {G.~A.}\ \bibnamefont {Worth}},\ eds.,\ \href
  {https://doi.org/10.1002/9783527627400.ch3} {\emph {\bibinfo {title}
  {Multidimensional quantum dynamics: MCTDH theory and applications}}}\
  (\bibinfo  {publisher} {Wiley-VCH},\ \bibinfo {year} {2009})\BibitemShut
  {NoStop}%
\bibitem [{\citenamefont {Hitchcock}(1927)}]{Hitchcock1927}%
  \BibitemOpen
  \bibfield  {author} {\bibinfo {author} {\bibfnamefont {F.~L.}\ \bibnamefont
  {Hitchcock}},\ }\bibfield  {title} {\bibinfo {title} {The expression of a
  tensor or a polyadic as a sum of products},\ }\href
  {https://doi.org/10.1002/sapm192761164} {\bibfield  {journal} {\bibinfo
  {journal} {Journal of Mathematics and Physics}\ }\textbf {\bibinfo {volume}
  {6}},\ \bibinfo {pages} {164} (\bibinfo {year} {1927})}\BibitemShut {NoStop}%
\bibitem [{\citenamefont {Tucker}(1964)}]{Tucker1964}%
  \BibitemOpen
  \bibfield  {author} {\bibinfo {author} {\bibfnamefont {L.~R.}\ \bibnamefont
  {Tucker}},\ }\bibfield  {title} {\bibinfo {title} {{T}he extension of factor
  analysis to three-dimensional matrices},\ }in\ \href@noop {} {\emph {\bibinfo
  {booktitle} {Contributions to mathematical psychology}}},\ \bibinfo {editor}
  {edited by\ \bibinfo {editor} {\bibfnamefont {H.}~\bibnamefont {Gulliksen}}\
  and\ \bibinfo {editor} {\bibfnamefont {N.}~\bibnamefont {Frederiksen}}}\
  (\bibinfo  {publisher} {Holt, Rinehart and Winston},\ \bibinfo {year}
  {1964})\ pp.\ \bibinfo {pages} {110--127}\BibitemShut {NoStop}%
\bibitem [{\citenamefont {Oseledets}(2011)}]{Oseledets2011}%
  \BibitemOpen
  \bibfield  {author} {\bibinfo {author} {\bibfnamefont {I.~V.}\ \bibnamefont
  {Oseledets}},\ }\bibfield  {title} {\bibinfo {title} {Tensor-train
  decomposition},\ }\href {https://doi.org/10.1137/090752286} {\bibfield
  {journal} {\bibinfo  {journal} {SIAM Journal on Scientific Computing}\
  }\textbf {\bibinfo {volume} {33}},\ \bibinfo {pages} {2295} (\bibinfo {year}
  {2011})}\BibitemShut {NoStop}%
\bibitem [{\citenamefont {Or{\'{u}}s}(2014)}]{Orus2014}%
  \BibitemOpen
  \bibfield  {author} {\bibinfo {author} {\bibfnamefont {R.}~\bibnamefont
  {Or{\'{u}}s}},\ }\bibfield  {title} {\bibinfo {title} {{A practical
  introduction to tensor networks: Matrix product states and projected
  entangled pair states}},\ }\href {https://doi.org/10.1016/j.aop.2014.06.013}
  {\bibfield  {journal} {\bibinfo  {journal} {Annals of Physics}\ }\textbf
  {\bibinfo {volume} {349}},\ \bibinfo {pages} {117} (\bibinfo {year}
  {2014})}\BibitemShut {NoStop}%
\bibitem [{\citenamefont {Borrelli}\ and\ \citenamefont
  {Gelin}(2016)}]{Borrelli2016}%
  \BibitemOpen
  \bibfield  {author} {\bibinfo {author} {\bibfnamefont {R.}~\bibnamefont
  {Borrelli}}\ and\ \bibinfo {author} {\bibfnamefont {M.~F.}\ \bibnamefont
  {Gelin}},\ }\bibfield  {title} {\bibinfo {title} {{Quantum
  electron-vibrational dynamics at finite temperature: Thermo field dynamics
  approach}},\ }\bibfield  {journal} {\bibinfo  {journal} {Journal of Chemical
  Physics}\ }\textbf {\bibinfo {volume} {145}},\ \href
  {https://doi.org/10.1063/1.4971211} {10.1063/1.4971211} (\bibinfo {year}
  {2016})\BibitemShut {NoStop}%
\bibitem [{\citenamefont {Bose}\ and\ \citenamefont
  {Walters}(2021)}]{Bose2021}%
  \BibitemOpen
  \bibfield  {author} {\bibinfo {author} {\bibfnamefont {A.}~\bibnamefont
  {Bose}}\ and\ \bibinfo {author} {\bibfnamefont {P.~L.}\ \bibnamefont
  {Walters}},\ }\bibfield  {title} {\bibinfo {title} {{A Multisite
  Decomposition of the Tensor Network Path Integrals}},\ }\href
  {https://doi.org/10.1063/5.0073234} {\bibfield  {journal} {\bibinfo
  {journal} {The Journal of Chemical Physics}\ }\textbf {\bibinfo {volume}
  {156}},\ \bibinfo {pages} {024101} (\bibinfo {year} {2021})}\BibitemShut
  {NoStop}%
\bibitem [{\citenamefont {Gelß}\ \emph {et~al.}(2022)\citenamefont {Gelß},
  \citenamefont {Klus}, \citenamefont {Shakibaei},\ and\ \citenamefont
  {Pokutta}}]{Gelss2022b}%
  \BibitemOpen
  \bibfield  {author} {\bibinfo {author} {\bibfnamefont {P.}~\bibnamefont
  {Gelß}}, \bibinfo {author} {\bibfnamefont {S.}~\bibnamefont {Klus}},
  \bibinfo {author} {\bibfnamefont {Z.}~\bibnamefont {Shakibaei}},\ and\
  \bibinfo {author} {\bibfnamefont {S.}~\bibnamefont {Pokutta}},\ }\bibfield
  {title} {\bibinfo {title} {Low-rank tensor decompositions of quantum
  circuits},\ }\href@noop {} {\bibfield  {journal} {\bibinfo  {journal}
  {arXiv:2205.09882}\ } (\bibinfo {year} {2022})}\BibitemShut {NoStop}%
\bibitem [{\citenamefont {Karahan}\ \emph {et~al.}(2015)\citenamefont
  {Karahan}, \citenamefont {Rojas-López}, \citenamefont {Bringas-Vega},
  \citenamefont {Valdés-Hernández},\ and\ \citenamefont
  {Valdes-Sosa}}]{Karahan2015}%
  \BibitemOpen
  \bibfield  {author} {\bibinfo {author} {\bibfnamefont {E.}~\bibnamefont
  {Karahan}}, \bibinfo {author} {\bibfnamefont {P.~A.}\ \bibnamefont
  {Rojas-López}}, \bibinfo {author} {\bibfnamefont {M.~L.}\ \bibnamefont
  {Bringas-Vega}}, \bibinfo {author} {\bibfnamefont {P.~A.}\ \bibnamefont
  {Valdés-Hernández}},\ and\ \bibinfo {author} {\bibfnamefont {P.~A.}\
  \bibnamefont {Valdes-Sosa}},\ }\bibfield  {title} {\bibinfo {title} {Tensor
  analysis and fusion of multimodal brain images},\ }\href
  {https://doi.org/10.1109/JPROC.2015.2455028} {\bibfield  {journal} {\bibinfo
  {journal} {Proceedings of the IEEE}\ }\textbf {\bibinfo {volume} {103}},\
  \bibinfo {pages} {1531} (\bibinfo {year} {2015})}\BibitemShut {NoStop}%
\bibitem [{\citenamefont {Chatzichristos}\ \emph {et~al.}(2019)\citenamefont
  {Chatzichristos}, \citenamefont {Kofidis}, \citenamefont {Morante},\ and\
  \citenamefont {Theodoridis}}]{Chatzichristos2019}%
  \BibitemOpen
  \bibfield  {author} {\bibinfo {author} {\bibfnamefont {C.}~\bibnamefont
  {Chatzichristos}}, \bibinfo {author} {\bibfnamefont {E.}~\bibnamefont
  {Kofidis}}, \bibinfo {author} {\bibfnamefont {M.}~\bibnamefont {Morante}},\
  and\ \bibinfo {author} {\bibfnamefont {S.}~\bibnamefont {Theodoridis}},\
  }\bibfield  {title} {\bibinfo {title} {Blind {fMRI} source unmixing via
  higher-order tensor decompositions},\ }\href
  {https://doi.org/10.1016/j.jneumeth.2018.12.007} {\bibfield  {journal}
  {\bibinfo  {journal} {Journal of Neuroscience Methods}\ }\textbf {\bibinfo
  {volume} {315}},\ \bibinfo {pages} {17} (\bibinfo {year} {2019})}\BibitemShut
  {NoStop}%
\bibitem [{\citenamefont {Erol}\ and\ \citenamefont
  {Hunyadi}(2022)}]{Erol2022}%
  \BibitemOpen
  \bibfield  {author} {\bibinfo {author} {\bibfnamefont {A.}~\bibnamefont
  {Erol}}\ and\ \bibinfo {author} {\bibfnamefont {B.}~\bibnamefont {Hunyadi}},\
  }\bibfield  {title} {\bibinfo {title} {Chapter 12 - tensors for neuroimaging:
  A review on applications of tensors to unravel the mysteries of the brain},\
  }in\ \href {https://doi.org/10.1016/B978-0-12-824447-0.00018-2} {\emph
  {\bibinfo {booktitle} {Tensors for Data Processing}}},\ \bibinfo {editor}
  {edited by\ \bibinfo {editor} {\bibfnamefont {Y.}~\bibnamefont {Liu}}}\
  (\bibinfo  {publisher} {Academic Press},\ \bibinfo {year} {2022})\ pp.\
  \bibinfo {pages} {427--482}\BibitemShut {NoStop}%
\bibitem [{\citenamefont {Klus}\ \emph {et~al.}(2018)\citenamefont {Klus},
  \citenamefont {Gel{\ss}}, \citenamefont {Peitz},\ and\ \citenamefont
  {Schütte}}]{Klus2018}%
  \BibitemOpen
  \bibfield  {author} {\bibinfo {author} {\bibfnamefont {S.}~\bibnamefont
  {Klus}}, \bibinfo {author} {\bibfnamefont {P.}~\bibnamefont {Gel{\ss}}},
  \bibinfo {author} {\bibfnamefont {S.}~\bibnamefont {Peitz}},\ and\ \bibinfo
  {author} {\bibfnamefont {C.}~\bibnamefont {Schütte}},\ }\bibfield  {title}
  {\bibinfo {title} {Tensor-based dynamic mode decomposition},\ }\href
  {https://doi.org/10.1088/1361-6544/aabc8f} {\bibfield  {journal} {\bibinfo
  {journal} {Nonlinearity}\ }\textbf {\bibinfo {volume} {31}},\ \bibinfo
  {pages} {3359} (\bibinfo {year} {2018})}\BibitemShut {NoStop}%
\bibitem [{\citenamefont {Gel{\ss}}\ \emph {et~al.}(2019)\citenamefont
  {Gel{\ss}}, \citenamefont {Klus}, \citenamefont {Eisert},\ and\ \citenamefont
  {Sch{\"u}tte}}]{Gelss2019}%
  \BibitemOpen
  \bibfield  {author} {\bibinfo {author} {\bibfnamefont {P.}~\bibnamefont
  {Gel{\ss}}}, \bibinfo {author} {\bibfnamefont {S.}~\bibnamefont {Klus}},
  \bibinfo {author} {\bibfnamefont {J.}~\bibnamefont {Eisert}},\ and\ \bibinfo
  {author} {\bibfnamefont {C.}~\bibnamefont {Sch{\"u}tte}},\ }\bibfield
  {title} {\bibinfo {title} {Multidimensional approximation of nonlinear
  dynamical systems},\ }\href {https://doi.org/10.1115/1.4043148} {\bibfield
  {journal} {\bibinfo  {journal} {Journal of Computational and Nonlinear
  Dynamics}\ }\textbf {\bibinfo {volume} {14}},\ \bibinfo {pages} {061006}
  (\bibinfo {year} {2019})}\BibitemShut {NoStop}%
\bibitem [{\citenamefont {Kargas}\ and\ \citenamefont
  {Sidiropoulos}(2020{\natexlab{a}})}]{Kargas2020a}%
  \BibitemOpen
  \bibfield  {author} {\bibinfo {author} {\bibfnamefont {N.}~\bibnamefont
  {Kargas}}\ and\ \bibinfo {author} {\bibfnamefont {N.~D.}\ \bibnamefont
  {Sidiropoulos}},\ }\bibfield  {title} {\bibinfo {title} {Nonlinear system
  identification via tensor completion},\ }in\ \href@noop {} {\emph {\bibinfo
  {booktitle} {Proceedings of the AAAI Conference on Artificial
  Intelligence}}},\ Vol.~\bibinfo {volume} {34}\ (\bibinfo {year} {2020})\ pp.\
  \bibinfo {pages} {4420--4427}\BibitemShut {NoStop}%
\bibitem [{\citenamefont {Stoudenmire}\ and\ \citenamefont
  {Schwab}(2016)}]{Stoudenmire2016}%
  \BibitemOpen
  \bibfield  {author} {\bibinfo {author} {\bibfnamefont {E.}~\bibnamefont
  {Stoudenmire}}\ and\ \bibinfo {author} {\bibfnamefont {D.~J.}\ \bibnamefont
  {Schwab}},\ }\bibfield  {title} {\bibinfo {title} {Supervised learning with
  tensor networks},\ }in\ \href@noop {} {\emph {\bibinfo {booktitle} {Advances
  in Neural Information Processing Systems}}},\ Vol.~\bibinfo {volume} {29},\
  \bibinfo {editor} {edited by\ \bibinfo {editor} {\bibfnamefont
  {D.}~\bibnamefont {Lee}}, \bibinfo {editor} {\bibfnamefont {M.}~\bibnamefont
  {Sugiyama}}, \bibinfo {editor} {\bibfnamefont {U.}~\bibnamefont {Luxburg}},
  \bibinfo {editor} {\bibfnamefont {I.}~\bibnamefont {Guyon}},\ and\ \bibinfo
  {editor} {\bibfnamefont {R.}~\bibnamefont {Garnett}}}\ (\bibinfo  {publisher}
  {Curran Associates, Inc.},\ \bibinfo {year} {2016})\BibitemShut {NoStop}%
\bibitem [{\citenamefont {Klus}\ and\ \citenamefont
  {Gel{\ss}}(2019)}]{Klus2019}%
  \BibitemOpen
  \bibfield  {author} {\bibinfo {author} {\bibfnamefont {S.}~\bibnamefont
  {Klus}}\ and\ \bibinfo {author} {\bibfnamefont {P.}~\bibnamefont
  {Gel{\ss}}},\ }\bibfield  {title} {\bibinfo {title} {Tensor-based algorithms
  for image classification},\ }\href {https://doi.org/10.3390/a12110240}
  {\bibfield  {journal} {\bibinfo  {journal} {Algorithms}\ }\textbf {\bibinfo
  {volume} {12}},\ \bibinfo {pages} {240} (\bibinfo {year} {2019})}\BibitemShut
  {NoStop}%
\bibitem [{\citenamefont {Kargas}\ and\ \citenamefont
  {Sidiropoulos}(2020{\natexlab{b}})}]{Kargas2020b}%
  \BibitemOpen
  \bibfield  {author} {\bibinfo {author} {\bibfnamefont {N.}~\bibnamefont
  {Kargas}}\ and\ \bibinfo {author} {\bibfnamefont {N.~D.}\ \bibnamefont
  {Sidiropoulos}},\ }\bibfield  {title} {\bibinfo {title} {Supervised learning
  via ensemble tensor completion},\ }in\ \href
  {https://doi.org/10.1109/IEEECONF51394.2020.9443399} {\emph {\bibinfo
  {booktitle} {2020 54th Asilomar Conference on Signals, Systems, and
  Computers}}}\ (\bibinfo {year} {2020})\ pp.\ \bibinfo {pages}
  {196--199}\BibitemShut {NoStop}%
\bibitem [{\citenamefont {Gel{\ss}}\ \emph {et~al.}(2022)\citenamefont
  {Gel{\ss}}, \citenamefont {Klein}, \citenamefont {Matera},\ and\
  \citenamefont {Schmidt}}]{Gelss2022a}%
  \BibitemOpen
  \bibfield  {author} {\bibinfo {author} {\bibfnamefont {P.}~\bibnamefont
  {Gel{\ss}}}, \bibinfo {author} {\bibfnamefont {R.}~\bibnamefont {Klein}},
  \bibinfo {author} {\bibfnamefont {S.}~\bibnamefont {Matera}},\ and\ \bibinfo
  {author} {\bibfnamefont {B.}~\bibnamefont {Schmidt}},\ }\bibfield  {title}
  {\bibinfo {title} {{Solving the time-independent Schr{\"{o}}dinger equation
  for chains of coupled excitons and phonons using tensor trains}},\ }\href
  {https://doi.org/10.1063/5.0074948} {\bibfield  {journal} {\bibinfo
  {journal} {The Journal of Chemical Physics}\ }\textbf {\bibinfo {volume}
  {156}},\ \bibinfo {pages} {024109} (\bibinfo {year} {2022})}\BibitemShut
  {NoStop}%
\bibitem [{\citenamefont {Gel{\ss}}\ \emph {et~al.}(2017)\citenamefont
  {Gel{\ss}}, \citenamefont {Klus}, \citenamefont {Matera},\ and\ \citenamefont
  {Sch{\"{u}}tte}}]{Gelss2017}%
  \BibitemOpen
  \bibfield  {author} {\bibinfo {author} {\bibfnamefont {P.}~\bibnamefont
  {Gel{\ss}}}, \bibinfo {author} {\bibfnamefont {S.}~\bibnamefont {Klus}},
  \bibinfo {author} {\bibfnamefont {S.}~\bibnamefont {Matera}},\ and\ \bibinfo
  {author} {\bibfnamefont {C.}~\bibnamefont {Sch{\"{u}}tte}},\ }\bibfield
  {title} {\bibinfo {title} {{Nearest-neighbor interaction systems in the
  tensor-train format}},\ }\href {https://doi.org/10.1016/j.jcp.2017.04.007}
  {\bibfield  {journal} {\bibinfo  {journal} {Journal of Computational
  Physics}\ }\textbf {\bibinfo {volume} {341}},\ \bibinfo {pages} {140}
  (\bibinfo {year} {2017})}\BibitemShut {NoStop}%
\bibitem [{\citenamefont {Hackbusch}(2012)}]{Hackbusch2012}%
  \BibitemOpen
  \bibfield  {author} {\bibinfo {author} {\bibfnamefont {W.}~\bibnamefont
  {Hackbusch}},\ }\href {https://doi.org/10.1007/978-3-642-28027-6} {\emph
  {\bibinfo {title} {Tensor Spaces and Numerical Tensor Calculus}}},\
  Vol.~\bibinfo {volume} {42}\ (\bibinfo {year} {2012})\BibitemShut {NoStop}%
\bibitem [{\citenamefont {Mainali}\ \emph {et~al.}(2021)\citenamefont
  {Mainali}, \citenamefont {Gatti}, \citenamefont {Iouchtchenko}, \citenamefont
  {Roy},\ and\ \citenamefont {Meyer}}]{Mainali2021}%
  \BibitemOpen
  \bibfield  {author} {\bibinfo {author} {\bibfnamefont {S.}~\bibnamefont
  {Mainali}}, \bibinfo {author} {\bibfnamefont {F.}~\bibnamefont {Gatti}},
  \bibinfo {author} {\bibfnamefont {D.}~\bibnamefont {Iouchtchenko}}, \bibinfo
  {author} {\bibfnamefont {P.-N.}\ \bibnamefont {Roy}},\ and\ \bibinfo {author}
  {\bibfnamefont {H.-D.}\ \bibnamefont {Meyer}},\ }\bibfield  {title} {\bibinfo
  {title} {{Comparison of the multi-layer multi-configuration time-dependent
  Hartree (ML-MCTDH) method and the density matrix renormalization group (DMRG)
  for ground state properties of linear rotor chains}},\ }\href
  {https://doi.org/10.1063/5.0047090} {\bibfield  {journal} {\bibinfo
  {journal} {The Journal of Chemical Physics}\ }\textbf {\bibinfo {volume}
  {154}},\ \bibinfo {pages} {174106} (\bibinfo {year} {2021})}\BibitemShut
  {NoStop}%
\bibitem [{\citenamefont {Serwatka}\ and\ \citenamefont
  {Roy}(2022)}]{Serwatka2022}%
  \BibitemOpen
  \bibfield  {author} {\bibinfo {author} {\bibfnamefont {T.}~\bibnamefont
  {Serwatka}}\ and\ \bibinfo {author} {\bibfnamefont {P.-N.}\ \bibnamefont
  {Roy}},\ }\bibfield  {title} {\bibinfo {title} {{Ground state of asymmetric
  tops with DMRG: water in one dimension}},\ }\href
  {https://doi.org/10.1063/5.0078770} {\bibfield  {journal} {\bibinfo
  {journal} {The Journal of Chemical Physics}\ }\textbf {\bibinfo {volume}
  {156}},\ \bibinfo {pages} {044116} (\bibinfo {year} {2022})}\BibitemShut
  {NoStop}%
\bibitem [{\citenamefont {Lubich}\ and\ \citenamefont
  {Oseledets}(2014)}]{Lubich2014}%
  \BibitemOpen
  \bibfield  {author} {\bibinfo {author} {\bibfnamefont {C.}~\bibnamefont
  {Lubich}}\ and\ \bibinfo {author} {\bibfnamefont {I.~V.}\ \bibnamefont
  {Oseledets}},\ }\bibfield  {title} {\bibinfo {title} {{A projector-splitting
  integrator for dynamical low-rank approximation}},\ }\href
  {https://doi.org/10.1007/s10543-013-0454-0} {\bibfield  {journal} {\bibinfo
  {journal} {BIT Numerical Mathematics}\ }\textbf {\bibinfo {volume} {54}},\
  \bibinfo {pages} {171} (\bibinfo {year} {2014})}\BibitemShut {NoStop}%
\bibitem [{\citenamefont {Lubich}\ \emph {et~al.}(2015)\citenamefont {Lubich},
  \citenamefont {Oseledets},\ and\ \citenamefont {Vandereycken}}]{Lubich2015}%
  \BibitemOpen
  \bibfield  {author} {\bibinfo {author} {\bibfnamefont {C.}~\bibnamefont
  {Lubich}}, \bibinfo {author} {\bibfnamefont {I.~V.}\ \bibnamefont
  {Oseledets}},\ and\ \bibinfo {author} {\bibfnamefont {B.}~\bibnamefont
  {Vandereycken}},\ }\bibfield  {title} {\bibinfo {title} {{Time Integration of
  Tensor Trains}},\ }\href {https://doi.org/10.1137/140976546} {\bibfield
  {journal} {\bibinfo  {journal} {SIAM Journal on Numerical Analysis}\ }\textbf
  {\bibinfo {volume} {53}},\ \bibinfo {pages} {917} (\bibinfo {year}
  {2015})}\BibitemShut {NoStop}%
\bibitem [{\citenamefont {Haegeman}\ \emph {et~al.}(2016)\citenamefont
  {Haegeman}, \citenamefont {Lubich}, \citenamefont {Oseledets}, \citenamefont
  {Vandereycken},\ and\ \citenamefont {Verstraete}}]{Haegeman2016}%
  \BibitemOpen
  \bibfield  {author} {\bibinfo {author} {\bibfnamefont {J.}~\bibnamefont
  {Haegeman}}, \bibinfo {author} {\bibfnamefont {C.}~\bibnamefont {Lubich}},
  \bibinfo {author} {\bibfnamefont {I.}~\bibnamefont {Oseledets}}, \bibinfo
  {author} {\bibfnamefont {B.}~\bibnamefont {Vandereycken}},\ and\ \bibinfo
  {author} {\bibfnamefont {F.}~\bibnamefont {Verstraete}},\ }\bibfield  {title}
  {\bibinfo {title} {{Unifying time evolution and optimization with matrix
  product states}},\ }\href {https://doi.org/10.1103/PhysRevB.94.165116}
  {\bibfield  {journal} {\bibinfo  {journal} {Physical Review B}\ }\textbf
  {\bibinfo {volume} {94}},\ \bibinfo {pages} {165116} (\bibinfo {year}
  {2016})}\BibitemShut {NoStop}%
\bibitem [{\citenamefont {Choi}\ and\ \citenamefont
  {Van{\'{i}}{\v{c}}ek}(2019)}]{Choi2019}%
  \BibitemOpen
  \bibfield  {author} {\bibinfo {author} {\bibfnamefont {S.}~\bibnamefont
  {Choi}}\ and\ \bibinfo {author} {\bibfnamefont {J.}~\bibnamefont
  {Van{\'{i}}{\v{c}}ek}},\ }\bibfield  {title} {\bibinfo {title} {{Efficient
  geometric integrators for nonadiabatic quantum dynamics. I. The adiabatic
  representation}},\ }\href {https://doi.org/10.1063/1.5092611} {\bibfield
  {journal} {\bibinfo  {journal} {The Journal of Chemical Physics}\ }\textbf
  {\bibinfo {volume} {150}},\ \bibinfo {pages} {204112} (\bibinfo {year}
  {2019})}\BibitemShut {NoStop}%
\bibitem [{\citenamefont {Feit}\ \emph {et~al.}(1982)\citenamefont {Feit},
  \citenamefont {Fleck},\ and\ \citenamefont {Steiger}}]{Feit1982}%
  \BibitemOpen
  \bibfield  {author} {\bibinfo {author} {\bibfnamefont {M.~D.}\ \bibnamefont
  {Feit}}, \bibinfo {author} {\bibfnamefont {J.~A.}\ \bibnamefont {Fleck}},\
  and\ \bibinfo {author} {\bibfnamefont {A.}~\bibnamefont {Steiger}},\
  }\bibfield  {title} {\bibinfo {title} {{Solution of the Schr{\"{o}}dinger
  equation by a spectral method}},\ }\href
  {https://doi.org/10.1016/0021-9991(82)90091-2} {\bibfield  {journal}
  {\bibinfo  {journal} {Journal of Computational Physics}\ }\textbf {\bibinfo
  {volume} {47}},\ \bibinfo {pages} {412} (\bibinfo {year} {1982})}\BibitemShut
  {NoStop}%
\bibitem [{\citenamefont {Roulet}\ \emph {et~al.}(2019)\citenamefont {Roulet},
  \citenamefont {Choi},\ and\ \citenamefont
  {Van{\'{i}}{\v{c}}ek}}]{Roulet2019}%
  \BibitemOpen
  \bibfield  {author} {\bibinfo {author} {\bibfnamefont {J.}~\bibnamefont
  {Roulet}}, \bibinfo {author} {\bibfnamefont {S.}~\bibnamefont {Choi}},\ and\
  \bibinfo {author} {\bibfnamefont {J.}~\bibnamefont {Van{\'{i}}{\v{c}}ek}},\
  }\bibfield  {title} {\bibinfo {title} {{Efficient geometric integrators for
  nonadiabatic quantum dynamics. II. the diabatic representation}},\ }\href
  {https://doi.org/10.1063/1.5094046} {\bibfield  {journal} {\bibinfo
  {journal} {Journal of Chemical Physics}\ }\textbf {\bibinfo {volume} {150}},\
  \bibinfo {pages} {204113} (\bibinfo {year} {2019})}\BibitemShut {NoStop}%
\bibitem [{\citenamefont {Volokitin}\ \emph {et~al.}(2019)\citenamefont
  {Volokitin}, \citenamefont {Vakulchyk}, \citenamefont {Kozinov},
  \citenamefont {Liniov}, \citenamefont {Meyerov}, \citenamefont {Ivanchenko},
  \citenamefont {Laptyeva},\ and\ \citenamefont {Denisov}}]{Volokitin2019}%
  \BibitemOpen
  \bibfield  {author} {\bibinfo {author} {\bibfnamefont {V.}~\bibnamefont
  {Volokitin}}, \bibinfo {author} {\bibfnamefont {I.}~\bibnamefont
  {Vakulchyk}}, \bibinfo {author} {\bibfnamefont {E.}~\bibnamefont {Kozinov}},
  \bibinfo {author} {\bibfnamefont {A.}~\bibnamefont {Liniov}}, \bibinfo
  {author} {\bibfnamefont {I.}~\bibnamefont {Meyerov}}, \bibinfo {author}
  {\bibfnamefont {M.}~\bibnamefont {Ivanchenko}}, \bibinfo {author}
  {\bibfnamefont {T.}~\bibnamefont {Laptyeva}},\ and\ \bibinfo {author}
  {\bibfnamefont {S.}~\bibnamefont {Denisov}},\ }\bibfield  {title} {\bibinfo
  {title} {{Propagating large open quantum systems towards their asymptotic
  states: cluster implementation of the time-evolving block decimation
  scheme}},\ }\href {https://doi.org/10.1088/1742-6596/1392/1/012061}
  {\bibfield  {journal} {\bibinfo  {journal} {Journal of Physics: Conference
  Series}\ }\textbf {\bibinfo {volume} {1392}},\ \bibinfo {pages} {012061}
  (\bibinfo {year} {2019})}\BibitemShut {NoStop}%
\bibitem [{\citenamefont {Paeckel}\ \emph {et~al.}(2019)\citenamefont
  {Paeckel}, \citenamefont {K{\"{o}}hler}, \citenamefont {Swoboda},
  \citenamefont {Manmana}, \citenamefont {Schollw{\"{o}}ck},\ and\
  \citenamefont {Hubig}}]{Paeckel2019}%
  \BibitemOpen
  \bibfield  {author} {\bibinfo {author} {\bibfnamefont {S.}~\bibnamefont
  {Paeckel}}, \bibinfo {author} {\bibfnamefont {T.}~\bibnamefont
  {K{\"{o}}hler}}, \bibinfo {author} {\bibfnamefont {A.}~\bibnamefont
  {Swoboda}}, \bibinfo {author} {\bibfnamefont {S.~R.}\ \bibnamefont
  {Manmana}}, \bibinfo {author} {\bibfnamefont {U.}~\bibnamefont
  {Schollw{\"{o}}ck}},\ and\ \bibinfo {author} {\bibfnamefont {C.}~\bibnamefont
  {Hubig}},\ }\bibfield  {title} {\bibinfo {title} {{Time-evolution methods for
  matrix-product states}},\ }\href {https://doi.org/10.1016/j.aop.2019.167998}
  {\bibfield  {journal} {\bibinfo  {journal} {Annals of Physics}\ }\textbf
  {\bibinfo {volume} {411}},\ \bibinfo {pages} {167998} (\bibinfo {year}
  {2019})}\BibitemShut {NoStop}%
\bibitem [{\citenamefont {Leforestier}\ \emph {et~al.}(1991)\citenamefont
  {Leforestier}, \citenamefont {Bisseling}, \citenamefont {Cerjan},
  \citenamefont {Feit}, \citenamefont {Friesner}, \citenamefont {Guldberg},
  \citenamefont {Hammerich}, \citenamefont {Jolicard}, \citenamefont
  {Karrlein}, \citenamefont {Meyer}, \citenamefont {Lipkin}, \citenamefont
  {Roncero},\ and\ \citenamefont {Kosloff}}]{Leforestier1991}%
  \BibitemOpen
  \bibfield  {author} {\bibinfo {author} {\bibfnamefont {C.}~\bibnamefont
  {Leforestier}}, \bibinfo {author} {\bibfnamefont {R.}~\bibnamefont
  {Bisseling}}, \bibinfo {author} {\bibfnamefont {C.}~\bibnamefont {Cerjan}},
  \bibinfo {author} {\bibfnamefont {M.}~\bibnamefont {Feit}}, \bibinfo {author}
  {\bibfnamefont {R.}~\bibnamefont {Friesner}}, \bibinfo {author}
  {\bibfnamefont {A.}~\bibnamefont {Guldberg}}, \bibinfo {author}
  {\bibfnamefont {A.}~\bibnamefont {Hammerich}}, \bibinfo {author}
  {\bibfnamefont {G.}~\bibnamefont {Jolicard}}, \bibinfo {author}
  {\bibfnamefont {W.}~\bibnamefont {Karrlein}}, \bibinfo {author}
  {\bibfnamefont {H.-D.}\ \bibnamefont {Meyer}}, \bibinfo {author}
  {\bibfnamefont {N.}~\bibnamefont {Lipkin}}, \bibinfo {author} {\bibfnamefont
  {O.}~\bibnamefont {Roncero}},\ and\ \bibinfo {author} {\bibfnamefont
  {R.}~\bibnamefont {Kosloff}},\ }\bibfield  {title} {\bibinfo {title} {{A
  comparison of different propagation schemes for the time dependent
  Schr{\"{o}}dinger equation}},\ }\href
  {https://doi.org/10.1016/0021-9991(91)90137-A} {\bibfield  {journal}
  {\bibinfo  {journal} {Journal of Computational Physics}\ }\textbf {\bibinfo
  {volume} {94}},\ \bibinfo {pages} {59} (\bibinfo {year} {1991})}\BibitemShut
  {NoStop}%
\bibitem [{\citenamefont {Holtz}\ \emph {et~al.}(2012)\citenamefont {Holtz},
  \citenamefont {Rohwedder},\ and\ \citenamefont {Schneider}}]{Holtz2012}%
  \BibitemOpen
  \bibfield  {author} {\bibinfo {author} {\bibfnamefont {S.}~\bibnamefont
  {Holtz}}, \bibinfo {author} {\bibfnamefont {T.}~\bibnamefont {Rohwedder}},\
  and\ \bibinfo {author} {\bibfnamefont {R.}~\bibnamefont {Schneider}},\
  }\bibfield  {title} {\bibinfo {title} {{The Alternating Linear Scheme for
  Tensor Optimization in the Tensor Train Format}},\ }\href
  {https://doi.org/10.1137/100818893} {\bibfield  {journal} {\bibinfo
  {journal} {SIAM Journal on Scientific Computing}\ }\textbf {\bibinfo {volume}
  {34}},\ \bibinfo {pages} {A683} (\bibinfo {year} {2012})}\BibitemShut
  {NoStop}%
\bibitem [{\citenamefont {Hedley}\ \emph {et~al.}(2017)\citenamefont {Hedley},
  \citenamefont {Ruseckas},\ and\ \citenamefont {Samuel}}]{Hedley2016}%
  \BibitemOpen
  \bibfield  {author} {\bibinfo {author} {\bibfnamefont {G.~J.}\ \bibnamefont
  {Hedley}}, \bibinfo {author} {\bibfnamefont {A.}~\bibnamefont {Ruseckas}},\
  and\ \bibinfo {author} {\bibfnamefont {I.~D.~W.}\ \bibnamefont {Samuel}},\
  }\bibfield  {title} {\bibinfo {title} {{Light Harvesting for Organic
  Photovoltaics}},\ }\href {https://doi.org/10.1021/acs.chemrev.6b00215}
  {\bibfield  {journal} {\bibinfo  {journal} {Chemical Reviews}\ }\textbf
  {\bibinfo {volume} {117}},\ \bibinfo {pages} {796} (\bibinfo {year}
  {2017})}\BibitemShut {NoStop}%
\bibitem [{\citenamefont {Dexter}(1953)}]{Dexter1953}%
  \BibitemOpen
  \bibfield  {author} {\bibinfo {author} {\bibfnamefont {D.~L.}\ \bibnamefont
  {Dexter}},\ }\bibfield  {title} {\bibinfo {title} {A theory of sensitized
  luminescence in solids},\ }\href {https://doi.org/10.1063/1.1699044}
  {\bibfield  {journal} {\bibinfo  {journal} {The Journal of Chemical Physics}\
  }\textbf {\bibinfo {volume} {21}},\ \bibinfo {pages} {836} (\bibinfo {year}
  {1953})}\BibitemShut {NoStop}%
\bibitem [{\citenamefont {Dexter}\ \emph {et~al.}(1969)\citenamefont {Dexter},
  \citenamefont {Knox},\ and\ \citenamefont {Förster}}]{Dexter1969}%
  \BibitemOpen
  \bibfield  {author} {\bibinfo {author} {\bibfnamefont {D.~L.}\ \bibnamefont
  {Dexter}}, \bibinfo {author} {\bibfnamefont {R.~S.}\ \bibnamefont {Knox}},\
  and\ \bibinfo {author} {\bibfnamefont {T.}~\bibnamefont {Förster}},\
  }\bibfield  {title} {\bibinfo {title} {The radiationless transfer of energy
  of electronic excitation between impurity molecules in crystals},\ }\href
  {https://doi.org/10.1002/pssb.19690340264} {\bibfield  {journal} {\bibinfo
  {journal} {physica status solidi (b)}\ }\textbf {\bibinfo {volume} {34}},\
  \bibinfo {pages} {K159} (\bibinfo {year} {1969})}\BibitemShut {NoStop}%
\bibitem [{\citenamefont {Scott}(1991)}]{Scott1991}%
  \BibitemOpen
  \bibfield  {author} {\bibinfo {author} {\bibfnamefont {A.~C.}\ \bibnamefont
  {Scott}},\ }\bibfield  {title} {\bibinfo {title} {{Davydov's soliton
  revisited}},\ }\href {https://doi.org/10.1016/0167-2789(91)90243-3}
  {\bibfield  {journal} {\bibinfo  {journal} {Physica D: Nonlinear Phenomena}\
  }\textbf {\bibinfo {volume} {51}},\ \bibinfo {pages} {333} (\bibinfo {year}
  {1991})}\BibitemShut {NoStop}%
\bibitem [{\citenamefont {Oseledets}(2009)}]{Oseledets2009a}%
  \BibitemOpen
  \bibfield  {author} {\bibinfo {author} {\bibfnamefont {I.~V.}\ \bibnamefont
  {Oseledets}},\ }\bibfield  {title} {\bibinfo {title} {A new tensor
  decomposition},\ }\href {https://doi.org/10.1134/S1064562409040115}
  {\bibfield  {journal} {\bibinfo  {journal} {Doklady Mathematics}\ }\textbf
  {\bibinfo {volume} {80}},\ \bibinfo {pages} {495} (\bibinfo {year}
  {2009})}\BibitemShut {NoStop}%
\bibitem [{\citenamefont {Oseledets}\ and\ \citenamefont
  {Tyrtyshnikov}(2009)}]{Oseledets2009b}%
  \BibitemOpen
  \bibfield  {author} {\bibinfo {author} {\bibfnamefont {I.~V.}\ \bibnamefont
  {Oseledets}}\ and\ \bibinfo {author} {\bibfnamefont {E.~E.}\ \bibnamefont
  {Tyrtyshnikov}},\ }\bibfield  {title} {\bibinfo {title} {Breaking the curse
  of dimensionality, or how to use {SVD} in many dimensions},\ }\href
  {https://doi.org/10.1137/090748330} {\bibfield  {journal} {\bibinfo
  {journal} {SIAM Journal on Scientific Computing}\ }\textbf {\bibinfo {volume}
  {31}},\ \bibinfo {pages} {3744} (\bibinfo {year} {2009})}\BibitemShut
  {NoStop}%
\bibitem [{\citenamefont {Kazeev}\ \emph {et~al.}(2013)\citenamefont {Kazeev},
  \citenamefont {Reichmann},\ and\ \citenamefont {Schwab}}]{Kazeev2013}%
  \BibitemOpen
  \bibfield  {author} {\bibinfo {author} {\bibfnamefont {V.}~\bibnamefont
  {Kazeev}}, \bibinfo {author} {\bibfnamefont {O.}~\bibnamefont {Reichmann}},\
  and\ \bibinfo {author} {\bibfnamefont {C.}~\bibnamefont {Schwab}},\
  }\bibfield  {title} {\bibinfo {title} {Low-rank tensor structure of linear
  diffusion operators in the {TT} and {QTT} formats},\ }\href
  {https://doi.org/10.1016/j.laa.2013.01.009} {\bibfield  {journal} {\bibinfo
  {journal} {Linear Algebra and its Applications}\ }\textbf {\bibinfo {volume}
  {438}},\ \bibinfo {pages} {4204} (\bibinfo {year} {2013})}\BibitemShut
  {NoStop}%
\bibitem [{\citenamefont {Askar}\ and\ \citenamefont
  {Cakmak}(1978)}]{Askar1978}%
  \BibitemOpen
  \bibfield  {author} {\bibinfo {author} {\bibfnamefont {A.}~\bibnamefont
  {Askar}}\ and\ \bibinfo {author} {\bibfnamefont {A.~S.}\ \bibnamefont
  {Cakmak}},\ }\bibfield  {title} {\bibinfo {title} {{Explicit integration
  method for the time-dependent Schr\"{o}dinger equation for collision
  problems}},\ }\href {https://doi.org/10.1063/1.436072} {\bibfield  {journal}
  {\bibinfo  {journal} {The Journal of Chemical Physics}\ }\textbf {\bibinfo
  {volume} {68}},\ \bibinfo {pages} {2794} (\bibinfo {year}
  {1978})}\BibitemShut {NoStop}%
\bibitem [{\citenamefont {Verlet}(1967)}]{Verlet1967}%
  \BibitemOpen
  \bibfield  {author} {\bibinfo {author} {\bibfnamefont {L.}~\bibnamefont
  {Verlet}},\ }\bibfield  {title} {\bibinfo {title} {{Computer "Experiments" on
  Classical Fluids. I. Thermodynamical Properties of Lennard-Jones
  Molecules}},\ }\href {https://doi.org/10.1103/PhysRev.159.98} {\bibfield
  {journal} {\bibinfo  {journal} {Physical Review}\ }\textbf {\bibinfo {volume}
  {159}},\ \bibinfo {pages} {98} (\bibinfo {year} {1967})}\BibitemShut
  {NoStop}%
\bibitem [{\citenamefont {Fleck}\ \emph {et~al.}(1976)\citenamefont {Fleck},
  \citenamefont {Morris},\ and\ \citenamefont {Feit}}]{Fleck1976}%
  \BibitemOpen
  \bibfield  {author} {\bibinfo {author} {\bibfnamefont {J.~A.}\ \bibnamefont
  {Fleck}}, \bibinfo {author} {\bibfnamefont {J.~R.}\ \bibnamefont {Morris}},\
  and\ \bibinfo {author} {\bibfnamefont {M.~D.}\ \bibnamefont {Feit}},\
  }\bibfield  {title} {\bibinfo {title} {{Time-dependent propagation of high
  energy laser beams through the atmosphere}},\ }\href
  {https://doi.org/10.1007/BF00896333} {\bibfield  {journal} {\bibinfo
  {journal} {Applied Physics}\ }\textbf {\bibinfo {volume} {10}},\ \bibinfo
  {pages} {129} (\bibinfo {year} {1976})}\BibitemShut {NoStop}%
\bibitem [{\citenamefont {Yoshida}(1990)}]{Yoshida1990}%
  \BibitemOpen
  \bibfield  {author} {\bibinfo {author} {\bibfnamefont {H.}~\bibnamefont
  {Yoshida}},\ }\bibfield  {title} {\bibinfo {title} {{Construction of higher
  order symplectic integrators}},\ }\href
  {https://doi.org/10.1016/0375-9601(90)90092-3} {\bibfield  {journal}
  {\bibinfo  {journal} {Physics Letters A}\ }\textbf {\bibinfo {volume}
  {150}},\ \bibinfo {pages} {262} (\bibinfo {year} {1990})}\BibitemShut
  {NoStop}%
\bibitem [{\citenamefont {Suzuki}(1990)}]{Suzuki1990}%
  \BibitemOpen
  \bibfield  {author} {\bibinfo {author} {\bibfnamefont {M.}~\bibnamefont
  {Suzuki}},\ }\bibfield  {title} {\bibinfo {title} {{Fractal decomposition of
  exponential operators with applications to many-body theories and Monte Carlo
  simulations}},\ }\href {https://doi.org/10.1016/0375-9601(90)90962-N}
  {\bibfield  {journal} {\bibinfo  {journal} {Physics Letters A}\ }\textbf
  {\bibinfo {volume} {146}},\ \bibinfo {pages} {319} (\bibinfo {year}
  {1990})}\BibitemShut {NoStop}%
\bibitem [{\citenamefont {Bandrauk}\ and\ \citenamefont
  {Shen}(1992)}]{Bandrauk1992}%
  \BibitemOpen
  \bibfield  {author} {\bibinfo {author} {\bibfnamefont {A.~D.}\ \bibnamefont
  {Bandrauk}}\ and\ \bibinfo {author} {\bibfnamefont {H.}~\bibnamefont
  {Shen}},\ }\bibfield  {title} {\bibinfo {title} {{Higher order exponential
  split operator method for solving time-dependent Schr{\"{o}}dinger
  equations}},\ }\href {https://doi.org/10.1139/v92-078} {\bibfield  {journal}
  {\bibinfo  {journal} {Canadian Journal of Chemistry}\ }\textbf {\bibinfo
  {volume} {70}},\ \bibinfo {pages} {555} (\bibinfo {year} {1992})}\BibitemShut
  {NoStop}%
\bibitem [{\citenamefont {Blanes}\ \emph {et~al.}(2006)\citenamefont {Blanes},
  \citenamefont {Casas},\ and\ \citenamefont {Murua}}]{Blanes2006}%
  \BibitemOpen
  \bibfield  {author} {\bibinfo {author} {\bibfnamefont {S.}~\bibnamefont
  {Blanes}}, \bibinfo {author} {\bibfnamefont {F.}~\bibnamefont {Casas}},\ and\
  \bibinfo {author} {\bibfnamefont {A.}~\bibnamefont {Murua}},\ }\bibfield
  {title} {\bibinfo {title} {{Symplectic splitting operator methods for the
  time-dependent Schr\"{o}dinger equation.}},\ }\href
  {https://doi.org/10.1063/1.2203609} {\bibfield  {journal} {\bibinfo
  {journal} {The Journal of Chemical Physics}\ }\textbf {\bibinfo {volume}
  {124}},\ \bibinfo {pages} {234105} (\bibinfo {year} {2006})}\BibitemShut
  {NoStop}%
\bibitem [{\citenamefont {Blanes}\ \emph {et~al.}(2015)\citenamefont {Blanes},
  \citenamefont {Casas},\ and\ \citenamefont {Murua}}]{Blanes2015}%
  \BibitemOpen
  \bibfield  {author} {\bibinfo {author} {\bibfnamefont {S.}~\bibnamefont
  {Blanes}}, \bibinfo {author} {\bibfnamefont {F.}~\bibnamefont {Casas}},\ and\
  \bibinfo {author} {\bibfnamefont {A.}~\bibnamefont {Murua}},\ }\bibfield
  {title} {\bibinfo {title} {{An efficient algorithm based on splitting for the
  time integration of the Schr\"{o}dinger equation}},\ }\href
  {https://doi.org/10.1016/j.jcp.2015.09.047} {\bibfield  {journal} {\bibinfo
  {journal} {Journal of Computational Physics}\ }\textbf {\bibinfo {volume}
  {303}},\ \bibinfo {pages} {396} (\bibinfo {year} {2015})}\BibitemShut
  {NoStop}%
\bibitem [{\citenamefont {Hairer}\ \emph {et~al.}(2006)\citenamefont {Hairer},
  \citenamefont {Lubich},\ and\ \citenamefont {Wanner}}]{Hairer2006}%
  \BibitemOpen
  \bibfield  {author} {\bibinfo {author} {\bibfnamefont {E.}~\bibnamefont
  {Hairer}}, \bibinfo {author} {\bibfnamefont {C.}~\bibnamefont {Lubich}},\
  and\ \bibinfo {author} {\bibfnamefont {G.}~\bibnamefont {Wanner}},\
  }\href@noop {} {\emph {\bibinfo {title} {{Geometric Numerical
  Integration}}}}\ (\bibinfo  {publisher} {Springer},\ \bibinfo {address} {New
  York},\ \bibinfo {year} {2006})\BibitemShut {NoStop}%
\bibitem [{\citenamefont {Lubich}(2008)}]{Lubich2008}%
  \BibitemOpen
  \bibfield  {author} {\bibinfo {author} {\bibfnamefont {C.}~\bibnamefont
  {Lubich}},\ }\href@noop {} {\emph {\bibinfo {title} {{From Quantum to
  Classical Molecular Dynamics: Reduced Models and Numerical Analysis}}}}\
  (\bibinfo  {publisher} {European Mathematical Society},\ \bibinfo {address}
  {Z{\"{u}}rich},\ \bibinfo {year} {2008})\BibitemShut {NoStop}%
\bibitem [{\citenamefont {Kahan}\ and\ \citenamefont {Li}(1997)}]{Kahan1997}%
  \BibitemOpen
  \bibfield  {author} {\bibinfo {author} {\bibfnamefont {W.}~\bibnamefont
  {Kahan}}\ and\ \bibinfo {author} {\bibfnamefont {R.-C.}\ \bibnamefont {Li}},\
  }\bibfield  {title} {\bibinfo {title} {Composition constants for raising the
  orders of unconventional schemes for ordinary differential equations},\
  }\href {https://doi.org/10.1090/S0025-5718-97-00873-9} {\bibfield  {journal}
  {\bibinfo  {journal} {Math. Comput.}\ }\textbf {\bibinfo {volume} {66}},\
  \bibinfo {pages} {1089} (\bibinfo {year} {1997})}\BibitemShut {NoStop}%
\bibitem [{\citenamefont {Hochbruck}\ and\ \citenamefont
  {Lubich}(1997)}]{Hochbruck1997}%
  \BibitemOpen
  \bibfield  {author} {\bibinfo {author} {\bibfnamefont {M.}~\bibnamefont
  {Hochbruck}}\ and\ \bibinfo {author} {\bibfnamefont {C.}~\bibnamefont
  {Lubich}},\ }\bibfield  {title} {\bibinfo {title} {On {K}rylov subspace
  approximations to the matrix exponential operator},\ }\href
  {https://doi.org/10.1137/S0036142995280572} {\bibfield  {journal} {\bibinfo
  {journal} {SIAM Journal on Numerical Analysis}\ }\textbf {\bibinfo {volume}
  {34}},\ \bibinfo {pages} {1911} (\bibinfo {year} {1997})}\BibitemShut
  {NoStop}%
\bibitem [{\citenamefont {Al-Mohy}\ and\ \citenamefont
  {Higham}(2011)}]{AlMohy2011}%
  \BibitemOpen
  \bibfield  {author} {\bibinfo {author} {\bibfnamefont {A.~H.}\ \bibnamefont
  {Al-Mohy}}\ and\ \bibinfo {author} {\bibfnamefont {N.~J.}\ \bibnamefont
  {Higham}},\ }\bibfield  {title} {\bibinfo {title} {Computing the action of
  the matrix exponential, with an application to exponential integrators},\
  }\href {https://doi.org/10.1137/100788860} {\bibfield  {journal} {\bibinfo
  {journal} {SIAM Journal on Scientific Computing}\ }\textbf {\bibinfo {volume}
  {33}},\ \bibinfo {pages} {488} (\bibinfo {year} {2011})}\BibitemShut
  {NoStop}%
\bibitem [{\citenamefont {Lanczos}(1950)}]{Lanczos1950}%
  \BibitemOpen
  \bibfield  {author} {\bibinfo {author} {\bibfnamefont {C.}~\bibnamefont
  {Lanczos}},\ }\bibfield  {title} {\bibinfo {title} {An iteration method for
  the solution of the eigenvalue problem of linear differential and integral
  operators},\ }\href {https://doi.org/10.6028/jres.045.026} {\bibfield
  {journal} {\bibinfo  {journal} {J. Res. Natl. Bur. Stand. B}\ }\textbf
  {\bibinfo {volume} {45}},\ \bibinfo {pages} {255} (\bibinfo {year}
  {1950})}\BibitemShut {NoStop}%
\bibitem [{\citenamefont {Blanes}\ \emph {et~al.}(2011)\citenamefont {Blanes},
  \citenamefont {Casas},\ and\ \citenamefont {Murua}}]{Blanes2011}%
  \BibitemOpen
  \bibfield  {author} {\bibinfo {author} {\bibfnamefont {S.}~\bibnamefont
  {Blanes}}, \bibinfo {author} {\bibfnamefont {F.}~\bibnamefont {Casas}},\ and\
  \bibinfo {author} {\bibfnamefont {A.}~\bibnamefont {Murua}},\ }\bibfield
  {title} {\bibinfo {title} {{Error Analysis of Splitting Methods for the Time
  Dependent Schrödinger Equation}},\ }\href
  {https://doi.org/10.1137/100794535} {\bibfield  {journal} {\bibinfo
  {journal} {SIAM Journal on Scientific Computing}\ }\textbf {\bibinfo {volume}
  {33}},\ \bibinfo {pages} {1525} (\bibinfo {year} {2011})}\BibitemShut
  {NoStop}%
\bibitem [{\citenamefont {Riedel}\ \emph {et~al.}(2023)\citenamefont {Riedel},
  \citenamefont {Gel{\ss}}, \citenamefont {Klein},\ and\ \citenamefont
  {Schmidt}}]{Riedel2023}%
  \BibitemOpen
  \bibfield  {author} {\bibinfo {author} {\bibfnamefont {J.}~\bibnamefont
  {Riedel}}, \bibinfo {author} {\bibfnamefont {P.}~\bibnamefont {Gel{\ss}}},
  \bibinfo {author} {\bibfnamefont {R.}~\bibnamefont {Klein}},\ and\ \bibinfo
  {author} {\bibfnamefont {B.}~\bibnamefont {Schmidt}},\ }\bibfield  {title}
  {\bibinfo {title} {{WaveTrain: A Python package for numerical quantum
  mechanics of chain-like systems based on tensor trains}},\ }\href
  {https://doi.org/10.1063/5.0147314} {\bibfield  {journal} {\bibinfo
  {journal} {The Journal of Chemical Physics}\ }\textbf {\bibinfo {volume}
  {158}},\ \bibinfo {pages} {164801} (\bibinfo {year} {2023})}\BibitemShut
  {NoStop}%
\bibitem [{\citenamefont {Gel{\ss}}\ \emph {et~al.}(2021)\citenamefont
  {Gel{\ss}}, \citenamefont {Klus}, \citenamefont {Scherer}, \citenamefont
  {N{\"u}ske},\ and\ \citenamefont {L{\"u}cke}}]{Gelss2021}%
  \BibitemOpen
  \bibfield  {author} {\bibinfo {author} {\bibfnamefont {P.}~\bibnamefont
  {Gel{\ss}}}, \bibinfo {author} {\bibfnamefont {S.}~\bibnamefont {Klus}},
  \bibinfo {author} {\bibfnamefont {M.}~\bibnamefont {Scherer}}, \bibinfo
  {author} {\bibfnamefont {F.}~\bibnamefont {N{\"u}ske}},\ and\ \bibinfo
  {author} {\bibfnamefont {M.}~\bibnamefont {L{\"u}cke}},\ }\href@noop {}
  {\bibinfo {title} {Scikit-{TT}}},\ \bibinfo {howpublished}
  {\url{https://github.com/PGelss/scikit_tt}} (\bibinfo {year}
  {2021})\BibitemShut {NoStop}%
\bibitem [{\citenamefont {Schmidt}\ and\ \citenamefont
  {Lorenz}(2017)}]{Schmidt2017}%
  \BibitemOpen
  \bibfield  {author} {\bibinfo {author} {\bibfnamefont {B.}~\bibnamefont
  {Schmidt}}\ and\ \bibinfo {author} {\bibfnamefont {U.}~\bibnamefont
  {Lorenz}},\ }\bibfield  {title} {\bibinfo {title} {{WavePacket: A Matlab
  package for numerical quantum dynamics. I: Closed quantum systems and
  discrete variable representations}},\ }\href
  {https://doi.org/10.1016/j.cpc.2016.12.007} {\bibfield  {journal} {\bibinfo
  {journal} {Computer Physics Communications}\ }\textbf {\bibinfo {volume}
  {213}},\ \bibinfo {pages} {223} (\bibinfo {year} {2017})}\BibitemShut
  {NoStop}%
\bibitem [{\citenamefont {Schmidt}\ and\ \citenamefont
  {Hartmann}(2018)}]{Schmidt2018}%
  \BibitemOpen
  \bibfield  {author} {\bibinfo {author} {\bibfnamefont {B.}~\bibnamefont
  {Schmidt}}\ and\ \bibinfo {author} {\bibfnamefont {C.}~\bibnamefont
  {Hartmann}},\ }\bibfield  {title} {\bibinfo {title} {{WavePacket: A Matlab
  package for numerical quantum dynamics.II: Open quantum systems, optimal
  control, and model reduction}},\ }\href
  {https://doi.org/10.1016/j.cpc.2018.02.022} {\bibfield  {journal} {\bibinfo
  {journal} {Computer Physics Communications}\ }\textbf {\bibinfo {volume}
  {228}},\ \bibinfo {pages} {229} (\bibinfo {year} {2018})}\BibitemShut
  {NoStop}%
\bibitem [{\citenamefont {Schmidt}\ \emph {et~al.}(2019)\citenamefont
  {Schmidt}, \citenamefont {Klein},\ and\ \citenamefont {{Cancissu
  Araujo}}}]{Schmidt2019}%
  \BibitemOpen
  \bibfield  {author} {\bibinfo {author} {\bibfnamefont {B.}~\bibnamefont
  {Schmidt}}, \bibinfo {author} {\bibfnamefont {R.}~\bibnamefont {Klein}},\
  and\ \bibinfo {author} {\bibfnamefont {L.}~\bibnamefont {{Cancissu
  Araujo}}},\ }\bibfield  {title} {\bibinfo {title} {{WavePacket: A Matlab
  package for numerical quantum dynamics. III. Quantum‐classical simulations
  and surface hopping trajectories}},\ }\href
  {https://doi.org/10.1002/jcc.26045} {\bibfield  {journal} {\bibinfo
  {journal} {Journal of Computational Chemistry}\ }\textbf {\bibinfo {volume}
  {40}},\ \bibinfo {pages} {2677} (\bibinfo {year} {2019})}\BibitemShut
  {NoStop}%
\bibitem [{\citenamefont {Kenkre}\ and\ \citenamefont
  {Phatak}(1984)}]{Kenkre1984}%
  \BibitemOpen
  \bibfield  {author} {\bibinfo {author} {\bibfnamefont {V.}~\bibnamefont
  {Kenkre}}\ and\ \bibinfo {author} {\bibfnamefont {S.}~\bibnamefont
  {Phatak}},\ }\bibfield  {title} {\bibinfo {title} {{Exact probability
  propagators for motion with arbitrary degree of transport coherence}},\
  }\href {https://doi.org/10.1016/0375-9601(84)90673-X} {\bibfield  {journal}
  {\bibinfo  {journal} {Physics Letters A}\ }\textbf {\bibinfo {volume}
  {100}},\ \bibinfo {pages} {101} (\bibinfo {year} {1984})}\BibitemShut
  {NoStop}%
\bibitem [{\citenamefont {Schleich}(2001)}]{Schleich2001}%
  \BibitemOpen
  \bibfield  {author} {\bibinfo {author} {\bibfnamefont {W.~P.}\ \bibnamefont
  {Schleich}},\ }\href@noop {} {\emph {\bibinfo {title} {{Quantum Optics in
  Phase Space}}}}\ (\bibinfo  {publisher} {Wiley--VCH},\ \bibinfo {address}
  {Berlin},\ \bibinfo {year} {2001})\BibitemShut {NoStop}%
\bibitem [{\citenamefont {Cohen-Tannoudji}\ \emph {et~al.}(1977)\citenamefont
  {Cohen-Tannoudji}, \citenamefont {Liu},\ and\ \citenamefont
  {Laloe}}]{Cohen1977}%
  \BibitemOpen
  \bibfield  {author} {\bibinfo {author} {\bibfnamefont {C.}~\bibnamefont
  {Cohen-Tannoudji}}, \bibinfo {author} {\bibfnamefont {B.}~\bibnamefont
  {Liu}},\ and\ \bibinfo {author} {\bibfnamefont {F.}~\bibnamefont {Laloe}},\
  }\href@noop {} {\emph {\bibinfo {title} {{Quantum Mechanics}}}}\ (\bibinfo
  {publisher} {Wiley},\ \bibinfo {address} {New York},\ \bibinfo {year}
  {1977})\BibitemShut {NoStop}%
\bibitem [{\citenamefont {Horenko}\ and\ \citenamefont
  {Weiser}(2003)}]{Horenko2003}%
  \BibitemOpen
  \bibfield  {author} {\bibinfo {author} {\bibfnamefont {I.}~\bibnamefont
  {Horenko}}\ and\ \bibinfo {author} {\bibfnamefont {M.}~\bibnamefont
  {Weiser}},\ }\bibfield  {title} {\bibinfo {title} {{Adaptive integration of
  molecular dynamics.}},\ }\href {https://doi.org/10.1002/jcc.10335} {\bibfield
   {journal} {\bibinfo  {journal} {Journal of Computational Chemistry}\
  }\textbf {\bibinfo {volume} {24}},\ \bibinfo {pages} {1921} (\bibinfo {year}
  {2003})}\BibitemShut {NoStop}%
\bibitem [{\citenamefont {Horenko}\ \emph {et~al.}(2004)\citenamefont
  {Horenko}, \citenamefont {Weiser}, \citenamefont {Schmidt},\ and\
  \citenamefont {Sch{\"{u}}tte}}]{Horenko2004}%
  \BibitemOpen
  \bibfield  {author} {\bibinfo {author} {\bibfnamefont {I.}~\bibnamefont
  {Horenko}}, \bibinfo {author} {\bibfnamefont {M.}~\bibnamefont {Weiser}},
  \bibinfo {author} {\bibfnamefont {B.}~\bibnamefont {Schmidt}},\ and\ \bibinfo
  {author} {\bibfnamefont {C.}~\bibnamefont {Sch{\"{u}}tte}},\ }\bibfield
  {title} {\bibinfo {title} {{Fully adaptive propagation of the
  quantum-classical Liouville equation.}},\ }\href
  {https://doi.org/10.1063/1.1691015} {\bibfield  {journal} {\bibinfo
  {journal} {The Journal of Chemical Physics}\ }\textbf {\bibinfo {volume}
  {120}},\ \bibinfo {pages} {8913} (\bibinfo {year} {2004})}\BibitemShut
  {NoStop}%
\end{thebibliography}%

\end{document}